\documentclass[arxiv, usenatbib]{iupartex}
\usepackage{newtxtext,newtxmath}
\usepackage[T1]{fontenc}
\usepackage{ae,aecompl}

\DeclareUnicodeCharacter{2212}{-} 

\usepackage{multicol}   
\usepackage{bm}		    
\usepackage{pdflscape}	

\title[An Up to date line list for spectroscopic analysis of F and G Stars]{An Updated Line List for Spectroscopic Investigation of G Stars - II: Refined Solar Abundances via Extended Wavelength Coverage to 10\,000 \AA}

\author[\c{S}ahin et. al]{T. \c{S}ahin$^{1\cc,2}$\orcid{0000-0002-0296-233X},
F. G\"{u}ney$^{2}$\orcid{0000-0003-3884-974X},
S.A. \c{S}ent\"{u}rk$^{2}$\orcid{0000-0003-3863-6525},
N. \c{C}ınar$^{2}$\orcid{0000-0002-5155-9280},
and M. Marı\c{s}mak$^{2}$\orcid{0000-0002-9397-2778}
\affsep \\
$^1$ Akdeniz University, Faculty of Science, Department of Space Sciences and Technologies 07058, Antalya, Türkiye\\
$^2$ Institute of Graduate Studies in Science, Akdeniz University, Türkiye
}
\corres{Timur \c{S}ahin}{timursahin@akdeniz.edu.tr}

\pubyear{2024}

\doiheader{XXXXXXX/PAR.20XX.00000}
\date{
	\pSubmit{05.11.2024} 
	\pAccept{27.12.2024}
	\pPubOnl{00.00.0000}
}
\volume{2}
\volnumber{2}

\begin{document}
\label{firstpage}
\pagerange{\pageref*{firstpage}--\pageref*{lastpage}}
\maketitle

\begin{abstract}
This study introduces a line list for the abundance analysis of F-and G-type stars across the 4\,080–9\,675 \AA\ wavelength range. A systematic search employing lower excitation potentials, accurate $\log gf$ values, and an updated multiplet table led to the identification of 592 lines across 33 species (25 elements), including C, O, Mg (ionized), Al, P, S, Cu, Zr (neutral), and La. To determine the uncertainties in $\log gf$ values, we assessed solar abundance using a very high-resolution ($R\approx$1\,000\,000) disk-integrated solar spectrum. These lines were confirmed to be blend-free in the solar spectrum. The line list was further validated by analyzing the metal-poor star HD\,218209 (G6V), which is notable for its well-documented and reliable abundance in literature. The abundances were obtained using the equivalent width (EW) method and further refined by applying the spectrum synthesis method. A comparative analysis with the {\it Gaia}-ESO line list v.6, provided by the {\it Gaia}-ESO collaboration, revealed additional neutral and ionized Fe lines. This extensively refined line list will facilitate precise stellar parameter determinations and accurate abundance analyses of spectra within the {\sc PolarBASE} spectral library.
\end{abstract}


\section{Introduction}
Advancements in spectroscopic methodologies for G-type stars have enabled more precise elemental abundance measurements. High-resolution spectroscopic techniques enable researchers to analyze stellar spectra in detail, providing insights into their atmospheric compositions and the underlying nucleosynthesis processes \citep{sharma2018, trevisan2021}. Analysis of G-dwarfs revealed discrepancies between the observed and predicted abundance patterns, challenging existing galactic chemical evolution models \citep{woolf2012}. These findings highlights the importance of combining improved modelling techniques with high-resolution spectroscopic data.

G-type stars, including the Sun, serve as fundamental benchmarks for understanding the stellar evolution and galactic chemical history \citep{bensby2003,heiter2015}. Their relatively long lifetimes allow them to retain the chemical signatures of the molecular clouds from which they form \citep{bensby2003,heiter2015,aoki2022}. Solar photospheric abundances, derived from spectroscopic observations, provide a reference point for abundance determination in metal-poor stars and are essential for understanding the processes that govern stellar and galactic evolution  \citep{lodders2003,pagel1975}. Recent studies have significantly advanced our understanding of solar abundance by incorporating various physical processes, such as gravitational settling, convective overshooting, solar wind mass loss, pre-main-sequence disk accretion, opacity, and helium abundance in the solar corona  \citep{wang2013,zhang2019,asplund2021,salmon2021,magg2022}. 

Migration complicates the interpretation of their origins because it can result in metal-poor stars being found in regions where they are not typically expected \citep{haywood2008}. \cite{zhang2019} explored the implications of convective overshoot, solar-wind mass loss, and pre-main-sequence disk accretion on solar models. Their findings indicate that incorporating additional physical processes significantly improves the alignment between solar models and helioseismic constraints, effectively addressing the solar abundance problem. \cite{magg2022} demonstrated how updated abundances can influence the internal solar structure. 

\cite{asplund2021} presented the updated solar photospheric and proto-solar abundances of 83 elements. Their work highlighted the so-called solar modelling problem, which refers to the persistent discrepancies between helioseismic observations and solar interior models constructed with low metallicity. This suggests that there may be shortcomings in the computed opacities or the treatment of mixing processes below the convection zone in the existing models. The updated abundances are essential for refining our understanding of the solar structure and evolution, as they provide a more accurate baseline for the solar modelling problem. 

Moreover, the variability in helium abundance in the solar corona, as discussed by \citet{ofman2024}, also plays a role in understanding solar atmospheric processes. This variability is crucial for interpreting solar observations and for understanding the dynamics of the solar atmosphere. This study presents a three-dimensional model that illustrates the influence of solar activity and coronal heating processes on helium abundance. 

These updates are essential for addressing the solar modelling problem and refining our understanding of the solar structure and evolution. Addressing this complex problem requires precise atmospheric modeling supported by comprehensive and accurate line lists.

The author's research team has been actively studying G-type stars, particularly those in solar neighborhoods. In our previous work \citep[hereafter Paper I]{Sahin2023}, we presented a line list covering the 4\,080-6\,780 \AA\ wavelength range designed for the spectroscopic analysis of more than 90 G-type metal-poor stars residing within the solar neighborhood. Previous studies by the research team, such as \cite{Marismak2024} and \cite{Senturk2024}, also utilized the line list presented in Paper I. For instance, \cite{Marismak2024} employed this line list to analyze two metal-poor high-proper motion stars, HD\,8724 and HD\,195633, whereas \cite{Senturk2024} used it for spectroscopic analysis of a solar analogue star in the optical region. 

Building on this foundation, we now extend the wavelength coverage of the line list to 10\,000 \AA, enabling a more comprehensive spectroscopic analysis of G-type stars, particularly in the near-infrared region. \cite{Senturk2024} presented a line list covering the 10\,000-25\,000 \AA\ range, which will be valuable for future spectroscopic studies of G-type stars, including solar analogue and solar twin stars in the $H$- and $K$-bands.

The remainder of this paper is organized as follows. Section 2 provides the observational data. Section 3 explains the methodology, including line identification and measurement procedures, the determination of model parameters, and the techniques for chemical abundance analysis of both HD\,218209 and the Sun. Section 4 presents the line list, including details on line identification, measurement, and the atomic data used in the analysis. Finally, Section 5 summarizes our findings and discusses their implications.

\section{Observations}
\begin{figure*}[h!]
\centering
\includegraphics[width=0.92\linewidth]{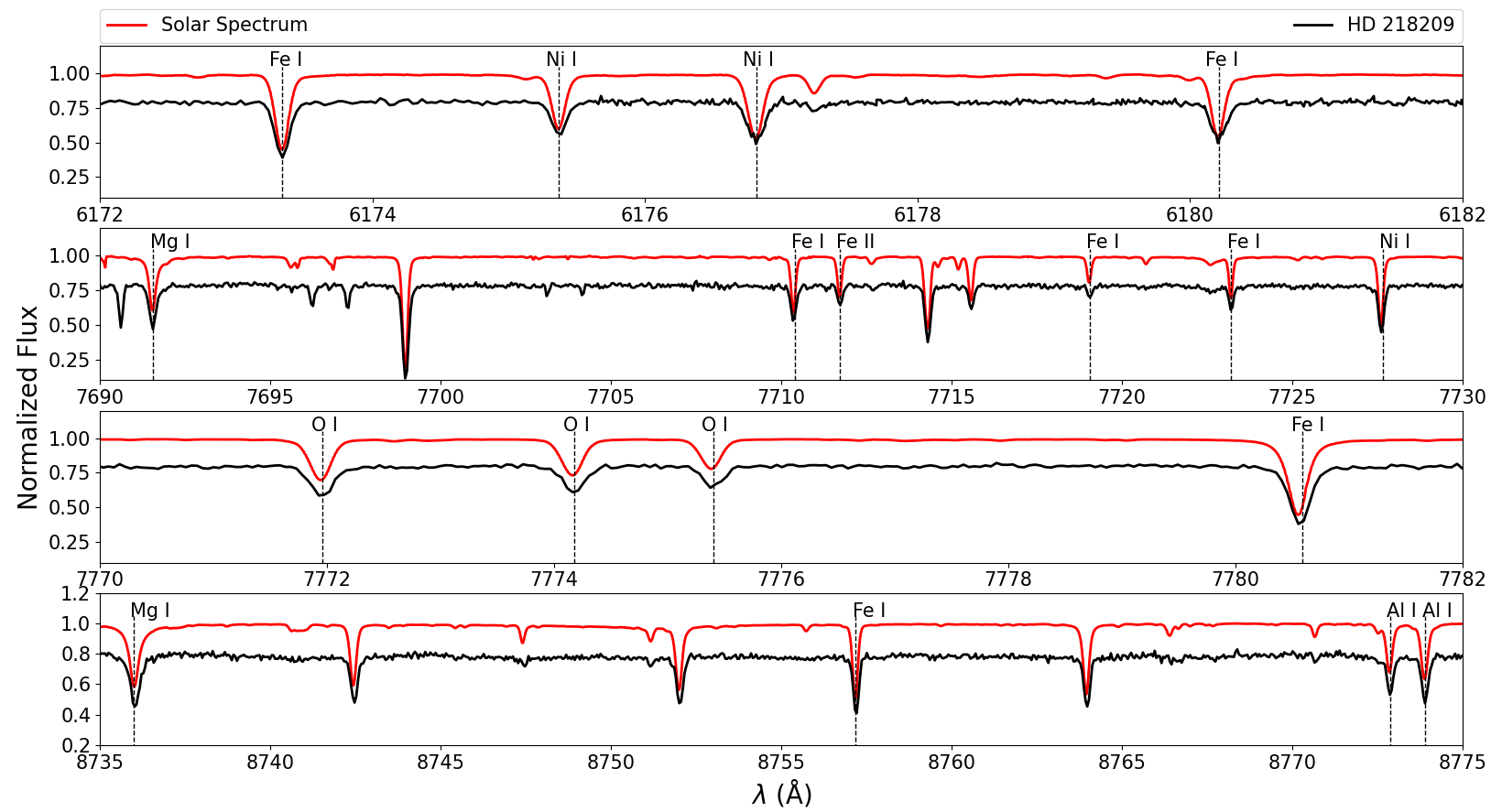}\vspace*{-4pt}
\caption{A small region of the KPNO solar spectrum and the {\sc PolarBASE} spectrum of HD\,218209. Identified lines are also indicated.}
\label{fig:all_spectra}
\end{figure*}

This study analyzes high-resolution spectra of the Sun and HD 218209 to develop and validate a line list. Compared with Paper I, this study significantly expands the scope of spectral analyses by extending the analysis to the near-infrared region. For HD\,218209, a high-resolution ($R \approx$ 76\,000) and high signal-to-noise ratio ($S/N = 156$) {\sc PolarBASE}\footnote{\url{http://polarbase.irap.omp.eu}} \citep{petit2014} Narval\footnote{Narval spectropolarimeter is adapted to the 2m Bernard Lyot telescope and provides high-resolution spectral and polarimetric data.} spectrum (HJD 2456232.48238; exposure time of 400 s) obtained from the PolarBASE archive. The characteristics of the HD\,218209’s spectrum and KPNO solar spectrum are displayed in Figure \ref{fig:all_spectra}. 

The spectrum was continuum-normalized and corrected for radial velocity ($V_{\rm Rad}$) before line measurements. The Python interface and synthetic Narval solar spectra, which include atomic transitions in the range of 3\,700–10\,048 \AA\, were used for RV correction ($V_{\rm Rad}=$ 16.03 km s$^{\rm -1}$), and the renormalization process was performed using the {\sc LIME} code developed in the IDL environment \citep{sahin2017}. Lines with equivalent widths (EW) below 5 m\AA\, or above 200 m\AA\, were excluded from the analysis.

\begin{figure*}[h!]
\centering
\includegraphics[width=0.92\linewidth]{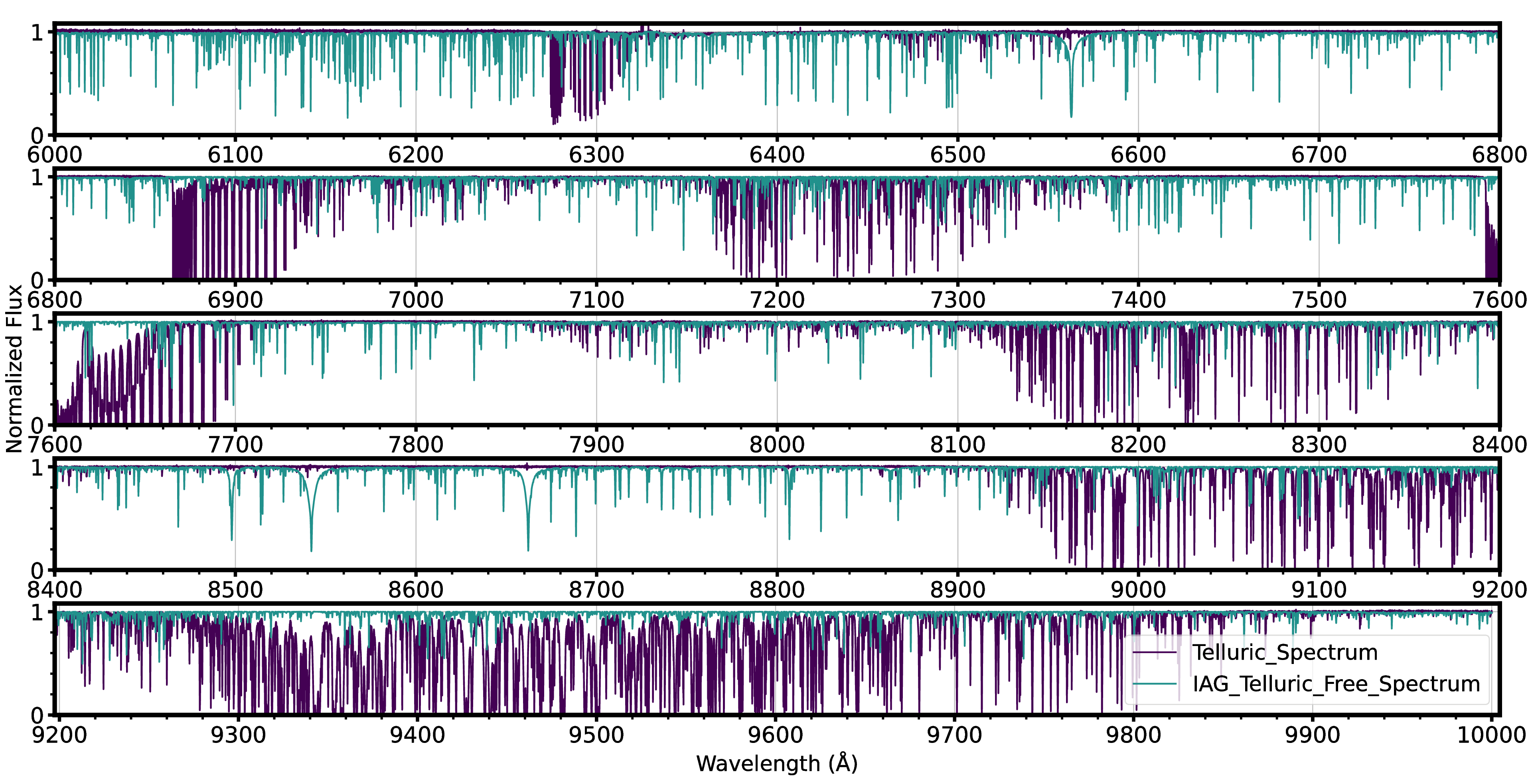}
\caption{The telluric corrected Göttingen (IAG) Solar Spectrum (BTFS). Telluric spectrum  (from \url{https://zenodo.org/records/3598136}) was also included to indicate the positions of the telluric lines. The telluric model shown is typical of the conditions at Göttingen (precipitable water vapour of $\approx$10 mm), where the VVT telescope resides.
}
\label{fig:telluric}
\end{figure*}

The solar spectrum serves as a fundamental reference for stellar astrophysics and analysis of physical processes in stars \citep{Molaro2012}. In this study, high-resolution ($R \approx$ 700\,000) Kitt Peak Fourier Transform Spectrometer (FTS) data (disk-integrated) obtained by \citet{Kurucz1984}, previously utilized by \citet{Sahin2023}, and a very high-resolution ($R \approx$ 1\,000\,000) disk-integrated Göttingen (IAG)\footnote{IAG: Institute for Astrophysics, Göttingen.} solar flux atlas\footnote{BTFS; \url{https://zenodo.org/records/3598136}} obtained by \citet{Baker2020} with Vacuum Vertical Telescope (VVT) were used. However, it should be noted that an alternative link\footnote{zenodo; \url{https://web.sas.upenn.edu/ashbaker/solar-atlas/}} was also provided by \citet{Baker2020}. Differences\footnote{Ashley Baker; private communication} were observed between the two spectra (see Appendix for Figure \ref{fig:zenodo_baker}). The KPNO solar spectrum was used for analyses in the 4\,000-5\,000 \AA\, range, while the telluric-free IAG solar spectrum (BTFS) was preferred for the 5000-10000 \AA\, range. Hence, both solar spectra have enabled line identification and other classical spectral analysis methods over the entire 4\,000-10\,000 \AA\ wavelength range. Although the KPNO spectrum is reliable, it contains telluric lines within the {\sc ELODIE} wavelength range; in particular, around 6\,000 \AA. In the longer wavelength regions, telluric bands caused by H$_{\rm 2}$O and molecular O$_{\rm 2}$ are prominent (see Figure \ref{fig:telluric} for details). In the KPNO solar spectrum, transitions outside the regions dominated by telluric lines were considered for the line list created in Paper I of the series, which covered 4\,000-6\,800 \AA\ range. The 5\,000-6\,800 \AA\ wavelength region is common between the KPNO and IAG (BTFS) solar spectra. We compared the equivalent widths (EW) of the lines in this region and found that the EW measurements of the two spectra were in good agreement [EW(KPNO) = (0.956$\pm$0.011)$\times$ EW(IAG)+(2.353$\pm$0.839)].

\section{The Abundance Analysis}
The elemental abundances were determined using the local thermodynamic equilibrium (LTE) line analysis code, {\sc MOOG} \citep{sneden1973}\footnote{The source code for {\sc MOOG} can be accessed at~\url{http://www.as.utexas.edu/chris/moog.html}}. Model atmospheres were generated using {\sc ATLAS9} code \citep{Castelli2003} with the LTE (ODFNEW) approach. Detailed descriptions of the abundance analysis procedure have been provided by \cite{sahin2009}, \cite{sahin2011}, \cite{sahin2016}, \cite{sahin2020}, and \cite{Sahin2023}. 
\begin{figure*}[h!]
\centering
\includegraphics[width=\columnwidth]{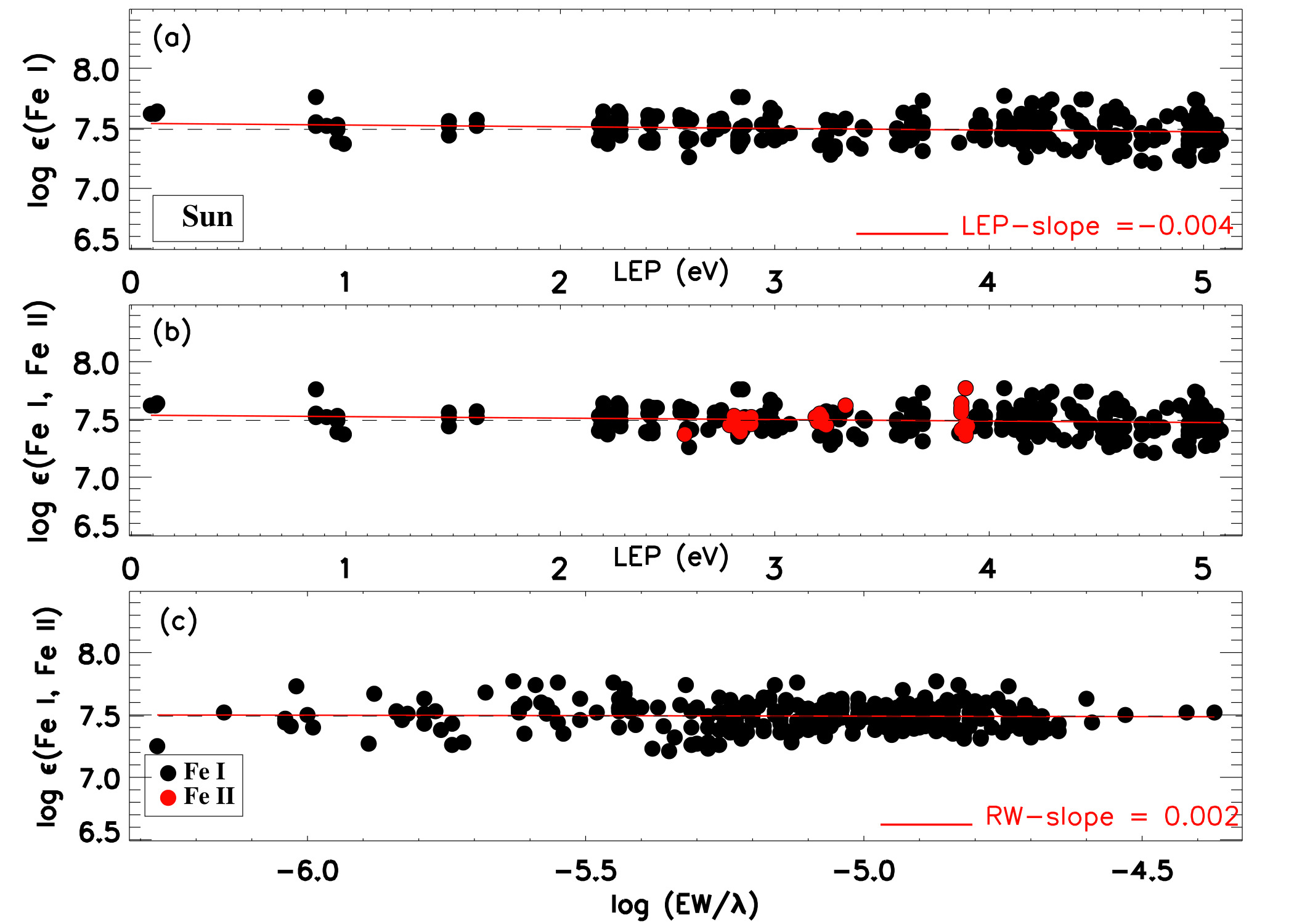}
\includegraphics[width=\columnwidth]{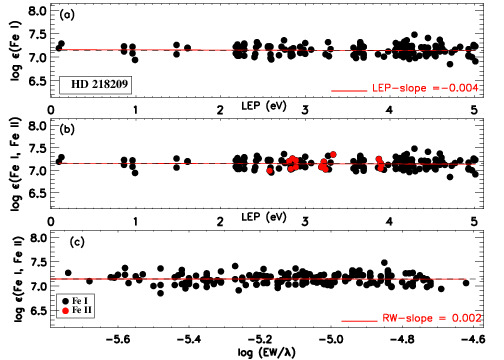}
\caption{An example for the determination of the atmospheric parameters $T_{\rm eff}$ and $\xi$ using abundance ($\log\epsilon$) as a function of both lower LEP (panels a and b) and reduced EW (panels c) for the Sun and HD\,218209. In all panels, the solid red line represents the least-squares fit to the data.}
\label{fig:sun_model_param}
\end{figure*}
The atmospheric parameters of the model, such as the effective temperature ($T_{\rm eff}$), surface gravity ($\log g$), metallicity ([Fe/H]), and microturbulent velocity ($\xi$), were determined using neutral (Fe\,{\sc i}) and ionized (Fe\,{\sc ii}) iron lines in an iterative process. The $T_{\rm eff}$ determination employed the excitation balance method (sensitive to neutral spectral lines with a broad range of excitation potentials) for Fe\,{\sc i}. $\xi$ represents the small-scale gas motion within the stellar atmosphere. $\xi$ was determined by ensuring that the abundance of Fe atoms (Fe\,{\sc i}) remained independent of the reduced equivalent width (EW/$\lambda$) under the assumption of LTE. These two conditions were simultaneously applied to a set of Fe\,{\sc i} lines (see Figure \ref{fig:sun_model_param}, upper and middle panels). In addition, $\xi$ is determined using a dispersion test for a given model atmosphere (Figure \ref{fig:kiel diagrams}). This involved computing the dispersion in abundance (Fe, Ti, Cr) over the range of 0.0 to 3.0 km s$^{\rm −1}$. By combining both methods, the measurement uncertainty for $\xi$ was estimated as 0.5 ~ km s$^{-1}$ (Figure \ref{fig:kiel diagrams}). In the same figure, an example Kiel diagram is included.

Surface gravity ($\log g$) was determined by analyzing Fe abundances calculated with {\sc MOOG}, ensuring ionization equilibrium where Fe\,{\sc i} and Fe\,{\sc ii} lines yield the same abundance. Notably, in the solar spectrum, ionization equilibrium is achieved between the neutral and ionized atoms of Mg, Sc, Ti, Cr, and Zr. Similarly, in the spectrum of HD 218209, in addition to Fe, ionization equilibrium is reached for Ti and Cr. Finally, the metallicity ([Fe/H]) was refined through an iterative process to achieve convergence between the derived Fe abundance and the initial abundance adopted for the model atmosphere construction. Convergence was achieved by adjusting $T_{\rm eff}$, $\log g$, and $\xi$ of the model. Figure \ref{fig:sun_model_param} illustrates a summary of the relationship between the physical parameters used to determine the stellar model parameters using the classical spectroscopic method (i.e., ionization and excitation equilibria of the Fe lines) for the Sun (left panel) and HD 218209 (right panel).
\begin{table}[h!]
\setlength{\tabcolsep}{3pt}
\renewcommand{\arraystretch}{1.3}
\caption{Model atmosphere parameters for HD\,218209, and the Sun.}
\label{tab:model_param}
\centering
\begin{tabular}{lcccc}
\hline
Star	&	$T_{\rm eff}$	&	$\log g$ 	&	[Fe/H] &  $\xi$ \\
    & (K) & (cgs) & (dex) & (km s$^{\rm -1}$) \\
    \hline
HD\,218209 & 5600$_{-177}^{+177}$ & 4.50$_{-0.24}^{+0.24}$ & -0.36	$_{-0.13}^{+0.13}$ & 0.44 $_{-0.50}^{+0.50}$ \\
Sun $^{\rm \dag}$        & 5770$_{-130}^{+130}$ &	4.40$_{-0.19}^{+0.19}$	&  0.00	$_{-0.09}^{+0.09}$ & 0.66 $_{-0.50}^{+0.50}$ \\
Sun $^{\rm \ast}$       & 5790$_{-45}^{+45}$ &	4.40$_{-0.09}^{+0.09}$	&  0.00	$_{-0.04}^{+0.04}$ & 0.66 $_{-0.50}^{+0.50}$ \\
\hline
\end{tabular}
\begin{list}{}{}
    \item ($^{\rm \dag}$): This study (TS), the solar spectrum was provided by \cite{Baker2020}.
     \item ($^{\rm \ast}$): The atmospheric parameters from \cite{Sahin2023}. The solar spectrum was obtained from \cite{Kurucz1984}.
\end{list}
\end{table}

The uncertainty in the derived $T_{\rm eff}$ originates from the error associated with the slope of the relationship between the Fe\,{\sc i} abundance and the LEPs of the lines. Additionally, a 1$\sigma$ difference in abundance ([X/H]) between the Fe\,{\sc i} and Fe\,{\sc ii} lines corresponds to a change in 0.19 dex in $\log g$. Table \ref{tab:model_param} summarizes the resulting model parameters for HD\,218209 and the Sun. The uncertainties in the atomic data ($\log gf$ values) were assessed by deriving solar abundances from the stellar spectral lines. 
\begin{table*}[h!]
\small
\setlength{\tabcolsep}{0.3pt}
\renewcommand{\arraystretch}{1.1}
\caption{The abundances of the observed species for Sun and HD\,218209. The solar abundances obtained in this study and those reported by \citet[][ASP09]{asplund2009}  and \citet[][ASP21]{asplund2021} are also provided. Abundances in bold are those calculated via the spectrum synthesis method.}
\label{table:abund}
\centering
\begin{tabular}{l|ccc|ccc|ccc|ccc|ccc}
\hline
     	& \multicolumn{3}{c}{HD\,218209}  & \multicolumn{12}{|c}{Sun}   \\
\cline{1-16}
Species   &  [$X$/Fe]$^{\rm \dag}$ & $\sigma$ & $n$ & $\log\epsilon_{\rm \odot}(X^{\rm \dag}$) & $\sigma$ & $n$ & $\log\epsilon_{\rm \odot}(X^{\rm \ast}$) & $\sigma$ & $n$ &$\log\epsilon_{\rm \odot}(X_{\rm ASP09}$) & $\sigma$ &   $\Delta\log\epsilon_{\rm \odot}(X_{1}$) &$\log\epsilon_{\rm \odot}(X_{\rm ASP21}$) & $\sigma$ &  $\Delta\log\epsilon_{\rm \odot}(X_{2}$) \\
\cline{1-16}
  &  (dex) &   &   & (dex)   &  &   & (dex) & & & (dex) & & (dex) &(dex) & &(dex) \\
\cline{1-16} 
\textbf{C\,{\sc i}}	&	\bf 0.14	&	\bf 0.22	&	\bf 2	&	\bf 8.48	&	\bf 0.11	&	\bf 2	&	--	&	--	&	--	&	8.43	&	0.05 	&	0.05 & 8.46 & 0.04	& 0.02\\
\textbf{O\,{\sc i}}	&	\bf 0.28	&	\bf 0.15	&	\bf 3	&	\bf 8.81	&	\bf 0.03	&	\bf 3	&	--	&	--	&	--	&	8.69	&	0.05 &	0.12 & 8.69 & 0.04 & 0.12\\
Na\,{\sc i}	&	-0.03	&	0.20	&	4	&	6.22	&	0.12	&	4	&	6.16	&	0.07	&	2	&	6.24	&	0.04	&	-0.02	& 6.22 & 0.03 & 0.00\\
\textbf{Mg\,{\sc i}}	&	\bf 0.24	&	\bf 0.16	&	\bf 5	&	\bf 7.62	&	\bf 0.03	&	\bf 5	&	7.60	&	0.08	&	2	&	7.60	&	0.04	&	0.02 & 7.55 & 0.03 & 0.07 \\
\textbf{Mg\,{\sc ii}}	&  --	&	 --	&	 --	&	\bf 7.63	&	\bf 0.00	&	\bf 1	&	--	&	--	&	--	&	7.60	&	0.04	&	0.03 & 7.55 & 0.03 &	0.08\\
\textbf{Al\,{\sc i}}	&	\bf 0.13	&	\bf 0.16	&	\bf 8	&	\bf 6.43	&	\bf 0.03	&	\bf 8	&	--	&	--	&	--	&	6.45	&	0.03	&	-0.02	& 6.43 & 0.03 & 0.00\\
Si\,{\sc i}	&	0.13	&	0.18	&	16	&	7.50	&	0.09	&	21	&	7.50	&	0.07	&	12	&	7.51	&	0.03	&	-0.01	& 7.51 & 0.03 & -0.01\\
\textbf{P\,{\sc i}}	&	--	&	--	&	--	&	\bf 5.51	&	\bf 0.06	&	\bf 3	&	--	&	--	&	--	&	5.41	&	0.03	&	0.10	& 5.41 & 0.03 & 0.10\\
\textbf{S\,{\sc i}}	&	--	&	--	&	--	&	\bf 7.15	&	\bf 0.00	&	\bf 2	&	--	&	--	&	--	&	7.12	&	0.03	&	0.03	& 7.12 & 0.03 & 0.03\\
Ca\,{\sc i}	&	0.15	&	0.20	&	15	&	6.29	&	0.09	&	21	&	6.34	&	0.08	&	18	&	6.34	&	0.04	&	-0.05	& 6.30 & 0.03 & -0.01\\
\textbf{Sc\,{\sc i}}	&	--	&	--	&	--	&	\bf 3.13	&	\bf 0.00	&	\bf 1	&	3.12	&	0.00	&	1	&	3.15	&	0.04	&	-0.02	& 3.14 & 0.04& -0.01\\
Sc\,{\sc ii}	&	0.06	&	0.14	&	2	&	3.14	&	0.02	&	12	&	3.23	&	0.08	&	7	&	3.15	&	0.04	&	-0.01	& 3.14 & 0.04& 0.00\\
Ti\,{\sc i}	&	0.21	&	0.21	&	44	&	4.93	&	0.09	&	63	&	4.96	&	0.09	&	43	&	4.95	&	0.05	&	-0.02	& 4.97 & 0.05 & -0.04\\
Ti\,{\sc ii}	&	0.20	&	0.21	&	7	&	5.01	&	0.11	&	11	&	4.99	&	0.08	&	12	&	4.95	&	0.05	&	0.06 & 4.97 & 0.05 &0.04	\\
\textbf{V\,{\sc i}}	&	\bf 0.01	&	\bf 0.15	&	\bf 3	&	\bf 3.90	&	\bf 0.03	&	\bf 5	&	3.99	&	0.05	&	5	&	3.93	&	0.08	&	-0.03	& 3.90 & 0.08 & 0.00\\
Cr\,{\sc i}	&	-0.02	&	0.19	&	17	&	5.68	&	0.09	&	29	&	5.71	&	0.07	&	19	&	5.64	&	0.04	&	0.04	& 5.62 & 0.04 & 0.06\\
Cr\,{\sc ii}	&	0.01	&	0.20	&	3	&	5.64	&	0.11	&	4	&	5.64	&	0.14	&	3	&	5.64	&	0.04	&	0.00	& 5.62 & 0.04 & 0.02\\
\textbf{Mn\,{\sc i}}	&	\bf -0.27	&	\bf 0.18	&	\bf 14	&	\bf 5.45	&	\bf 0.08	&	\bf 14	&	5.62	&	0.13	&	13	&	5.43	&	0.05	&	0.02	&  5.42 & 0.06 &0.03\\
Fe\,{\sc i}	&	0.01	&	0.21	&	152	&	7.50	&	0.11	&	252	&	7.54	&	0.09	&	132	&	7.50	&	0.04	&	0.00	& 7.46 &0.04 & 0.04\\
Fe\,{\sc ii}	&	0.00	&	0.20	&	17	&	7.50	&	0.09	&	32	&	7.51	&	0.04	&	17	&	7.50	&	0.04	&	0.00 & 7.46 & 0.04 & 0.04	\\
\textbf{Co\,{\sc i}}	&	\bf -0.10	&	\bf 0.17	&	\bf 6	&	\bf 4.95	&	\bf 0.06	&	\bf 8	&	--	&	--	&	--	&	4.99	&	0.07	&	-0.04 & 4.94 & 0.05 &	0.01\\
Ni\,{\sc i}	&	-0.02	&	0.20	&	45	&	6.25	&	0.10	&	66	&	6.28	&	0.09	&	54	&	6.22	&	0.04	&	0.03	& 6.20 & 0.04 & 0.05\\
\textbf{Cu\,{\sc i}}	&	\bf -0.13	&	\bf 0.20	&	\bf 3	&	\bf 4.20	&	\bf 0.06	&	\bf 4	&	--	&	--	&	--	&	4.19	&	0.04	&	0.01	& 4.18 & 0.05 & 0.02\\
\textbf{Zn\,{\sc i}}	&	\bf 0.20	&	\bf 0.15	&	\bf 2	&	\bf 4.63	&	\bf 0.02	&	\bf 2	&	4.68	&	0.03	&	2	&	4.56	&	0.05	&	0.07	& 4.56 & 0.05 & 0.07\\
\textbf{Sr\,{\sc i}}	&	\bf -0.18 &	\bf 0.14	&	\bf 1	&	\bf 2.84	&	\bf 0.00	&	\bf 1	&	2.91	&	0.00	&	1	&	2.87	&	0.07	&	-0.03 & 2.83 & 0.06 & 0.01	\\
\textbf{Y\,{\sc ii}}	&	\bf -0.14	&	\bf 0.15	&	\bf 2	&	\bf 2.28	&	\bf 0.02	&	\bf 2	&	2.29	&	0.05	&	2	&	2.21	&	0.05	&	0.07	& 2.21 & 0.05 & 0.07\\
\textbf{Zr\,{\sc i}}	&	--	&	--	&	--	&	2.53	&	0.00	&	1	&	--	&	--	&	--	&	2.58	&	0.04	&	-0.05	& 2.59 & 0.04 &-0.06\\
\textbf{Zr\,{\sc ii}}	&	0.04	&	0.14	&	1	&	2.61	&	0.02	&	2	&	2.68	&	0.00	&	1	&	2.58	&	0.04	&	0.03	& 2.59 & 0.04 & 0.02\\
\textbf{Ba\,{\sc ii}}	&	\bf 0.04	&	\bf 0.14	&	\bf 2	&	\bf 2.32	&	\bf 0.02	&	\bf 2	&	2.24	&	0.06	&	4	&	2.18	&	0.09	&	0.14	& 2.27 & 0.05 & 0.05\\
\textbf{La\,{\sc ii}}	&	\bf 0.03	&	\bf 0.16	&	\bf 2	&	\bf 1.14	&	\bf 0.05	&	\bf 3	&	--	&	--	&	--	&	1.10	&	0.04	&	0.04 & 1.11 & 0.04 &0.03	\\
\textbf{Ce\,{\sc ii}}	&	\bf 0.26	&	\bf 0.15	&	\bf 1	&	\bf 1.60	&	\bf 0.04	&	\bf 3	&	1.64	&	0.02	&	2	&	1.58	&	0.04	&	0.02 & 1.58 & 0.04 & 0.02	\\
\textbf{Nd\,{\sc i}}	&	\bf 0.08	&	\bf 0.15	&	\bf 1	&	\bf 1.36	&	\bf 0.03	&	\bf 3	&	1.42	&	0.05	&	3	&	1.42	&	0.04	&	-0.06 & 1.42 & 0.04 & -0.06	\\
\textbf{Sm\,{\sc ii}}	&	\bf 0.14	&	\bf 0.14	&	\bf 1	&	\bf 0.95 &	\bf 0.02	&	\bf 2	&	0.96	&	0.00	&	1	&	0.96	&	0.04	&	-0.01 & 0.95 & 0.04 &	0.00\\
\hline
\end{tabular}
\\
($^{\dag}$): This study, ($^{\ast}$): \citet{Sahin2023}, 
      $\Delta \log \epsilon_{\odot}(X_{1})=\log \epsilon_{\odot}(X^{\rm \dag}) - \log \epsilon_{\odot}(X_{\rm ASP09})$, $\Delta \log \epsilon_{\odot}(X_{2})=\log \epsilon_{\odot}(X^{\rm {\dag}}) - \log \epsilon_{\odot}(X_{\rm ASP21})$
\end{table*}
The solar model derived from our analysis yielded the following atmospheric parameters: $T_{\rm eff}=$ 5770 K, $\log g =$ 4.40 cgs, [Fe/H] = 0.00 dex, and  $\xi$ = 0.66 km s$^{\rm -1}$. These values are in good agreement with the standard solar models. The abundances obtained for the solar photosphere as a result of solar analysis were calculated using these model parameters (Table \ref{tab:model_param}). In Table \ref{tab:model_param}, the solar abundances reported by \citet{asplund2009, asplund2021} are also included. In Table \ref{table:abund}, we provide a summary of element abundances based on the model parameters in LTE. log $\epsilon$ is the logarithm of abundance. The errors reported in log $\epsilon$ abundances are represented by 1$\sigma$ line-to-line scatter in abundance. [X/H] is the logarithmic abundance ratio of hydrogen to the corresponding solar values and [X/Fe] is the logarithmic abundance considering the abundance of Fe\, {\sc i}. The error in [X/Fe] is the square root of the sum of the quadratures of the errors in [X/H] and [Fe/H]. Table \ref{table:abund} presents the abundances obtained using PolarBase spectrum of the star as a function of the [X/Fe] ratio. 
\begin{figure*}[h!]
\centering
\includegraphics[width=0.9\linewidth]{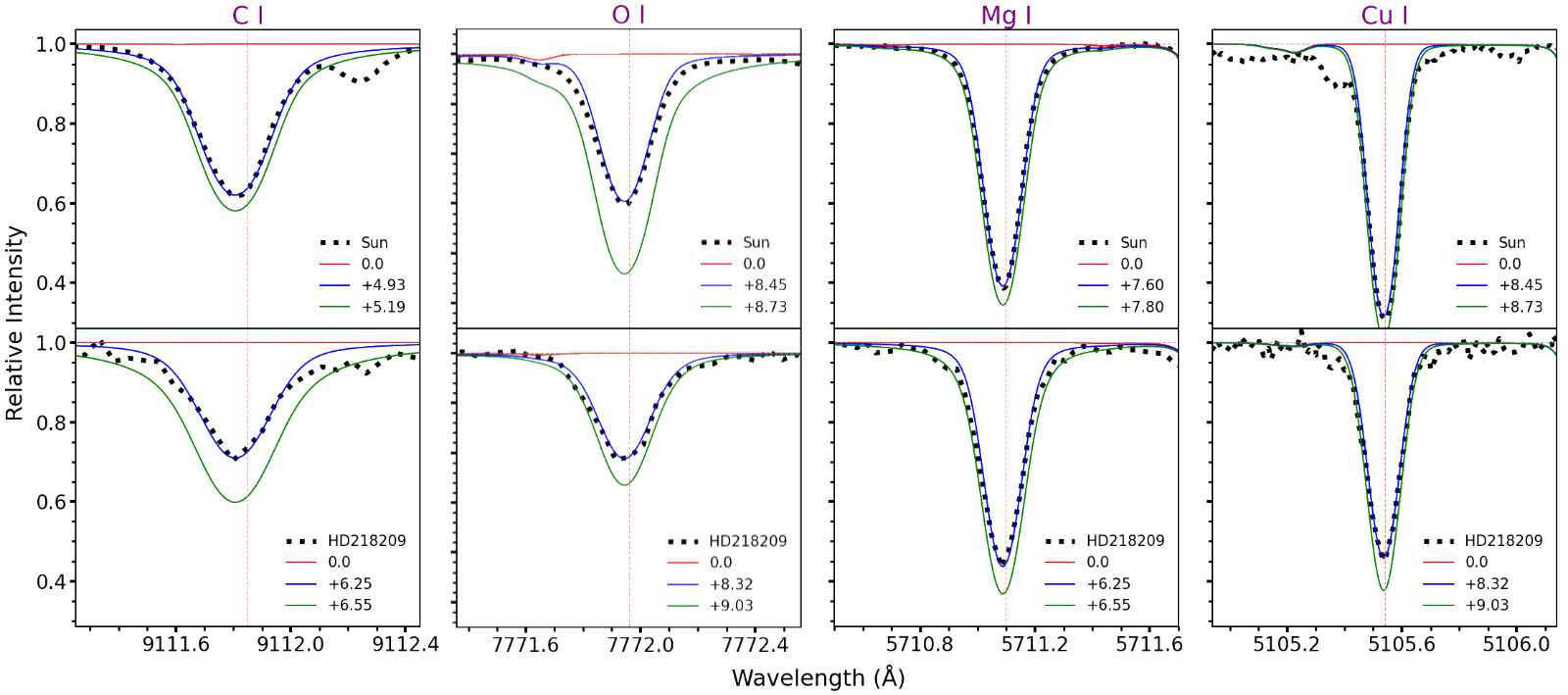}
\caption{The figure presents observed (filled circles) and computed (full blue line) line profiles for C\,{\sc i} 9\,111 \AA, O\,{\sc i} 7\,772 \AA, Mg\,{\sc i} 5\,711 \AA, and Cu\,{\sc i} 5\,105 \AA\, in both the Sun (upper panels) and HD\,218209 (bottom panels). The computed profiles represent the synthetic spectra derived from three logarithmic abundances. The red lines depict the spectra computed without considering the contributions from ionized metal lines.}
\label{fig:cu_o_al}
\end{figure*}

An analysis of the chemical abundances of 33 species belonging to 27 elements, as presented in Table \ref{table:abund}, was consistent with the solar chemical abundances established by \cite{asplund2009,asplund2021}. The abundances of C, O, Mg, Al, P, S, Sc, Ti, V, Mn, Co, Cu, Zn, Sr, Y, Zr, Ba, La, Ce, Nd, and Sm reported in Table \ref{table:abund} were determined using both the equivalent width (EW) method and spectrum synthesis techniques. The synthetic spectra calculated for some sample lines (C\,{\sc i} 9\,111 \AA, O\,{\sc i} 7\,772 \AA, Mg\,{\sc i} 5\,711 \AA, and Cu\,{\sc i} 5\,105 \AA\,), whose elemental abundances were checked using the spectrum synthesis technique, are shown in Figure \ref{fig:cu_o_al}. On the other hand, when compared to the solar abundances reported by \cite{asplund2009}, the scatter among the elements ranges from -0.07 dex for Na to 0.16 dex for O. For the remaining 31 species, the average scatter in abundance ($\log\epsilon_{\rm \odot}(X_{\rm ASP09}$)) is 0.02$\pm$0.04 dex. \cite{asplund2021} presented a revised solar chemical composition, with notable changes observed in the abundance of elements such as Ba, Mg, Co, Sr, Fe, and Ca. For instance, the abundance value obtained for Ba is 0.11 dex higher than that reported by \cite{asplund2009} but shows better agreement with the values presented by \cite{asplund2021}. Similarly, a lower scatter was observed for Na compared to the results of  \cite{asplund2009}.

The results can be affected by various systematic uncertainties, including those related to the correction of non-LTE effects on the formation of convection and atomic transitions. To investigate the potential convective effect, two different mixing length parameters ($\alpha$) were calculated in this study using equations based on 2D hydrodynamic models from \citet{Ludwig1999} and 3D hydrodynamic models from \citet{Magic2015}. The formula by \citet{Magic2015} yielded an $\alpha$ value of 1.99, whereas the formula by \citet{Ludwig1999} yielded an $\alpha$ value of 1.60. Two different {\sc ATLAS9} models were constructed for the two mixing-length parameters. The synthetic spectra calculated using these models were compared to the observed spectrum of HD\,218209. Although no significant difference was observed, the synthetic spectrum derived from the mixing length parameter obtained by \citet{Magic2015} was found to be in slightly better agreement with the observed spectrum.

Given that the Fe\,{\sc i} and Fe\,{\sc ii} abundances were used to constrain the model atmospheric parameters in this study, we must consider the non-LTE effects on Fe. These effects were found to be negligible (0.00 dex) for both solar and stellar Fe\,{\sc ii} lines \citep{Bergemann2012Fe,Lind2012, Bensby2014}. For Fe\,{\sc i} lines with low excitation potentials ($<$8 eV) and metallicities [Fe/H] > -3.0 dex, the non-LTE deviations were minimal according to K \citep{Lind2012}. The non-LTE corrections \citep{Bergemann2012FeTi} for 66 Fe\,{\sc i} lines in the IAG solar spectrum and 56 Fe\,{\sc i} lines in HD\,218209 were found to be 0.01 dex. Similar trends were observed for the other elements in both the Sun and Star. For example, the non-LTE corrections (Sun/Star) for Si\,{\sc i} (-0.01/0.00), Ca\,{\sc i} (-0.01/-0.01), Ti\,{\sc i} (0.10/0.13), Ti\,{\sc ii} (-0.01/0.00), Cr\,{\sc i} (0.05/0.08), Mn\,{\sc i} (0.05/0.12), and Co\,{\sc i} (0.11/0.15) were generally small, with the largest corrections found for Ti and Co

\subsection{Notes on the errors for model atmospheric parameters of the Sun}
The solar spectrum is used as a standard reference spectrum for the spectroscopic analysis of F-G-K-type stars, in both the optical and NIR regions \citep{sahin2020, Sahin2023, Senturk2024}. This is mainly due to the well-characterized atmosphere of the Sun and extensive observational data in the optical and IR regions. Many published NIR line lists include lines with poorly defined or calibrated oscillator strengths, often relying on theoretical calculations \citep[e.g.,][]{ryde2009}. In particular, a recent spectroscopic study of a solar analogue star, HD\,76151, in the $Y$, $J$, $H$, and $K$ bands by \citet{Senturk2024} provides a detailed review of the line libraries published in the infrared region over the last 40 years in terms of $\log gf$ values and atomic data. 

In the first paper of the series \citep{Sahin2023}, the effective temperature obtained from the solar atmosphere analysis differed by 20 K from the effective temperature value obtained in this study. This difference is consistent with the error values. Similarly, a significant difference in Paper I is the increase in the reported errors for $T_{\rm eff}$, $\log g$ because of the increase in the error for metallicity ($\Delta \sigma {\rm [Fe/H]}$ = 0.05 dex). For $T_{\rm eff}$, $\Delta \sigma T_{\rm eff}$ = 85 K and for $\log g$, $\Delta \sigma \log g$ = 0.10 cgs. In this study, we obtained an additional 187 atomic transitions in the near-IR region. In addition, two different solar spectra were preferred for the solar abundance analysis. The KPNO solar spectrum is in the 4\,000-5\,000 \AA\ region and the IAG solar spectrum is in the 5\,000-10\,000 \AA\ region.

The following subsections provide details of the line list and atomic data. 

\section{Line List: Identification, Line Measurement, and Atomic Data}
Initially, the centers of the lines exhibiting Gaussian profiles appropriate for equivalent width analysis within the range of 4\,000-10\,000 \AA\ were identified in the KPNO \citep{Kurucz1984} and IAG solar spectra \citep[][BTFS]{Baker2020}. The established line centers for the selected isolated lines were compared with the wavelengths identified in the laboratory environment within the Revised Multiplet Table (RMT) \citep{moore1966}. Subsequently, a multiplet \citep[cf.][]{moore1954} analysis technique was applied. The relative intensities of the lines within a multiplet are generally insensitive to variations in the excitation conditions in most spectroscopic sources. A standard approach involves verifying the presence of multiple members with expected relative intensities. Subsequent analyses focused on identifying lines that exhibited similar excitation and laboratory strengths.

The common wavelength range of the first article of the series and this study was 4\,024-6\,772 \AA. In this range, 54 atomic transitions from 19 species of 17 elements were added to the first report on this series. The distributions of these transitions are Na\,{\sc i} (one line), Al\,{\sc i} (two lines), Si\,{\sc i} (two lines), Ca\,{\sc i} (two lines), Sc\,{\sc ii} (five lines), Ti\,{\sc i} (four lines), V\,{\sc i} (one line), Cr\,{\sc ii} (one line), Mn\,{\sc i} (one line), Fe\,{\sc i} (17 lines), Fe\,{\sc ii} (six lines), Co\,{\sc i} (two lines), Cu\,{\sc i} (two lines), Zr\,{\sc i} (one line), Zr\,{\sc ii} (one line), La\,{\sc ii} (three lines), Ce\,{\sc ii} (one line), Nd\,{\sc ii} (one line) and Sm\,{\sc ii} (one line). In the region 6\,772\footnote{Upper wavelength limit from \citet{Sahin2023} is 6\,780 \AA.}-9\,944 \AA, 189 atomic transitions from 27 species of 23 elements were added to the line list. The distributions of these transitions are C\,{\sc i} (two lines), O\,{\sc i} (three lines), Na\,{\sc i} (two lines), Mg\,{\sc i} (three lines), Mg\,{\sc ii} (one line), AI\,{\sc i} (eight lines), Si\,{\sc i} (nine lines), P\,{\sc i} (three lines), S\,{\sc i} (two lines), Ca\,{\sc i} (six lines), Sc\,{\sc ii} (five lines), Ti\,{\sc i} (20 lines), V\,{\sc i} (one line), Cr\,{\sc i} (10 lines), Cr\,{\sc ii} (one line), Mn\,{\sc i} (one line), Fe\,{\sc i} (123 lines), Fe\,{\sc ii} (15 lines), Co\,{\sc i} (three lines), Ni\,{\sc i} (13 lines), Cu\,{\sc i} (four lines), Zr\,{\sc i} (one line), Zr\,{\sc ii} (one line), La\,{\sc ii} (three lines), Ce\,{\sc ii} (one line), Nd\,{\sc i} (two lines), and Sm\,{\sc ii} (one line). In total, 13 atomic transitions from seven species of seven elements were included in the first article of the series but were not included in this study. The statistics of these transitions are as follows: Ca\,{\sc i} (three lines), Ti\,{\sc ii} (one line), Fe\,{\sc i} (three lines), Co\,{\sc i} (two lines), Ni\,{\sc i} (one line), Zr\,{\sc ii} (one line), and Nd\,{\sc ii} (two lines).
Lower-level excitation potential (L.E.P) values for the new line list were obtained from the MOORE catalogue \citep{moore1966}.

Accurate determination of elemental abundances in stars requires precise knowledge of the atomic transition probability, quantified by the $\log gf$ value. This study utilized a comprehensive compilation of $\log gf$ values from recent literature, including \cite{Biemont1980,Biemont1981}, \cite{Hannaford1982}, \cite{Klose2002}, \cite{Takeda2003}, \cite{fuhr2006}, \cite{kelleher2008}, \cite{Lawler2009}, \cite{DenHartog2011}, \cite{shi2011},  \cite{Hansen2013}, \cite{Lawler2006,Lawler2013, Lawler2015,Lawler2017,Lawler2019}, \cite{Rhodin2017}, and \cite{DenHartog2021}. For transitions not documented in these sources, data from the NIST\footnote{\url{http://physics.nist.gov/PhysRefData/ASD}} and VALD\footnote{\url{https://vald.astro.uu.se/}} atomic line databases were used. When multiple sources were available, the $\log gf$ value that yielded the most consistent abundance with solar abundance values reported by \cite{asplund2009, asplund2021} was prioritized. References for the adopted $\log gf$ values and corresponding RMT numbers for each line are tabulated in Tables \ref{tab:lineslit_fe_lines}, \ref{tab:lineslit_fe_lines_2}, \ref{tab:lineslit_other_lines}, \ref{tab:lineslit_other_lines_2}, and \ref{tab:lineslit_other_lines_3}.

Further verification of the $\log gf$ values was performed by comparing the $\log gf$ values used in this study with those in the {\it Gaia}-ESO line list v.6 provided by GES collaboration \citep{heiter2021}. Note that the $gf$ values for the chosen lines of Fe\,{\sc i} and Fe\,{\sc ii} in this study were obtained from \citet {fuhr2006}. The GES line list contains the recommended lines and atomic data (i.e., $gf$ values corrected for the hyperfine structure) for the analysis of FGK stars. Notably, several lines in the spectra of FGK stars have not yet been identified \citep{heiter2015}.  

The GES line list (v.6) comprises 141\,233 lines spanning a 4\,200-9\,200\AA.~A total of 561 lines were analyzed in this study, of which 548 were common to the GES line list. These 592 atomic transitions involve 30 species from 26 elements. A total of 40 atomic transitions were included in this study's line list in the regions outside the GES line list boundaries (lower limit: 4\,021-4\,200 \AA\ and upper limit: 9\,200-9\,944 \AA). In the spectral region overlapping with the GES line list (4\,200-9\,200 \AA), additional Fe\,{\sc i} (8958.88 \AA), and Fe\,{\sc ii} (6806.85 \AA, 6810.28 \AA, 6820.43 \AA) atomic transitions were found compared to the GES line list. Of the 55 lines identified in this study within the same wavelength range, 51 were found in the GES line list. This wavelength range aligns with the PolarBASE spectrum of HD\,218209 used in this analysis.
\begin{table}
\setlength{\tabcolsep}{2.2pt}
\renewcommand{\arraystretch}{0.9}
\centering
\caption{Comparison of $\log gf$ values for common lines in GESv6. The number of common lines (n) was also reported. The mean of the $\log gf$ differences ($\Delta\log gf$) for each element is also reported.}
\label{tab:loggf_comparison}
\begin{tabular}{lccc|lccc}
\hline
Element & n & $\Delta\log (gf)$  & $\sigma$ & Element & n & $\Delta\log (gf)$ & $\sigma$ \\ 
\cline{3-4}
\cline{7-8}
 &  &  (dex)            &  (dex)        &        &            &     (dex)            &  (dex)        \\
\hline
C\,{\sc i}   & 2  & -0.02 & 0.02 & Mn\,{\sc i}  & 12  & 0.68  & 0.83 \\
O\,{\sc i}   & 3  & 0.00  & 0.00 & Fe\,{\sc i}  & 236 & 0.00  & 0.16 \\
Na\,{\sc i}  & 4  & 0.01  & 0.02 & Fe\,{\sc ii} & 28  & 0.00  & 0.07 \\
Mg\,{\sc i}  & 5  & 0.32  & 0.54 & Co\,{\sc i}  & 7   & 1.33  & 1.03 \\
Mg\,{\sc ii} & 1  & -0.01 & 0.00 & Ni\,{\sc i}  & 66  & -0.04 & 0.10 \\
Al\,{\sc i}  & 8  & 0.29  & 0.58 & Cu\,{\sc i}  & 4   & 0.27  & 0.26 \\
Si\,{\sc i}  & 19 & -0.01 & 0.11 & Zn\,{\sc i}  & 2   & -0.03 & 0.02 \\
S\,{\sc i}   & 2  & -0.29 & 0.24 & Sr\,{\sc i}  & 1   & 0.00  & 0.00 \\
Ca\,{\sc i}  & 20 & 0.00  & 0.04 & Y\,{\sc ii}  & 2   & -0.07 & 0.05 \\
Sc\,{\sc ii} & 12 & 0.02  & 0.06 & Zr\,{\sc i}  & 1   & 0.00  & 0.00 \\
Ti\,{\sc i}  & 56 & 0.00  & 0.03 & Ba\,{\sc ii} & 2   & -0.02  & 0.01 \\
Ti\,{\sc ii} & 11 & 0.04  & 0.11 & La\,{\sc ii} & 2   & -0.01 & 0.01 \\
V\,{\sc i}   & 5  & 0.71  & 0.61 & Ce\,{\sc ii} & 2   & 0.00  & 0.00 \\
Cr\,{\sc i}  & 26 & 0.07  & 0.49 & Nd\,{\sc ii} & 2   & 0.00  & 0.00 \\
Cr\,{\sc ii} & 4  & 0.12  & 0.20 & Sm\,{\sc ii} & 2   & 0.00  & 0.00 \\
\hline
\end{tabular}
\begin{list}{}{}
 
    \item $\Delta\log (gf)$ = $\log (gf)_{\rm This~Study}$ - $\log (gf)_{\rm GESv6}$
\end{list}
\end{table}

For the 236 common Fe\,{\sc i} lines in the GES line list, the difference in the $\log gf$ value was 0.00 $\pm$ 0.16 dex. For the 28 Fe\,{\sc ii} lines, the difference in the $\log gf$ values was 0.00 $\pm$ 0.07 dex. A detailed comparison of the $\log gf$ values was performed for the 548 lines common to both line lists, as listed in Table \ref{tab:loggf_comparison} which summarises the mean difference in $\log gf$ values and 

\begin{figure*}[h!]
\centering
\includegraphics[width=0.96\linewidth]{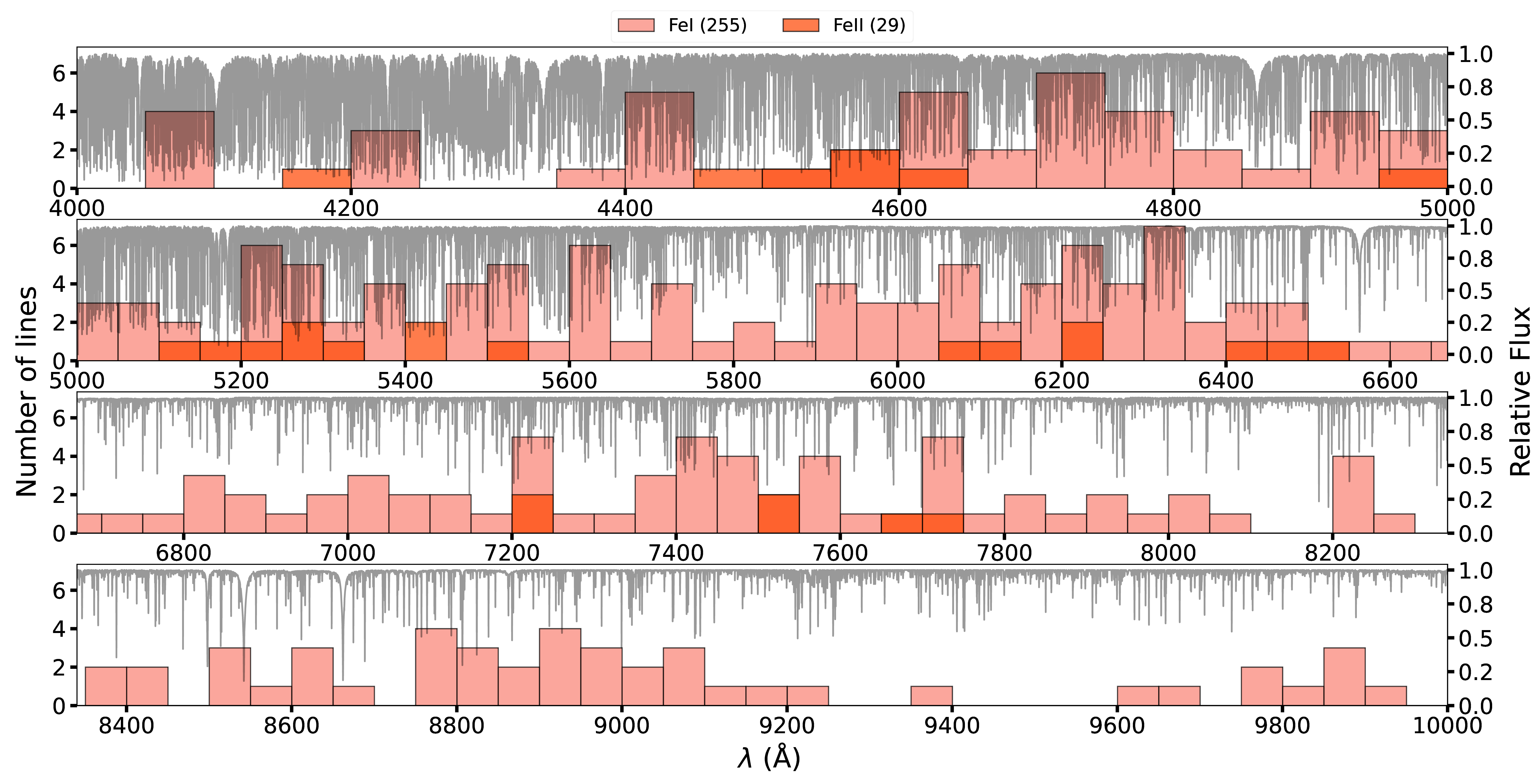}
\caption{The figure displays the telluric-free Solar spectrum obtained from \citet{Baker2020}, along with the number of identified lines within each 50 \AA\, region of the spectrum. The 4\,000 - 5\,000 \AA\, spectral range is based on solar data from \citet{Kurucz1984}, while the 5\,000 - 10\,000 \AA\, range utilizes the telluric-free Solar spectrum (BTFS) provided by \citet{Baker2020}.}
\label{fig:lines_ident}
\end{figure*}

the corresponding standard deviations for each element with at least two common lines. The results show overall good agreement between the two line lists, though significant differences were observed for certain elements, such as Co and Mn. These discrepancies can be attributed to various factors including differences in the atomic data used to construct the line lists, uncertainties in the line identification process, and the presence of non-LTE effects.

Figure \ref{fig:lines_ident} presents the numerical statistics for the final line list generated in this study are shown in Figure \ref{fig:lines_ident}. The same figure shows the number of lines in the spectral region of 50 \AA\, each.

\section{Conclusion}
This study presents an expanded line list covering the wavelength range of 4\,080–10\,000 \AA\ for abundance analyses of F- and G-type stars. Although Paper I  reported 363 atomic transitions, only 592 lines were reported in this study. The line list was compared with the existing {\it Gaia}-ESO v6 line list (Table \ref{tab:loggf_comparison}), and a 93$\%$ overlap was found, with 548 of the 592 line matches.

Utilizing high-resolution solar spectra from IAG (5\,000-10\,000 \AA, $R\approx$1\,000\,000) and KPNO (4\,000-6\,780 \AA, R$\approx$700\,000), 592 spectral lines belonging to 33 chemical species were identified and included in the abundance analysis. Compared to the previous paper in this series, not only has the wavelength range extended, but elements such as C, O, Al, P, S, Co, Cu, Zr, and La have also been added to the list.

Additionally, the abundances of C, O, Mg, Al, P, S, Sc, V, Mn, Co, Cu, Zn, Sr, Y, Zr, Ba, La, Ce, Nd, and Sm were determined using the synthesis method. To calculate the reported abundances, it was assumed that the solar spectrum was disk-integrated\footnote{At this point, the flux/int switch in the abfind and synth drivers of the {\sc MOOG} code, which we used to determine model atmosphere parameters and abundance calculations under LTE conditions, was set to zero.}. 

A comparison of the elemental abundances ([X/Fe]) reported in this study for HD\,218209 with those presented by \citep{Sahin2023} reveals several differences. No significant differences were observed for Cr\,{\sc ii}, Ti\,{\sc i}, V\,{\sc i}, Sr\,{\sc i}, and Zr\,{\sc ii} ($\Delta \log_{\epsilon} = 0.00$ dex). Elements exhibiting a difference of -0.01 dex include Fe\,{\sc i}, Ni\,{\sc i}, Cr\,{\sc i}, Ca\,{\sc i}, and Nd\,{\sc ii}. A difference of 0.06 dex was observed for Ce\,{\sc ii}, Ba\,{\sc ii}, and Sc\,{\sc ii}. Other notable differences include -0.02 dex for Na\,{\sc i} and Ti\,{\sc ii}, 0.01 dex for Si\,{\sc i}, 0.07 dex for Y\,{\sc ii}, 0.05 dex for Mn\,{\sc i}, 0.02 dex for Zn\,{\sc i}, 0.09 dex for Co\,{\sc i}, and 0.04 dex for Mg\,{\sc i}.

In this study, we employed both equivalent width (EW) measurements and spectrum synthesis techniques to determine the elemental abundances in the solar and HD\, 218209 spectra. The resulting abundances were compared to those reported by \citet{asplund2009} and \citet{asplund2021} as well as other solar abundance values found in the literature (Table \ref{table:abund-lit}). Our results are in excellent agreement with those of the previous studies. Notably, the revision of Ba abundance in \citet{asplund2021} significantly reduced the discrepancy between the two studies.

Having accurately determined the solar abundances using a constructed line list, we applied a similar methodology to the star, HD\,218209. Table \ref{table:abund-lit-star} presents a comparison of the effective temperature, surface gravity, metallicity, and derived chemical abundances of this star. A thorough examination of the available abundance data for HD\,218209 revealed a scarcity of literature regarding the abundance of several elements (C, O, Cr, Co, Cu, Zn, Sr, Y, Zr, Ba, La, Ce, Nd, and Sm). This highlights the significant contributions of our study to this field. A detailed element-by-element literature analysis is provided in Appendix A1.

\begin{description}
  \item[Peer Review:] Externally peer-reviewed.
  \item[Author Contribution:] Conception/Design of study - T.Ş.; Data Acquisition - T.Ş., F.G., M.M., S.A.Ş; Data Analysis/Interpretation - T.Ş., F.G., S.A.Ş., M.M., N.Ç.; ~Drafting Manuscript - T.Ş.;~Critical Revision of Manuscript - T.Ş.;~Final Approval and Accountability - T.Ş., F.G., S.A.Ş., M.M., N.Ç.;~Technical or Material - T.Ş., M.M., F.G.; Support Supervision - T.Ş.
  \item[Conflict of Interest:] Authors declared no conflict of interest.
  \item[Financial Disclosure:] This study was supported by the Scientific and Technological Research Council of Turkey (TUBITAK) under Grant Number 121F265. The authors thank TUBITAK for their support.
\end{description}

\section*{Acknowledgements}
  This study used NASA’s Astrophysics Data System and the SIMBAD database operated at CDS, Strasbourg, France. The nonpublic data underlying this article will be made available upon reasonable request from the authors.

Software: LIME \citep{sahin2017}, SPECTRE \citep{sneden1973}, MOOG \citep{sneden1973}
\spacebref{-2pt}{-20pt}
\bibliographystyle{mnras}
\bibliography{Sahin_et_al_arxiv}

\appendix
\renewcommand{\thetable}{A\arabic{table}}
\section{Appendix}

\subsection{Literature Review for HD\,218209} 
This section presents a comprehensive literature review of the elemental abundances of the star, focusing on studies conducted over the past four decades. Table \ref{table:abund-lit-star} summarizes the literature values for each element and compares our results with those of previous studies.

Carbon abundance for star has been reported in the literature over the last decade by \citet[DA15]{dasilva2015}, \citet[RI20]{rice2020}, and \citet[TA23]{takeda2023}. The carbon abundance ([C/Fe]=0.14 dex) reported in this study is in good agreement with that of \citet[RI20]{rice2020} ([C/Fe]=0.18 dex), differing by only 0.04 dex. The largest discrepancy is found for \citet[TA23]{takeda2023}, with a difference of 0.22 dex.

The literature values for [O/Fe] exhibited a scatter of approximately 0.3 dex. Our value ($\approx$ 0.3 dex) agrees well with \citet[MI13]{mishenina2013} ($\Delta=0.06$ dex), but shows a larger discrepancy than \citet[TA23]{takeda2023} ($\Delta=0.20$ dex) and \citet[RI20]{rice2020} ($\Delta =0.14$ dex).

The [Na/Fe] ratio of -0.03 dex shows good agreement with \citet[MI11]{mishenina2011} ($\Delta=-0.01$ dex), \citet[RI20]{rice2020} ($\Delta=-0.06$ dex), \citet[LU17]{luck2017} ($\Delta=-0.09$ dex), and \citet[VA05]{valenti2005} ($\Delta=-0.13$ dex). However, a significant discrepancy ($\Delta=-0.26$ dex) was observed compared to in that \citet[GE04]{gehren2004}.

Moving on to magnesium, our [Mg/Fe] value of 0.24 dex is consistent with the values reported in \citet[MI04]{Mishenina2004}, \citet[MI13]{mishenina2013} ($\Delta=0.05$ dex), \citet[DA15]{dasilva2015} ($\Delta=0.06$ dex), \citet[RI20]{rice2020} ($\Delta=0.07$ dex), and \citet[LU17]{luck2017} ($\Delta=-0.05$ dex). However, a significant discrepancy of -0.17 dex was observed compared to \citet[GE04]{gehren2004}.

The reported [Al/Fe] ratio in this study is consistent with the values reported by \citet[MI11]{mishenina2011}, \citet[DA15]{dasilva2015}, \citet[LU17]{luck2017}, and \citet[RI20]{rice2020}, except for the abundance ratio reported by \citet[AB88]{Abia1988}, which shows a significant discrepancy ($\Delta=-0.32$ dex).

The literature values for [Si/Fe] exhibited a relatively homogeneous distribution. Our value of 0.13 dex agrees well with \citet[DA15]{dasilva2015} and \citet[LU17]{luck2017} ($\Delta=0.02$ dex). The largest discrepancy was observed in \citet[TA07]{Takeda2007} ($\Delta=-0.13$ dex).

Our [Ca/Fe] value of 0.15 dex is in good agreement with \citet[RI20]{rice2020} ($\Delta=0.03$ dex), \citet[DA15]{dasilva2015} ($\Delta=0.02$ dex), and \citet[LU17]{luck2017} ($\Delta=-0.04$ dex). A significant discrepancy is observed with \citet[MI11]{mishenina2011} ($\Delta=0.50$ dex).
 
Our [Sc/Fe] value of 0.06 dex shows a discrepancy of 0.09 dex compared to \citet[LU17]{luck2017}.

Our [Ti/Fe] value of 0.21 dex agrees well with \citet[LU17]{luck2017} and shows good agreement with \citet[DA15]{dasilva2015} ($\Delta=0.01$ dex), \citet[VA05]{valenti2005} ($\Delta=-0.02$ dex), and \citet[RI20]{rice2020}  ($\Delta=-0.03$ dex). A significant discrepancy is observed with \citet[TA07]{Takeda2007} ($\Delta=0.18$ dex).

Our [V/Fe] value of -0.02 dex aligns well with the findings of \citet[TA07]{Takeda2007} ($\Delta=0.05$ dex) but shows discrepancies of 0.19 dex, 0.18 dex, and 0.15 dex when compared to \citet[RI20]{rice2020}, \citet[LU17]{luck2017}, and \citet[DA15]{dasilva2015}, respectively.

The [Cr/Fe] value determined in this study agrees well with previous findings, with discrepancies of approximately $\pm$0.05 dex observed when compared to \citet[RI20]{rice2020} and \citet[LU17]{luck2017}.

Our [Mn/Fe] value of -0.27 dex precisely matches the value reported by \citet[RI20]{rice2020} and demonstrates good agreement with \citet[DA15]{dasilva2015} ($\Delta=-0.09$ dex) and \citet[LU17]{luck2017} ($\Delta=-0.03$ dex).

The [Co/Fe] value determined in this study exhibits discrepancies of -0.18 dex compared to \citet[LU17]{luck2017} and -0.23 dex compared to \citet[TA07]{Takeda2007}.

The [Ni/Fe] value determined in this study aligns well with the literature values, with the exception of a significant discrepancy ($\Delta=-0.21$ dex) observed in the work of \citet[AB88]{Abia1988}. The smallest discrepancy is found with \citet[LU17]{luck2017} ($\Delta=-0.01$ dex), followed by \citet[RI20]{rice2020} ($\Delta=-0.03$ dex), \citet[MI13]{mishenina2013} and \citet[MI04]{Mishenina2004} ($\Delta=-0.06$ dex), and \citet[TA07]{Takeda2007} ($\Delta=-0.02$ dex).

The [Cu/Fe] value determined in this study shows discrepancies of -0.10 dex compared to \citet[LU17]{luck2017}, -0.06 dex compared to \citet[DA15]{dasilva2015}, and -0.11 dex compared to \citet[MI11]{mishenina2011}.

The [Zn/Fe] value determined in this study is in good agreement with literature values, with a difference of 0.08 dex compared to \citet[LU17]{luck2017} and 0.06 dex compared to \citet[MI13]{mishenina2013}.

The [Sr/Fe] value of 0.10 dex determined in this study exhibits a discrepancy of -0.28 dex compared to \citet[LU17]{luck2017}.

The [Y/Fe] value determined in this study shows discrepancies of -0.16 dex compared to \citet[RI20]{rice2020}, -0.22 dex compared to \citet[LU17]{luck2017}, and -0.10 dex compared to \citet[MI11]{mishenina2011}.

The [Zr/Fe] value determined in this study agrees well with \citet[MI13]{mishenina2013} ($\Delta=0.04$ dex), but shows a discrepancy of -0.21 dex compared to \citet[LU17]{luck2017}.

The [Ba/Fe] value determined in this study precisely matches that reported by \citet[LU17]{luck2017} ([Ba/Fe]=0.04 dex), while a difference of 0.03 dex is observed compared to \citet[MI13]{mishenina2013}.

The [La/Fe] value determined in this study shows a discrepancy of -0.60 dex compared to \citet[LU17]{luck2017}, while a difference of -0.06 dex is observed compared to \citet[MI13]{mishenina2013}.

The [Ce/Fe] value determined in this study shows a discrepancy of -0.02 dex compared to \citet[LU17]{luck2017}, while a difference of -0.28 dex is observed compared to \citet[MI13]{mishenina2013}.

The difference in neodymium abundance compared to \citet[LU17]{luck2017} is -0.24 dex, while the difference compared to \citet[MI13]{mishenina2013} is -0.07 dex.

The [Sm/Fe] value determined in this study shows a discrepancy of -0.12 dex compared to \citet[LU17]{luck2017}, while a difference of 0.01 dex is observed compared to \citet[MI13]{mishenina2013}.

\begin{table*}
\setlength{\tabcolsep}{1.2pt}
\renewcommand{\arraystretch}{0.8}
\centering\small
\caption{Fe\,{\sc i} and Fe\,{\sc ii} lines. The abundances were obtained for a model with $T_{\rm eff}=5770$ K, $\log g =4.40$ cgs, and $\xi=$ 0.66 km s$^{\rm −1}$ for the solar spectrum. $T_{\rm eff}=$5600 K, $\log g=4.50$ cgs, and $\xi=$ 0.44 km s$^{\rm −1}$ for the HD\,218209 spectrum.}
\label{tab:lineslit_fe_lines}
\resizebox{\linewidth}{!}{%
\begin{tabular}{lccccccccc||ccccccccccc}
\hline
 & & & & \multicolumn{2}{c}{Sun} & \multicolumn{2}{c}{HD\,218209}  & & & & & &\multicolumn{2}{c}{Sun} &\multicolumn{2}{c}{HD\,218209} & \\
\cline{2-4}
\cline{7-8}
\cline{11-13}
\cline{16-17}
 Spec.	&   $\lambda$ & LEP  &  $\log(gf)$ & EW  & $\log \epsilon$(X) & EW	& $\log\epsilon$(X) & RMT &Ref. &Spec. & $\lambda$ & LEP  &  $\log(gf)$ & EW  & $\log\epsilon$(X) & EW  & $\log \epsilon$(X) & RMT & Ref. \\
\cline{2-4}
\cline{7-8}
\cline{11-13}
\cline{16-17}
   & (\AA)  & (eV) & (dex)   &(m\AA) & (dex) &(m\AA) & (dex)&   & & & (\AA)  & (eV) & (dex)   &(m\AA) & (dex)&(m\AA) & (dex) & \\
\hline
Fe\,{\sc i} 	&	 4080.22 	&	 3.28 	&	 -1.23 	&	 80.9 	&	 7.32 	&	 - 	&	 - 	&	 558 	&	 1 	&	Fe\,{\sc i} 	&	 5501.48 	&	 0.96 	&	 -3.05 	&	 115.3 	&	 7.50 	&	 104.5 	&	 7.08 	&	 15 	&	 1 \\
Fe\,{\sc i} 	&	 4082.11 	&	 3.42 	&	 -1.51 	&	 68.2 	&	 7.49 	&	 - 	&	 - 	&	 698 	&	 1 	&	Fe\,{\sc i} 	&	 5506.79 	&	 0.99 	&	 -2.80 	&	 123.2 	&	 7.37 	&	 111.4 	&	 6.94 	&	 15 	&	 1 \\
Fe\,{\sc i} 	&	 4088.56 	&	 3.64 	&	 -1.50 	&	 52.3 	&	 7.41 	&	 - 	&	 - 	&	 906 	&	 1 	&	 Fe\,{\sc i} 	&	 5525.55 	&	 4.23 	&	 -1.08 	&	 53.0 	&	 7.35 	&	 39.4 	&	 6.96 	&	 1062 	&	 1 \\
Fe\,{\sc i} 	&	 4090.96 	&	 3.37 	&	 -1.73 	&	 55.8 	&	 7.37 	&	 - 	&	 - 	&	 695 	&	 1 	&	 Fe\,{\sc i} 	&	 5543.94 	&	 4.22 	&	 -1.11 	&	 61.5 	&	 7.55 	&	 49.1 	&	 7.17 	&	 1062 	&	 1 \\
Fe\,{\sc i} 	&	 4204.00 	&	 2.84 	&	 -1.01 	&	 125.1 	&	 7.50 	&	 - 	&	 - 	&	 355 	&	 1 	&	 Fe\,{\sc i} 	&	 5546.51 	&	 4.37 	&	 -1.28 	&	 50.8 	&	 7.63 	&	 35.9 	&	 7.21 	&	 1145 	&	 1 \\
Fe\,{\sc i} 	&	 4207.13 	&	 2.83 	&	 -1.41 	&	 82.5 	&	 7.39 	&	 77.3 	&	 7.12 	&	 352 	&	 1 	&	 Fe\,{\sc i} 	&	 5560.22 	&	 4.43 	&	 -1.16 	&	 50.2 	&	 7.55 	&	 36.4 	&	 7.15 	&	 1164 	&	 1 \\
Fe\,{\sc i} 	&	 4220.35 	&	 3.07 	&	 -1.31 	&	 83.6 	&	 7.46 	&	 77.4 	&	 7.16 	&	 482 	&	 1 	&	 Fe\,{\sc i} 	&	 5618.64 	&	 4.21 	&	 -1.28 	&	 50.0 	&	 7.47 	&	 36.1 	&	 7.06 	&	 1107 	&	 1 \\
Fe\,{\sc i} 	&	 4365.90 	&	 2.99 	&	 -2.25 	&	 51.4 	&	 7.48 	&	 40.7 	&	 7.09 	&	 415 	&	 1 	&	 Fe\,{\sc i} 	&	 5624.03 	&	 4.39 	&	 -1.20 	&	 50.4 	&	 7.56 	&	 37.2 	&	 7.18 	&	 1160 	&	 1 \\
Fe\,{\sc i} 	&	 4432.58 	&	 3.57 	&	 -1.56 	&	 51.9 	&	 7.37 	&	 42.9 	&	 7.04 	&	 797 	&	 1 	&	 Fe\,{\sc i} 	&	 5633.95 	&	 4.99 	&	 -0.32 	&	 65.3 	&	 7.57 	&	 53.3 	&	 7.23 	&	 1314 	&	 1 \\
Fe\,{\sc i} 	&	 4439.89 	&	 2.28 	&	 -3.00 	&	 52.1 	&	 7.56 	&	 36.7 	&	 7.03 	&	 116 	&	 1 	&	 Fe\,{\sc i} 	&	 5636.71 	&	 3.64 	&	 -2.56 	&	 20.7 	&	 7.53 	&	 13.6 	&	 7.18 	&	 868 	&	 1 \\
Fe\,{\sc i} 	&	 4442.35 	&	 2.20 	&	 -1.25 	&	 187.7 	&	 7.52 	&	 - 	&	 - 	&	 68 	&	 1 	&	 Fe\,{\sc i} 	&	 5638.27 	&	 4.22 	&	 -0.84 	&	 76.6 	&	 7.53 	&	 67.7 	&	 7.23 	&	 1087 	&	 1 \\
Fe\,{\sc i} 	&	 4447.14 	&	 2.20 	&	 -2.73 	&	 66.6 	&	 7.64 	&	 57.3 	&	 7.27 	&	 69 	&	 1 	&	 Fe\,{\sc i} 	&	 5641.45 	&	 4.26 	&	 -1.15 	&	 66.7 	&	 7.70 	&	 49.0 	&	 7.24 	&	 1087 	&	 1 \\
Fe\,{\sc i} 	&	 4447.73 	&	 2.22 	&	 -1.34 	&	 171.0 	&	 7.52 	&	 - 	&	 - 	&	 68 	&	 1 	&	 Fe\,{\sc i} 	&	 5662.52 	&	 4.18 	&	 -0.57 	&	 91.2 	&	 7.57 	&	 81.4 	&	 7.25 	&	 1087 	&	 1 \\
Fe\,{\sc i} 	&	 4502.60 	&	 3.57 	&	 -2.31 	&	 28.6 	&	 7.50 	&	 - 	&	 - 	&	 796 	&	 1 	&	 Fe\,{\sc i} 	&	 5701.56 	&	 2.56 	&	 -2.22 	&	 84.8 	&	 7.61 	&	 72.7 	&	 7.21 	&	 209 	&	 1 \\
Fe\,{\sc i} 	&	 4556.93 	&	 3.25 	&	 -2.66 	&	 26.8 	&	 7.49 	&	 - 	&	 - 	&	 638 	&	 1 	&	 Fe\,{\sc i} 	&	 5705.47 	&	 4.30 	&	 -1.36 	&	 37.8 	&	 7.37 	&	 25.4 	&	 6.98 	&	 1087 	&	 1 \\
Fe\,{\sc i} 	&	 4593.53 	&	 3.94 	&	 -2.03 	&	 28.3 	&	 7.53 	&	 - 	&	 - 	&	 971 	&	 1 	&	 Fe\,{\sc i} 	&	 5717.84 	&	 4.28 	&	 -1.10 	&	 62.1 	&	 7.58 	&	 - 	&	 -   	&	 1107 	&	 1 \\
Fe\,{\sc i} 	&	 4602.01 	&	 1.61 	&	 -3.15 	&	 70.9 	&	 7.53 	&	 63.9 	&	 7.20 	&	 39 	&	 1 	&	 Fe\,{\sc i} 	&	 5741.86 	&	 4.26 	&	 -1.67 	&	 32.6 	&	 7.52 	&	 - 	&	 -   	&	 1086 	&	 1 \\
Fe\,{\sc i} 	&	 4602.95 	&	 1.48 	&	 -2.22 	&	 118.8 	&	 7.44 	&	 110.7 	&	 7.06 	&	 39 	&	 1 	&	 Fe\,{\sc i} 	&	 5778.46 	&	 2.59 	&	 -3.43 	&	 22.4 	&	 7.42 	&	 15.5 	&	 7.06 	&	 209 	&	 1 \\
Fe\,{\sc i} 	&	 4619.30 	&	 3.60 	&	 -1.08 	&	 83.9 	&	 7.43 	&	 68.9 	&	 6.98 	&	 821 	&	 1 	&	 Fe\,{\sc i} 	&	 5806.73 	&	 4.61 	&	 -1.03 	&	 51.6 	&	 7.58 	&	 41.9 	&	 7.28 	&	 1180 	&	 1 \\
Fe\,{\sc i} 	&	 4630.13 	&	 2.28 	&	 -2.59 	&	 73.2 	&	 7.62 	&	 66.5 	&	 7.29 	&	 115 	&	 1 	&	 Fe\,{\sc i} 	&	 5814.80 	&	 4.26 	&	 -1.94 	&	 22.0 	&	 7.53 	&	 12.2 	&	 7.10 	&	 1086 	&	 1 \\
Fe\,{\sc i} 	&	 4635.85 	&	 2.84 	&	 -2.36 	&	 54.1 	&	 7.50 	&	 42.5 	&	 7.08 	&	 349 	&	 1 	&	 Fe\,{\sc i} 	&	 5881.28 	&	 4.59 	&	 -1.70 	&	 14.3 	&	 7.35 	&	 - 	&	 - 	&	 1178 	&	 1 \\
Fe\,{\sc i} 	&	 4661.54 	&	 4.54 	&	 -1.26 	&	 37.9 	&	 7.54 	&	 25.1 	&	 7.14 	&	 1207 	&	 1 	&	 Fe\,{\sc i} 	&	 5905.67 	&	 4.63 	&	 -0.77 	&	 57.0 	&	 7.44 	&	 - 	&	 - 	&	 1181 	&	 1 \\
Fe\,{\sc i} 	&	4678.85	&	3.60	&	-0.83	&	syn	&	7.54	&	syn	&	7.06	&	821	&	1	&	 Fe\,{\sc i} 	&	 5916.26 	&	 2.45 	&	 -2.99 	&	 54.4 	&	 7.60 	&	 45.2 	&	 7.26 	&	 170 	&	 1 \\
Fe\,{\sc i} 	&	 4704.95 	&	 3.69 	&	 -1.53 	&	 62.7 	&	 7.56 	&	 53.5 	&	 7.22 	&	 821 	&	 1 	&	 Fe\,{\sc i} 	&	 5929.68 	&	 4.55 	&	 -1.38 	&	 39.5 	&	 7.65 	&	 24.8 	&	 7.21 	&	 1176 	&	 1 \\
Fe\,{\sc i} 	&	 4728.55 	&	 3.65 	&	 -1.17 	&	 81.3 	&	 7.63 	&	 - 	&	 - 	&	 822 	&	 1 	&	 Fe\,{\sc i} 	&	 5934.67 	&	 3.93 	&	 -1.12 	&	 76.0 	&	 7.44 	&	 59.8 	&	 7.01 	&	 982 	&	 1 \\
Fe\,{\sc i} 	&	 4733.60 	&	 1.48 	&	 -2.99 	&	 83.9 	&	 7.58 	&	 77.8 	&	 7.26 	&	 38 	&	 1 	&	 Fe\,{\sc i} 	&	 5952.73 	&	 3.98 	&	 -1.39 	&	 59.6 	&	 7.48 	&	 - 	&	 -   	&	 959 	&	 1 \\
Fe\,{\sc i} 	&	 4735.85 	&	 4.07 	&	 -1.32 	&	 64.1 	&	 7.77 	&	 - 	&	 - 	&	 1042 	&	 1 	&	 Fe\,{\sc i} 	&	 5956.71 	&	 0.86 	&	 -4.61 	&	 50.9 	&	 7.55 	&	 44.4 	&	 7.23 	&	 14 	&	 1 \\
Fe\,{\sc i} 	&	 4741.53 	&	 2.83 	&	 -1.76 	&	 71.3 	&	 7.36 	&	 66.0 	&	 7.11 	&	 346 	&	 1 	&	 Fe\,{\sc i} 	&	 5983.70 	&	 4.53 	&	 -0.49 	&	 67.1 	&	 7.35 	&	 57.5 	&	 7.04 	&	 1175 	&	 3 \\
Fe\,{\sc i} 	&	 4745.81 	&	 3.65 	&	 -1.27 	&	 73.9 	&	 7.59 	&	 68.3 	&	 7.32 	&	 821 	&	 1 	&	 Fe\,{\sc i} 	&	 6003.03 	&	 3.86 	&	 -1.03 	&	 81.3 	&	 7.38 	&	 72.4 	&	 7.06 	&	 959 	&	 2 \\
Fe\,{\sc i} 	&	 4779.44 	&	 3.40 	&	 -2.02 	&	 40.7 	&	 7.34 	&	 - 	&	 - 	&	 720 	&	 1 	&	 Fe\,{\sc i} 	&	 6005.53 	&	 2.58 	&	 -3.60 	&	 21.8 	&	 7.56 	&	 15.8 	&	 7.23 	&	 959 	&	 3 \\
Fe\,{\sc i} 	&	 4788.77 	&	 3.24 	&	 -1.76 	&	 65.4 	&	 7.58 	&	 - 	&	 - 	&	 588 	&	 1 	&	 Fe\,{\sc i} 	&	 6027.06 	&	 4.07 	&	 -1.09 	&	 63.5 	&	 7.47 	&	 - 	&	 - 	&	 1018 	&	 1 \\
Fe\,{\sc i} 	&	 4793.96 	&	 3.03 	&	 -3.47 	&	 8.7 	&	 7.43 	&	 - 	&	 - 	&	 512 	&	 1 	&	 Fe\,{\sc i} 	&	 6065.49 	&	 2.61 	&	 -1.53 	&	 117.0 	&	 7.41 	&	 - 	&	 - 	&	 207 	&	 1 \\
Fe\,{\sc i} 	&	 4794.36 	&	 2.41 	&	 -4.05 	&	 11.5 	&	 7.55 	&	 - 	&	 - 	&	 115 	&	 2 	&	 Fe\,{\sc i} 	&	 6078.50 	&	 4.77 	&	 -0.32 	&	 75.9 	&	 7.53 	&	 62.3 	&	 7.16 	&	 1259 	&	 3 \\
Fe\,{\sc i} 	&	 4802.89 	&	 3.64 	&	 -1.51 	&	 59.7 	&	 7.49 	&	 47.6 	&	 7.10 	&	 888 	&	 1 	&	 Fe\,{\sc i} 	&	 6079.02 	&	 4.65 	&	 -1.10 	&	 44.4 	&	 7.55 	&	 28.9 	&	 7.12 	&	 1176 	&	 1 \\
Fe\,{\sc i} 	&	 4839.55 	&	 3.27 	&	 -1.82 	&	 61.8 	&	 7.57 	&	 55.8 	&	 7.32 	&	 588 	&	 1 	&	 Fe\,{\sc i} 	&	 6082.72 	&	 2.22 	&	 -3.57 	&	 34.3 	&	 7.48 	&	 24.1 	&	 7.08 	&	 64 	&	 1 \\
Fe\,{\sc i} 	&	 4875.88 	&	 3.33 	&	 -1.97 	&	 61.0 	&	 7.58 	&	 - 	&	 - 	&	 687 	&	 1 	&	 Fe\,{\sc i} 	&	 6096.67 	&	 3.98 	&	 -1.88 	&	 36.9 	&	 7.53 	&	 24 	&	 7.12 	&	 959 	&	 1 \\
Fe\,{\sc i} 	&	 4917.23 	&	 4.19 	&	 -1.16 	&	 62.8 	&	 7.60 	&	 51.9 	&	 7.25 	&	 1066 	&	 1 	&	 Fe\,{\sc i} 	&	 6127.91 	&	 4.14 	&	 -1.40 	&	 47.5 	&	 7.49 	&	 36.8 	&	 7.15 	&	 1017 	&	 1 \\
Fe\,{\sc i} 	&	 4918.02 	&	 4.23 	&	 -1.34 	&	 52.0 	&	 7.60 	&	 40.0 	&	 7.23 	&	 1070 	&	 1 	&	 Fe\,{\sc i} 	&	 6137.70 	&	 2.59 	&	 -1.40 	&	 129.4 	&	 7.40 	&	 114.3 	&	 6.98 	&	 207 	&	 1 \\
Fe\,{\sc i} 	&	 4924.78 	&	 2.28 	&	 -2.11 	&	 92.6 	&	 7.50 	&	 85.5 	&	 7.16 	&	 114 	&	 1 	&	 Fe\,{\sc i} 	&	 6157.73 	&	 4.07 	&	 -1.22 	&	 61.5 	&	 7.55 	&	 48.8 	&	 7.17 	&	 1015 	&	 1 \\
Fe\,{\sc i} 	&	 4939.69 	&	 0.86 	&	 -3.34 	&	 98.4 	&	 7.53 	&	 - 	&	 - 	&	 16 	&	 1 	&	 Fe\,{\sc i} 	&	 6165.36 	&	 4.14 	&	 -1.47 	&	 43.9 	&	 7.48 	&	 30.5 	&	 7.07 	&	 1018 	&	 1 \\
Fe\,{\sc i} 	&	 4961.92 	&	 3.63 	&	 -2.25 	&	 26.2 	&	 7.40 	&	 - 	&	 - 	&	 845 	&	 1 	&	 Fe\,{\sc i} 	&	 6173.34 	&	 2.22 	&	 -2.88 	&	 67.7 	&	 7.57 	&	 56.9 	&	 7.18 	&	 62 	&	 1 \\
Fe\,{\sc i} 	&	 4962.58 	&	 4.18 	&	 -1.18 	&	 53.2 	&	 7.48 	&	 37.8 	&	 7.03 	&	 66 	&	 1 	&	 Fe\,{\sc i} 	&	 6180.21 	&	 2.73 	&	 -2.65 	&	 53.3 	&	 7.50 	&	 - 	&	 - 	&	 269 	&	 1 \\
Fe\,{\sc i} 	&	 4973.10 	&	 3.96 	&	 -0.92 	&	 87.3 	&	 7.61 	&	 - 	&	 - 	&	 173 	&	 1 	&	 Fe\,{\sc i} 	&	 6200.32 	&	 2.61 	&	 -2.44 	&	 72.2 	&	 7.58 	&	 - 	&	 - 	&	 207 	&	 1 \\
Fe\,{\sc i} 	&	 5022.24 	&	 3.98 	&	 -0.56 	&	 97.1 	&	 7.40 	&	 - 	&	 - 	&	 965 	&	 1 	&	 Fe\,{\sc i} 	&	 6213.44 	&	 2.22 	&	 -2.48 	&	 81.0 	&	 7.45 	&	 - 	&	 - 	&	 62 	&	 1 \\
Fe\,{\sc i} 	&	 5029.62 	&	 3.41 	&	 -2.00 	&	 48.6 	&	 7.52 	&	 - 	&	  	&	 718 	&	 1 	&	 Fe\,{\sc i} 	&	 6219.29 	&	 2.20 	&	 -2.43 	&	 89.5 	&	 7.55 	&	 - 	&	 - 	&	 62 	&	 1 \\
Fe\,{\sc i} 	&	 5044.22 	&	 2.84 	&	 -2.02 	&	 71.9 	&	 7.39 	&	 64.7 	&	 7.06 	&	 318 	&	 1 	&	 Fe\,{\sc i} 	&	 6232.65 	&	 3.65 	&	 -1.22 	&	 81.0 	&	 7.58 	&	 73.6 	&	 7.29 	&	 816 	&	 1 \\
Fe\,{\sc i} 	&	 5074.75 	&	 4.22 	&	 -0.23 	&	 113.7 	&	 7.43 	&	 97.1 	&	 7.02 	&	 1094 	&	 1 	&	 Fe\,{\sc i} 	&	 6240.65 	&	 2.22 	&	 -3.17 	&	 47.6 	&	 7.38 	&	 38.7 	&	 7.03 	&	 64 	&	 1 \\
Fe\,{\sc i} 	&	 5083.35 	&	 0.96 	&	 -2.96 	&	 109.7 	&	 7.40 	&	 - 	&	 - 	&	 16 	&	 1 	&	 Fe\,{\sc i} 	&	 6246.33 	&	 3.59 	&	 -0.88 	&	 111.4 	&	 7.36 	&	 106.6 	&	 7.07 	&	 816 	&	 1 \\
Fe\,{\sc i} 	&	 5088.16 	&	 4.15 	&	 -1.75 	&	 37.0 	&	 7.63 	&	 - 	&	 - 	&	 1066 	&	 1 	&	 Fe\,{\sc i} 	&	 6252.56 	&	 2.40 	&	 -1.69 	&	 119.2 	&	 7.40 	&	 106.8 	&	 7.00 	&	 169 	&	 1 \\
Fe\,{\sc i} 	&	 5141.75 	&	 2.42 	&	 -2.24 	&	 86.1 	&	 7.61 	&	 75.6 	&	 7.21 	&	 114 	&	 1 	&	 Fe\,{\sc i} 	&	 6265.14 	&	 2.18 	&	 -2.55 	&	 84.0 	&	 7.54 	&	 76.5 	&	 7.22 	&	 62 	&	 1 \\
Fe\,{\sc i} 	&	 5145.10 	&	 2.20 	&	 -3.08 	&	 53.8 	&	 7.50 	&	 42.4 	&	 7.09 	&	 66 	&	 1 	&	 Fe\,{\sc i} 	&	 6270.23 	&	 2.86 	&	 -2.61 	&	 51.0 	&	 7.52 	&	 42.0 	&	 7.19 	&	 342 	&	 1 \\
Fe\,{\sc i} 	&	 5198.72 	&	 2.22 	&	 -2.13 	&	 95.1 	&	 7.46 	&	 - 	&	 - 	&	 66 	&	 1 	&	 Fe\,{\sc i} 	&	 6297.80 	&	 2.22 	&	 -2.74 	&	 74.4 	&	 7.56 	&	 65.5 	&	 7.22 	&	 62 	&	 1 \\
Fe\,{\sc i} 	&	 5217.40 	&	 3.21 	&	 -1.16 	&	 108.7 	&	 7.37 	&	 - 	&	 - 	&	 66 	&	 1 	&	 Fe\,{\sc i} 	&	 6301.51 	&	 3.65 	&	 -0.72 	&	 113.9 	&	 7.57 	&	 106.4 	&	 7.27 	&	 816 	&	 1 \\
Fe\,{\sc i} 	&	 5225.53 	&	 0.11 	&	 -4.79 	&	 73.7 	&	 7.63 	&	 - 	&	 - 	&	 1 	&	 1 	&	 Fe\,{\sc i} 	&	 6315.81 	&	 4.07 	&	 -1.66 	&	 40.3 	&	 7.51 	&	 24.6 	&	 7.03 	&	 1014 	&	 1 \\
Fe\,{\sc i} 	&	5228.38	&	4.22	&	-1.26	&	syn	&	7.79	&	syn	&	7.18	&	1091	&	1	&	 Fe\,{\sc i} 	&	 6322.69 	&	 2.59 	&	 -2.43 	&	 75.3 	&	 7.60 	&	 64.3 	&	 7.22 	&	 207 	&	 1 \\
Fe\,{\sc i} 	&	 5242.50 	&	 3.63 	&	 -0.97 	&	 85.3 	&	 7.48 	&	 75.2 	&	 7.13 	&	 843 	&	 1 	&	 Fe\,{\sc i} 	&	 6330.86 	&	 4.71 	&	 -0.97 	&	 33.2 	&	 7.23 	&	 21.1 	&	 6.85 	&	 1254 	&	 1 \\
Fe\,{\sc i} 	&	 5243.78 	&	 4.26 	&	 -1.12 	&	 61.1 	&	 7.59 	&	 49 	&	 7.22 	&	 1089 	&	 1 	&	 Fe\,{\sc i} 	&	 6335.34 	&	 2.20 	&	 -2.18 	&	 97.0 	&	 7.42 	&	 85.9 	&	 7.03 	&	 62 	&	 1 \\
Fe\,{\sc i} 	&	 5247.06 	&	 0.09 	&	 -4.95 	&	 68.1 	&	 7.63 	&	 57.5 	&	 7.19 	&	 1 	&	 1 	&	 Fe\,{\sc i} 	&	 6336.83 	&	 3.69 	&	 -0.86 	&	 102.2 	&	 7.32 	&	 93.9 	&	 6.99 	&	 816 	&	 1 \\
Fe\,{\sc i} 	&	 5250.22 	&	 0.12 	&	 -4.94 	&	 68.2 	&	 7.65 	&	 60.5 	&	 7.29 	&	 66 	&	 1 	&	 Fe\,{\sc i} 	&	 6344.15 	&	 2.43 	&	 -2.92 	&	 50.2 	&	 7.38 	&	 - 	&	 - 	&	 169 	&	 1 \\
Fe\,{\sc i} 	&	 5250.65 	&	 2.20 	&	 -2.18 	&	 101.5 	&	 7.59 	&	 92.4 	&	 7.22 	&	 66 	&	 1 	&	 Fe\,{\sc i} 	&	 6392.53 	&	 2.27 	&	 -4.03 	&	 17.1 	&	 7.51 	&	 - 	&	 - 	&	 109 	&	 3 \\
Fe\,{\sc i} 	&	 5253.47 	&	 3.28 	&	 -1.57 	&	 75.4 	&	 7.36 	&	 66.9 	&	 7.02 	&	 553 	&	 1 	&	 Fe\,{\sc i} 	&	 6393.61 	&	 2.43 	&	 -1.58 	&	 130.4 	&	 7.42 	&	 - 	&	 - 	&	 168 	&	 1 \\
Fe\,{\sc i} 	&	 5288.53 	&	 3.69 	&	 -1.51 	&	 57.3 	&	 7.47 	&	 44.8 	&	 7.07 	&	 929 	&	 1 	&	 Fe\,{\sc i} 	&	6408.03	&	3.69	&	-1.02	&	syn	&	7.65	&	syn	&	7.23	&	816	&	 1 \\
Fe\,{\sc i} 	&	 5298.78 	&	 3.64 	&	 -2.02 	&	 42.2 	&	 7.55 	&	 - 	&	 - 	&	 875 	&	 1 	&	 Fe\,{\sc i} 	&	 6419.96 	&	 4.73 	&	 -0.27 	&	 80.5 	&	 7.41 	&	 70.2 	&	 7.10 	&	 1258 	&	 1 \\
Fe\,{\sc i} 	&	 5307.37 	&	 1.61 	&	 -2.99 	&	 86.0 	&	 7.58 	&	 76.2 	&	 7.19 	&	 36 	&	 1 	&	 Fe\,{\sc i} 	&	 6430.86 	&	 2.18 	&	 -2.01 	&	 109.5 	&	 7.41 	&	 101.8 	&	 7.08 	&	 62 	&	 1 \\
Fe\,{\sc i} 	&	 5322.05 	&	 2.28 	&	 -2.80 	&	 60.3 	&	 7.44 	&	 - 	&	 - 	&	 112 	&	 1 	&	 Fe\,{\sc i} 	&	 6469.19 	&	 4.83 	&	 -0.81 	&	 55.0 	&	 7.61 	&	 40.5 	&	 7.22 	&	 1258 	&	 1 \\
Fe\,{\sc i} 	&	 5365.41 	&	 3.57 	&	 -1.22 	&	 76.9 	&	 7.49 	&	 69.8 	&	 7.2 	&	 786 	&	 1 	&	 Fe\,{\sc i} 	&	 6481.88 	&	 2.28 	&	 -2.98 	&	 63.7 	&	 7.59 	&	 55.0 	&	 7.26 	&	 109 	&	 1 \\
Fe\,{\sc i} 	&	 5373.71 	&	 4.47 	&	 -0.84 	&	 61.6 	&	 7.47 	&	 48.1 	&	 7.09 	&	 1166 	&	 1 	&	 Fe\,{\sc i} 	&	 6498.94 	&	 0.96 	&	 -4.69 	&	 44.7 	&	 7.54 	&	 37.9 	&	 7.22 	&	 13 	&	 1 \\
Fe\,{\sc i} 	&	 5379.58 	&	 3.69 	&	 -1.51 	&	 60.8 	&	 7.54 	&	 44.3 	&	 7.04 	&	 928 	&	 1 	&	 Fe\,{\sc i} 	&	 6518.37 	&	 2.83 	&	 -2.46 	&	 56.0 	&	 7.44 	&	 49.1 	&	 7.16 	&	 342 	&	 1 \\
Fe\,{\sc i} 	&	 5398.29 	&	 4.44 	&	 -0.71 	&	 72.5 	&	 7.51 	&	 59.4 	&	 7.13 	&	 553 	&	 1 	&	 Fe\,{\sc i} 	&	 6593.88 	&	 2.43 	&	 -2.42 	&	 85.0 	&	 7.61 	&	 79.0 	&	 7.32 	&	 168 	&	 1 \\
Fe\,{\sc i} 	&	 5461.54 	&	 4.43 	&	 -1.88 	&	 26.0 	&	 7.74 	&	 - 	&	 - 	&	 1145 	&	 1 	&	 Fe\,{\sc i} 	&	 6609.12 	&	 2.56 	&	 -2.69 	&	 64.2 	&	 7.57 	&	 54.9 	&	 7.23 	&	 206 	&	 1 \\
Fe\,{\sc i} 	&	 5473.91 	&	 4.15 	&	 -0.79 	&	 76.9 	&	 7.45 	&	 - 	&	 - 	&	 1062 	&	 1 	&	 Fe\,{\sc i} 	&	 6678.00 	&	 2.69 	&	 -1.42 	&	 122.8 	&	 7.41 	&	 111.7 	&	 7.04 	&	 268 	&	 1 \\
Fe\,{\sc i} 	&	 5483.11 	&	 4.15 	&	 -1.41 	&	 46.5 	&	 7.47 	&	 - 	&	 - 	&	 1061 	&	 1 	&	 Fe\,{\sc i} 	&	 6703.58 	&	 2.76 	&	 -3.06 	&	 36.6 	&	 7.52 	&	 26.9 	&	 7.16 	&	 268 	&	 1 \\
Fe\,{\sc i} 	&	 5487.15 	&	 4.41 	&	 -1.51 	&	 35.6 	&	 7.57 	&	 24.4 	&	 7.21 	&	 1143 	&	 1 	&	 Fe\,{\sc i} 	&	 6750.16 	&	 2.42 	&	 -2.62 	&	 73.1 	&	 7.54 	&	 - 	&	 - 	&	 111 	&	 1 \\
\hline
\end{tabular}}
\end{table*}

\begin{table*}
\setlength{\tabcolsep}{1.2pt}
\renewcommand{\arraystretch}{0.8}
\centering\small
\caption{Fe\,{\sc i} and Fe\,{\sc ii} lines. The abundances were obtained for a model with $T_{\rm eff}=5770$ K, $\log g = 4.40$ cgs, and $\xi=$0.66 km s$^{\rm −1}$ for the solar spectrum. $T_{\rm eff}=5600$ K, $\log g = 4.50$ cgs, and $\xi=$ 0.44 km s$^{\rm −1}$ for the HD\,218209 spectrum.}
\label{tab:lineslit_fe_lines_2}
\resizebox{\linewidth}{!}{%
\begin{tabular}{lccccccccc||ccccccccccc}
\hline
 & & & & \multicolumn{2}{c}{Sun} & \multicolumn{2}{c}{HD\,218209}  & & & & & &\multicolumn{2}{c}{Sun} &\multicolumn{2}{c}{HD\,218209} & \\

\cline{2-4}
\cline{7-8}
\cline{11-13}
\cline{16-17}
 Spec.	&   $\lambda$ & LEP  &  $\log(gf)$ & EW  & $\log \epsilon$(X) & EW	& $\log\epsilon$(X) & RMT &Ref. &Spec. & $\lambda$ & LEP  &  $\log(gf)$ & EW  & $\log\epsilon$(X) & EW  & $\log \epsilon$(X) & RMT & Ref. \\
\cline{2-4}
\cline{7-8}
\cline{11-13}
\cline{16-17}
   & (\AA)  & (eV) & (dex)   &(m\AA) & (dex) &(m\AA) & (dex)&   & & & (\AA)  & (eV) & (dex)   &(m\AA) & (dex)&(m\AA) & (dex) & \\
\hline
Fe\,{\sc i} 	&	 6806.85 	&	 2.72 	&	 -3.21 	&	 34.0 	&	 7.57 	&	 25.2 	&	 7.22 	&	 268 	&	1	&	Fe\,{\sc i} 	&	 8526.68 	&	 4.89 	&	 -0.76 	&	 58.3 	&	 7.62 	&	 - 	&	 - 	&	 1270 	&	 1 \\
Fe\,{\sc i} 	&	 6810.28 	&	 4.59 	&	 -0.99 	&	 48.7 	&	 7.44 	&	 38.1 	&	 7.12 	&	 1197 	&	1	&	 Fe\,{\sc i} 	&	 8582.27 	&	 2.98 	&	 -2.13 	&	 77.6 	&	 7.57 	&	 63.0 	&	 7.16 	&	 401 	&	 1 \\
Fe\,{\sc i} 	&	 6820.43 	&	 4.62 	&	 -1.29 	&	 39.7 	&	 7.63 	&	 24.6 	&	 7.18 	&	 1197 	&	2	&	 Fe\,{\sc i} 	&	 8611.81 	&	 2.83 	&	 -1.85 	&	 98.5 	&	 7.41 	&	 83.9 	&	 7.02 	&	 339 	&	 1 \\
Fe\,{\sc i} 	&	 6862.48 	&	 4.54 	&	 -1.57 	&	 29.3 	&	 7.57 	&	 20.2 	&	 7.24 	&	 1191 	&	 3 	&	 Fe\,{\sc i} 	&	 8613.93 	&	 4.97 	&	 -1.25 	&	 31.5 	&	 7.63 	&	 - 	&	 - 	&	 1272 	&	 3 \\
Fe\,{\sc i} 	&	 6898.31 	&	 4.20 	&	 -2.23 	&	 16.6 	&	 7.55 	&	 - 	&	 - 	&	 1078 	&	 3 	&	 Fe\,{\sc i} 	&	 8616.27 	&	 4.89 	&	 -0.71 	&	 43.4 	&	 7.27 	&	 - 	&	 - 	&	 1266 	&	 3 \\
Fe\,{\sc i} 	&	 6916.70 	&	 4.14 	&	 -1.40 	&	 57.1 	&	 7.55 	&	 45.4 	&	 7.20 	&	 1052 	&	 1 	&	 Fe\,{\sc i} 	&	 8699.43 	&	 4.93 	&	 -0.38 	&	 65.7 	&	 7.40 	&	 - 	&	 - 	&	 1267 	&	 1 \\
Fe\,{\sc i} 	&	 6977.44 	&	 4.57 	&	 -1.56 	&	 19.9 	&	 7.35 	&	 - 	&	 - 	&	 1225 	&	 3 	&	 Fe\,{\sc i} 	&	 8757.19 	&	 2.83 	&	 -1.92 	&	 92.7 	&	 7.39 	&	 82.9 	&	 7.06 	&	 339 	&	 1 \\
Fe\,{\sc i} 	&	 6999.90 	&	 4.09 	&	 -1.51 	&	 54.0 	&	 7.56 	&	 43.4 	&	 7.23 	&	 1051 	&	 2 	&	 Fe\,{\sc i} 	&	 8793.38 	&	 4.59 	&	 -0.09 	&	 107.9 	&	 7.45 	&	 98.3 	&	 7.16 	&	 1172 	&	 3 \\
Fe\,{\sc i} 	&	 7016.07 	&	 2.41 	&	 -3.21 	&	 50.8 	&	 7.62 	&	 - 	&	 - 	&	 109 	&	 3 	&	 Fe\,{\sc i} 	&	 8796.42 	&	 4.93 	&	 -1.23 	&	 27.3 	&	 7.46 	&	 - 	&	 - 	&	 1266 	&	 3 \\
Fe\,{\sc i} 	&	 7022.98 	&	 4.17 	&	 -1.20 	&	 63.5 	&	 7.48 	&	 50.6 	&	 7.11 	&	 1051 	&	 1 	&	 Fe\,{\sc i} 	&	 8798.05 	&	 4.96 	&	 -1.89 	&	 8.0 	&	 7.47 	&	 - 	&	 - 	&	 1286 	&	 3 \\
Fe\,{\sc i} 	&	 7038.25 	&	 4.20 	&	 -1.25 	&	 60.0 	&	 7.49 	&	 - 	&	 - 	&	 1051 	&	 1 	&	 Fe\,{\sc i} 	&	 8834.04 	&	 4.20 	&	 -2.59 	&	 8.0 	&	 7.44 	&	 - 	&	 - 	&	 1050 	&	 3 \\
Fe\,{\sc i} 	&	 7071.88 	&	 4.59 	&	 -1.70 	&	 26.5 	&	 7.67 	&	 18.6 	&	 7.37 	&	 1194 	&	 3 	&	 Fe\,{\sc i} 	&	 8838.43 	&	 2.85 	&	 -1.87 	&	 97.5 	&	 7.42 	&	 83.9 	&	 7.04 	&	 339 	&	 1 \\
Fe\,{\sc i} 	&	 7090.40 	&	 4.21 	&	 -1.16 	&	 64.5 	&	 7.49 	&	 50.0 	&	 7.10 	&	 1051 	&	 1 	&	 Fe\,{\sc i} 	&	 8846.82 	&	 4.99 	&	 -0.78 	&	 48.5 	&	 7.52 	&	 39.3 	&	 7.26 	&	 1267 	&	 3 \\
Fe\,{\sc i} 	&	 7130.94 	&	 4.20 	&	 -0.80 	&	 87.3 	&	 7.46 	&	 73.0 	&	 7.08 	&	 1051 	&	 1 	&	 Fe\,{\sc i} 	&	 8878.26 	&	 2.98 	&	 -3.83 	&	 11.7 	&	 7.67 	&	 - 	&	 - 	&	 401 	&	 3 \\
Fe\,{\sc i} 	&	 7132.99 	&	 4.06 	&	 -1.63 	&	 41.6 	&	 7.47 	&	 31.3 	&	 7.13 	&	 1002 	&	 1 	&	 Fe\,{\sc i} 	&	 8887.10 	&	 4.93 	&	 -1.94 	&	 4.8 	&	 7.25 	&	 - 	&	 - 	&	 1265 	&	 3 \\
Fe\,{\sc i} 	&	 7180.02 	&	 1.48 	&	 -4.78 	&	 20.0 	&	 7.52 	&	 - 	&	 - 	&	 1 	&	 3 	&	 Fe\,{\sc i} 	&	 8902.94 	&	 4.97 	&	 -2.11 	&	 8.6 	&	 7.73 	&	 - 	&	 - 	&	 1266 	&	 3 \\
Fe\,{\sc i} 	&	 7212.47 	&	 4.93 	&	 -0.83 	&	 30.4 	&	 7.23 	&	 - 	&	 - 	&	 1273 	&	 3 	&	 Fe\,{\sc i} 	&	 8922.66 	&	 4.97 	&	 -1.70 	&	 12.9 	&	 7.53 	&	 - 	&	 - 	&	 1298 	&	 3 \\
Fe\,{\sc i} 	&	 7219.69 	&	 4.07 	&	 -1.69 	&	 45.0 	&	 7.61 	&	 35.6 	&	 7.30 	&	 1001 	&	 3 	&	 Fe\,{\sc i} 	&	 8945.20 	&	 5.01 	&	 -0.22 	&	 72.2 	&	 7.40 	&	 49.6 	&	 6.91 	&	 1301 	&	 3 \\
Fe\,{\sc i} 	&	 7221.22 	&	 4.54 	&	 -1.18 	&	 40.5 	&	 7.44 	&	 26.1 	&	 7.02 	&	 1189 	&	 3 	&	 Fe\,{\sc i} 	&	 8950.20 	&	 4.14 	&	 -2.43 	&	 13.1 	&	 7.46 	&	 - 	&	 - 	&	 1050 	&	 3 \\
Fe\,{\sc i} 	&	 7222.88 	&	 4.59 	&	 -2.04 	&	 15.2 	&	 7.68 	&	 - 	&	 - 	&	 1187 	&	 3 	&	 Fe\,{\sc i} 	&	 8959.88 	&	 5.00 	&	 -1.84 	&	 8.9 	&	 7.50 	&	 - 	&	 - 	&	 1320 	&	 3 \\
Fe\,{\sc i} 	&	 7228.70 	&	 2.75 	&	 -3.38 	&	 27.2 	&	 7.58 	&	 23.7 	&	 7.36 	&	 2 	&	 3 	&	 Fe\,{\sc i} 	&	 8975.41 	&	 2.98 	&	 -2.22 	&	 77.3 	&	 7.60 	&	 - 	&	 - 	&	 400 	&	 1 \\
Fe\,{\sc i} 	&	 7284.84 	&	 4.12 	&	 -1.57 	&	 41.6 	&	 7.46 	&	 27.8 	&	 7.04 	&	 1004 	&	 3 	&	 Fe\,{\sc i} 	&	 8984.87 	&	 5.08 	&	 -0.92 	&	 32.5 	&	 7.40 	&	 - 	&	 - 	&	 1301 	&	 3 \\
Fe\,{\sc i} 	&	 7306.61 	&	 4.16 	&	 -1.44 	&	 42.0 	&	 7.37 	&	 30.3 	&	 7.00 	&	 1077 	&	 3 	&	 Fe\,{\sc i} 	&	 9010.55 	&	 2.60 	&	 -2.95 	&	 44.1 	&	 7.27 	&	 35.2 	&	 6.95 	&	 202 	&	 1 \\
Fe\,{\sc i} 	&	 7351.16 	&	 4.97 	&	 -0.84 	&	 36.2 	&	 7.40 	&	 23.0 	&	 7.00 	&	 1275 	&	 3 	&	 Fe\,{\sc i} 	&	 9030.67 	&	 2.83 	&	 -3.64 	&	 25.5 	&	 7.77 	&	 - 	&	 - 	&	 338 	&	 1 \\
Fe\,{\sc i} 	&	 7351.56 	&	 4.93 	&	 -0.64 	&	 45.5 	&	 7.36 	&	 34.3 	&	 7.04 	&	 1275 	&	 3 	&	 Fe\,{\sc i} 	&	 9070.42 	&	 4.20 	&	 -2.05 	&	 33.7 	&	 7.71 	&	 - 	&	 - 	&	 1076 	&	 3 \\
Fe\,{\sc i} 	&	 7396.50 	&	 4.97 	&	 -1.64 	&	 12.6 	&	 7.53 	&	 - 	&	 - 	&	 1278 	&	 3 	&	 Fe\,{\sc i} 	&	 9079.60 	&	 4.63 	&	 -0.81 	&	 54.1 	&	 7.31 	&	 - 	&	 - 	&	 1172 	&	 3 \\
Fe\,{\sc i} 	&	 7401.69 	&	 4.17 	&	 -1.35 	&	 40.9 	&	 7.26 	&	 - 	&	 - 	&	 1004 	&	 2 	&	 Fe\,{\sc i} 	&	 9089.41 	&	 2.94 	&	 -1.68 	&	 99.8 	&	 7.41 	&	 88.9 	&	 7.09 	&	 400 	&	 1 \\
Fe\,{\sc i} 	&	 7411.18 	&	 4.26 	&	 -0.30 	&	 101.9 	&	 7.43 	&	 - 	&	 - 	&	 4 	&	 3 	&	 Fe\,{\sc i} 	&	 9117.10 	&	 2.85 	&	 -3.46 	&	 32.3 	&	 7.76 	&	 - 	&	 - 	&	 338 	&	 3 \\
Fe\,{\sc i} 	&	 7418.67 	&	 4.12 	&	 -1.38 	&	 48.7 	&	 7.41 	&	 - 	&	 - 	&	 4 	&	 1 	&	 Fe\,{\sc i} 	&	 9156.23 	&	 3.00 	&	 -3.67 	&	 9.3 	&	 7.40 	&	 - 	&	 - 	&	 400 	&	 3 \\
Fe\,{\sc i} 	&	 7443.03 	&	 4.17 	&	 -1.82 	&	 34.7 	&	 7.59 	&	 28.0 	&	 7.33 	&	 1309 	&	 1 	&	 Fe\,{\sc i} 	&	 9210.03 	&	 2.83 	&	 -2.40 	&	 65.0 	&	 7.37 	&	 - 	&	 - 	&	 83 	&	 2 \\
Fe\,{\sc i} 	&	 7447.43 	&	 4.93 	&	 -0.85 	&	 34.1 	&	 7.32 	&	 24.8 	&	 7.02 	&	 1273 	&	 3 	&	 Fe\,{\sc i} 	&	 9382.93 	&	 4.96 	&	 -1.59 	&	 24.3 	&	 7.74 	&	 - 	&	 - 	&	 1284 	&	 3 \\
Fe\,{\sc i} 	&	 7454.02 	&	 4.17 	&	 -2.41 	&	 12.0 	&	 7.51 	&	 - 	&	 - 	&	 5 	&	 3 	&	 Fe\,{\sc i} 	&	 9602.07 	&	 4.99 	&	 -1.74 	&	 15.4 	&	 7.64 	&	 - 	&	 - 	&	 1283 	&	 3 \\
Fe\,{\sc i} 	&	 7473.56 	&	 4.59 	&	 -1.87 	&	 18.2 	&	 7.60 	&	 - 	&	 - 	&	 1188 	&	 3 	&	 Fe\,{\sc i} 	&	 9653.14 	&	 4.71 	&	 -0.68 	&	 68.4 	&	 7.46 	&	 - 	&	 - 	&	 1247 	&	 3 \\
Fe\,{\sc i} 	&	 7491.68 	&	 4.28 	&	 -0.90 	&	 64.9 	&	 7.42 	&	 53.1 	&	 7.08 	&	 1077 	&	 3 	&	 Fe\,{\sc i} 	&	 9753.13 	&	 4.77 	&	 -0.78 	&	 56.9 	&	 7.40 	&	 - 	&	 - 	&	 1247 	&	 3 \\
Fe\,{\sc i} 	&	 7498.56 	&	 4.12 	&	 -2.25 	&	 18.0 	&	 7.52 	&	 13.5 	&	 7.27 	&	 1001 	&	 3 	&	 Fe\,{\sc i} 	&	 9786.62 	&	 4.59 	&	 -1.68 	&	 18.5 	&	 7.28 	&	 - 	&	 - 	&	 1171 	&	 3 \\
Fe\,{\sc i} 	&	 7511.05 	&	 4.16 	&	 0.09 	&	 151.6 	&	 7.46 	&	 141.1 	&	 7.14 	&	 1077 	&	 1 	&	 Fe\,{\sc i} 	&	 9800.34 	&	 5.06 	&	 -0.45 	&	 59.7 	&	 7.38 	&	 - 	&	 - 	&	 1292 	&	 3 \\
Fe\,{\sc i} 	&	 7540.44 	&	 2.72 	&	 -3.85 	&	 11.5 	&	 7.51 	&	 - 	&	 - 	&	 266 	&	 3 	&	 Fe\,{\sc i} 	&	 9861.79 	&	 5.04 	&	 -0.14 	&	 73.8 	&	 7.28 	&	 - 	&	 - 	&	 1296 	&	 1 \\
Fe\,{\sc i} 	&	 7551.10 	&	 5.06 	&	 -1.63 	&	 11.0 	&	 7.53 	&	 - 	&	 - 	&	 1303 	&	 3 	&	 Fe\,{\sc i} 	&	 9881.51 	&	 4.56 	&	 -1.71 	&	 18.1 	&	 7.26 	&	 - 	&	 - 	&	 1209 	&	 3 \\
Fe\,{\sc i} 	&	 7568.93 	&	 4.26 	&	 -0.77 	&	 74.3 	&	 7.45 	&	 64.4 	&	 7.15 	&	 1077 	&	 3 	&	 Fe\,{\sc i} 	&	 9889.08 	&	 5.01 	&	 -0.45 	&	 75.4 	&	 7.59 	&	 - 	&	 - 	&	 1296 	&	 1 \\
Fe\,{\sc i} 	&	 7583.80 	&	 3.00 	&	 -1.89 	&	 82.3 	&	 7.46 	&	 69.1 	&	 7.07 	&	 402 	&	 1 	&	 Fe\,{\sc i} 	&	 9944.13 	&	 4.99 	&	 -1.34 	&	 30.8 	&	 7.63 	&	 - 	&	 - 	&	 1285 	&	 3 \\
Fe\,{\sc i} 	&	 7586.04 	&	 4.29 	&	 -0.47 	&	 112.3 	&	 7.74 	&	 106 	&	 7.48 	&	 1137 	&	 3 	&	 Fe\,{\sc ii} 	&	 4178.86 	&	 2.58 	&	 -2.51 	&	 83.8 	&	 7.38 	&	 67.3 	&	 6.99 	&	 28 	&	 1 \\
Fe\,{\sc i} 	&	 7620.54 	&	 4.71 	&	 -0.66 	&	 55.4 	&	 7.38 	&	 - 	&	 - 	&	 1250 	&	 3 	&	 Fe\,{\sc ii} 	&	 4491.40 	&	 2.84 	&	 -2.64 	&	 74.8 	&	 7.50 	&	 - 	&	 - 	&	 37 	&	 1 \\
Fe\,{\sc i} 	&	 7653.78 	&	 4.77 	&	 -0.89 	&	 34.4 	&	 7.21 	&	 - 	&	 - 	&	 1250 	&	 1 	&	 Fe\,{\sc ii} 	&	 4508.29 	&	 2.85 	&	 -2.44 	&	 85.5 	&	 7.52 	&	 73.7 	&	 7.26 	&	 38 	&	 1 \\
Fe\,{\sc i} 	&	 7710.39 	&	 4.20 	&	 -1.11 	&	 64.9 	&	 7.54 	&	 61.0 	&	 7.36 	&	 1077 	&	 1 	&	 Fe\,{\sc ii} 	&	 4576.34 	&	 2.84 	&	 -2.92 	&	 63.8 	&	 7.48 	&	 - 	&	 - 	&	 38 	&	 1 \\
Fe\,{\sc i} 	&	 7719.05 	&	 5.01 	&	 -1.15 	&	 28.3 	&	 7.55 	&	 19.2 	&	 7.23 	&	 1304 	&	 3 	&	 Fe\,{\sc ii} 	&	 4582.83 	&	 2.84 	&	 -3.06 	&	 55.9 	&	 7.40 	&	 - 	&	 - 	&	 37 	&	 1 \\
Fe\,{\sc i} 	&	 7723.20 	&	 2.27 	&	 -3.62 	&	 42.2 	&	 7.65 	&	 - 	&	 - 	&	 108 	&	 2 	&	 Fe\,{\sc ii} 	&	 4620.52 	&	 2.83 	&	 -3.19 	&	 52.0 	&	 7.41 	&	 37.1 	&	 7.03 	&	 38 	&	 1 \\
Fe\,{\sc i} 	&	 7745.48 	&	 5.06 	&	 -1.17 	&	 21.7 	&	 7.44 	&	 - 	&	 - 	&	 1305 	&	 3 	&	 Fe\,{\sc ii} 	&	 4993.36 	&	 2.79 	&	 -3.68 	&	 37.7 	&	 7.46 	&	 - 	&	 - 	&	 36 	&	 1 \\
Fe\,{\sc i} 	&	 7748.28 	&	 2.94 	&	 -1.75 	&	 100.6 	&	 7.54 	&	 91.9 	&	 7.22 	&	 402 	&	 1 	&	 Fe\,{\sc ii} 	&	 5132.67 	&	 2.81 	&	 -4.09 	&	 25.1 	&	 7.53 	&	 14.6 	&	 7.20 	&	 35 	&	 1 \\
Fe\,{\sc i} 	&	 7780.59 	&	 4.45 	&	 0.03 	&	 114.7 	&	 7.39 	&	 103.4 	&	 7.07 	&	 1154 	&	 3 	&	 Fe\,{\sc ii} 	&	 5197.58 	&	 3.23 	&	 -2.22 	&	 79.8 	&	 7.47 	&	 - 	&	 - 	&	 49 	&	 1 \\
Fe\,{\sc i} 	&	 7832.22 	&	 4.42 	&	 0.11 	&	 118.0 	&	 7.31 	&	 112.0 	&	 7.06 	&	 1154 	&	 3 	&	 Fe\,{\sc ii} 	&	 5234.63 	&	 3.22 	&	 -2.21 	&	 82.1 	&	 7.49 	&	 67.7 	&	 7.19 	&	 49 	&	 1 \\
Fe\,{\sc i} 	&	 7844.55 	&	 4.81 	&	 -1.70 	&	 12.8 	&	 7.43 	&	 - 	&	 - 	&	 1250 	&	 3 	&	 Fe\,{\sc ii} 	&	 5264.81 	&	 3.33 	&	 -3.13 	&	 45.8 	&	 7.63 	&	 33.9 	&	 7.35 	&	 48 	&	 1 \\
Fe\,{\sc i} 	&	 7879.75 	&	 5.01 	&	 -1.47 	&	 10.2 	&	 7.27 	&	 - 	&	 - 	&	 1306 	&	 3 	&	 Fe\,{\sc ii} 	&	5284.11	&	2.89	&	-3.11	&	syn	&	7.5	&	syn	&	6.98	&	41	&	 1 \\
Fe\,{\sc i} 	&	 7912.87 	&	 0.86 	&	 -4.84 	&	 48.3 	&	 7.56 	&	 36.0 	&	 7.12 	&	 12 	&	 1 	&	 Fe\,{\sc ii} 	&	 5325.56 	&	 3.21 	&	 -3.26 	&	 42.5 	&	 7.56 	&	 - 	&	 - 	&	 49 	&	 1 \\
Fe\,{\sc i} 	&	 7941.09 	&	 3.26 	&	 -2.29 	&	 41.7 	&	 7.28 	&	 - 	&	 - 	&	 623 	&	 1 	&	 Fe\,{\sc ii} 	&	 5414.07 	&	 3.22 	&	 -3.58 	&	 28.2 	&	 7.49 	&	 15.7 	&	 7.12 	&	 48 	&	 1 \\
Fe\,{\sc i} 	&	 7998.97 	&	 4.35 	&	 0.15 	&	 129.7 	&	 7.33 	&	 - 	&	 - 	&	 1136 	&	 3 	&	 Fe\,{\sc ii} 	&	 5425.26 	&	 3.20 	&	 -3.22 	&	 41.8 	&	 7.48 	&	 26.5 	&	 7.09 	&	 49 	&	 1 \\
Fe\,{\sc i} 	&	 8028.34 	&	 4.45 	&	 -0.69 	&	 68.2 	&	 7.39 	&	 64.0 	&	 7.20 	&	 1154 	&	 3 	&	 Fe\,{\sc ii} 	&	 5534.85 	&	 3.23 	&	 -2.75 	&	 57.4 	&	 7.46 	&	 39.3 	&	 7.02 	&	 55 	&	 1 \\
Fe\,{\sc i} 	&	 8047.60 	&	 0.86 	&	 -4.79 	&	 60.4 	&	 7.77 	&	 - 	&	 - 	&	 12 	&	 3 	&	 Fe\,{\sc ii} 	&	 6084.11 	&	 3.19 	&	 -3.88 	&	 20.3 	&	 7.52 	&	 8.9 	&	 7.07 	&	 46 	&	 1 \\
Fe\,{\sc i} 	&	 8096.87 	&	 4.06 	&	 -1.78 	&	 35.1 	&	 7.41 	&	 - 	&	 - 	&	 999 	&	 1 	&	 Fe\,{\sc ii} 	&	 6149.24 	&	 3.87 	&	 -2.84 	&	 35.9 	&	 7.57 	&	 23.2 	&	 7.25 	&	 74 	&	 1 \\
Fe\,{\sc i} 	&	 8204.10 	&	 0.91 	&	 -6.05 	&	 5.8 	&	 7.52 	&	 - 	&	 - 	&	 12 	&	 3 	&	 Fe\,{\sc ii} 	&	 6238.38 	&	 3.87 	&	 -2.75 	&	 42.1 	&	 7.64 	&	 - 	&	 - 	&	 74 	&	 1 \\
Fe\,{\sc i} 	&	 8207.77 	&	 4.43 	&	 -0.86 	&	 65.4 	&	 7.48 	&	 47.4 	&	 7.02 	&	 1136 	&	 3 	&	 Fe\,{\sc ii} 	&	 6247.56 	&	 3.89 	&	 -2.30 	&	 52.0 	&	 7.46 	&	 38.2 	&	 7.16 	&	 74 	&	 1 \\
Fe\,{\sc i} 	&	 8239.13 	&	 2.41 	&	 -3.18 	&	 44.9 	&	 7.38 	&	 36.4 	&	 7.07 	&	 108 	&	 3 	&	 Fe\,{\sc ii} 	&	 6432.68 	&	 2.89 	&	 -3.57 	&	 40.3 	&	 7.47 	&	 24.8 	&	 7.07 	&	 40 	&	 1 \\
Fe\,{\sc i} 	&	 8248.15 	&	 4.35 	&	 -0.89 	&	 60.5 	&	 7.32 	&	 50.7 	&	 7.03 	&	 1136 	&	 3 	&	 Fe\,{\sc ii} 	&	 6456.39 	&	 3.90 	&	 -2.05 	&	 62.1 	&	 7.45 	&	 46.6 	&	 7.12 	&	 74 	&	 1 \\
Fe\,{\sc i} 	&	 8293.53 	&	 3.30 	&	 -2.14 	&	 57.6 	&	 7.50 	&	 46.8 	&	 7.17 	&	 623 	&	 1 	&	 Fe\,{\sc ii} 	&	 6516.08 	&	 2.89 	&	 -3.31 	&	 53.4 	&	 7.53 	&	 40.0 	&	 7.22 	&	 40 	&	 1 \\
Fe\,{\sc i} 	&	 8360.82 	&	 4.45 	&	 -1.29 	&	 57.2 	&	 7.74 	&	 45.3 	&	 7.41 	&	 1153 	&	 3 	&	 Fe\,{\sc ii} 	&	 7222.39 	&	 3.87 	&	 -3.40 	&	 18.9 	&	 7.60 	&	 - 	&	 - 	&	 73 	&	 1 \\
Fe\,{\sc i} 	&	 8365.64 	&	 3.24 	&	 -1.91 	&	 69.0 	&	 7.45 	&	 55.4 	&	 7.06 	&	 623 	&	 1 	&	 Fe\,{\sc ii} 	&	 7224.51 	&	 3.87 	&	 -3.36 	&	 19.4 	&	 7.58 	&	 - 	&	 - 	&	 73 	&	 1 \\
Fe\,{\sc i} 	&	 8424.14 	&	 4.93 	&	 -1.16 	&	 33.4 	&	 7.56 	&	 20.7 	&	 7.17 	&	 1272 	&	 3 	&	 Fe\,{\sc ii} 	&	 7515.88 	&	 3.89 	&	 -3.39 	&	 13.1 	&	 7.38 	&	 - 	&	 - 	&	 73 	&	 1 \\
Fe\,{\sc i} 	&	 8439.60 	&	 4.53 	&	 -0.59 	&	 73.2 	&	 7.41 	&	 61.6 	&	 7.09 	&	 1172 	&	 3 	&	 Fe\,{\sc ii} 	&	 7533.42 	&	 3.89 	&	 -3.60 	&	 17.7 	&	 7.77 	&	 - 	&	 - 	&	 72 	&	 3 \\
Fe\,{\sc i} 	&	 8514.08 	&	 2.19 	&	 -2.23 	&	 116.1 	&	 7.46 	&	 102.4 	&	 7.07 	&	 60 	&	 1 	&	 Fe\,{\sc ii} 	&	 7655.47 	&	 3.87 	&	 -3.77 	&	 7.1 	&	 7.41 	&	 - 	&	 - 	&	 73 	&	 3 \\
Fe\,{\sc i} 	&	 8515.08 	&	 3.00 	&	 -2.07 	&	 83.0 	&	 7.64 	&	 71.4 	&	 7.30 	&	 401 	&	 2 	&	 Fe\,{\sc ii} 	&	 7711.71 	&	 3.89 	&	 -2.45 	&	 44.7 	&	 7.36 	&	 30.4 	&	 7.04 	&	 73 	&	 1 \\
\hline
\end{tabular}}
\end{table*}

\begin{table*}
\setlength{\tabcolsep}{1.2pt}
\renewcommand{\arraystretch}{0.8}
\centering\small
\caption{The abundances were obtained for a model with $T_{\rm eff}=5770$ K, $\log g =4.40$ cgs, and $\xi=$ 0.66 km s$^{\rm −1}$ for the solar spectrum. $T_{\rm eff}=5600$ K, $\log g = 4.50$ cgs, and $\xi=$ 0.44 km s$^{\rm −1}$ for the HD\,218209 spectrum.}
\label{tab:lineslit_other_lines}
\resizebox{\linewidth}{!}{%
\begin{tabular}{lccccccccc||cccccccccc}
\hline
 & & & & \multicolumn{2}{c}{Sun} & \multicolumn{2}{c}{HD\,218209}  & & & &  &&   &\multicolumn{2}{c}{Sun} &\multicolumn{2}{c}{HD\,218209}&  & \\
\cline{2-4}
\cline{7-8}
\cline{12-14}
\cline{17-18}
 Spec.& $\lambda$ & LEP& $\log(gf)$ & EW & $\log \epsilon$(X) & EW& $\log\epsilon$(X)&RMT&Ref.&Spec.&$\lambda$ & LEP&  $\log(gf)$ & EW     & $\log \epsilon$(X) & EW	& $\log\epsilon$(X) & RMT &Ref.\\
\cline{2-4}
\cline{7-8}
\cline{12-14}
\cline{17-18}
   & (\AA)  & (eV) & (dex)   &(m\AA) & (dex) &(m\AA) & (dex)&    &    && (\AA)  & (eV) & (dex)   &(m\AA) & (dex) &(m\AA) & (dex)&\\
\hline
C\,{\sc i} 	&	 8335.19 	&	 7.65 	&	 -0.44 	&	 syn 	&	 8.41 	&	 syn 	&	 8.16 	&	 10 	&	 2 	&	Sc\,{\sc ii} 	&	 5526.82 	&	 1.77 	&	 -0.01 	&	 75.3 	&	 3.32 	&	 68.0 	&	 3.13 	&	 18 	&	 9 \\
C\,{\sc i} 	&	 9111.85 	&	 7.46 	&	 -0.34 	&	 syn 	&	 8.56 	&	 syn 	&	 8.34 	&	 3 	&	 4 	&	Sc\,{\sc ii} 	&	 5640.99 	&	1.5	&	-0.99	&	 syn 	&	3.12	&	 syn 	&	2.84	&	29	&	9\\
O\,{\sc i} 	&	 7771.96 	&	 9.11 	&	 0.37 	&	 syn 	&	 8.83 	&	 syn 	&	 8.73 	&	 1 	&	 2 	&	Sc\,{\sc ii} 	&	 5657.88 	&	 1.51 	&	 -0.54 	&	 67.1 	&	 3.39 	&	 57.4 	&	 3.14 	&	 29 	&	9\\ 
O\,{\sc i} 	&	 7774.18 	&	 9.11 	&	 0.22 	&	 syn 	&	 8.83 	&	 syn 	&	 8.74 	&	 1 	&	 2 	&	Sc\,{\sc ii} 	&	 5667.15 	&	 1.50 	&	 -1.21 	&	 32.8 	&	 3.24 	&	 26.5 	&	 3.04 	&	 29 	&	9\\ 
O\,{\sc i} 	&	 7775.40 	&	 9.11 	&	 0.00 	&	 syn 	&	 8.77 	&	 syn 	&	 8.69 	&	 1 	&	 2 	&	Sc\,{\sc ii} 	&	 5669.04 	&	 1.50 	&	 -1.10 	&	 34.2 	&	 3.16 	&	 28.1 	&	 2.97 	&	 29 	&	 9 \\
Na\,{\sc i} 	&	 5682.65 	&	2.1	&	-0.7	&	 syn 	&	6.36	&	 syn 	&	5.87	&	6	&	2	&	 Sc\,{\sc ii} 	&	 6245.63 	&	 1.50 	&	 -1.02 	&	 34.0 	&	 3.05 	&	 30.4 	&	 2.92 	&	 28 	&	 2 \\
Na\,{\sc i} 	&	 5688.22 	&	 2.10 	&	 -0.37 	&	 122.1 	&	 6.11 	&	 101.1 	&	 5.73 	&	 6 	&	 4 	&	 Sc\,{\sc ii} 	&	 6279.76 	&	 1.49 	&	 -1.33 	&	 28.6 	&	 3.22 	&	 21.0 	&	 2.97 	&	 28 	&	 3 \\
Na\,{\sc i} 	&	 6154.23 	&	 2.09 	&	 -1.55 	&	 36.9 	&	 6.27 	&	 22.9 	&	 5.89 	&	 5 	&	 2 	&	 Sc\,{\sc ii} 	&	 6300.70 	&	 1.50 	&	 -1.90 	&	 7.1 	&	 3.02 	&	 -- 	&	 -- 	&	 28 	&	 3 \\
Na\,{\sc i} 	&	 8183.26 	&	 2.09 	&	 0.22 	&	 201.9 	&	 6.13 	&	 183.1 	&	 5.79 	&	 4 	&	 3 	&	 Sc\,{\sc ii} 	&	 6320.85 	&	 1.49 	&	 -1.82 	&	 9.0 	&	 3.05 	&	 -- 	&	 -- 	&	 28 	&	 3 \\
Mg\,{\sc i} 	&	 4571.10 	&	 0.00 	&	 -5.40 	&	 syn 	&	 7.57 	&	 syn 	&	 7.58 	&	 1 	&	 5 	&	 Sc\,{\sc ii} 	&	 6604.60 	&	 1.35 	&	 -1.31 	&	 35.3 	&	 3.21 	&	 30.9 	&	 3.06 	&	 19 	&	 2 \\
Mg\,{\sc i} 	&	 5711.10 	&	 4.34 	&	 -1.74 	&	 syn 	&	 7.63 	&	 syn 	&	 7.55 	&	 8 	&	 5 	&	 Ti\,{\sc i} 	&	4060.27	&	1.05	&	-0.69	&	 syn 	&	4.85	&	 syn 	&	4.51	&	80	&	 10\\
Mg\,{\sc i} 	&	 7657.60 	&	 5.09 	&	 -1.27 	&	 syn 	&	 7.64 	&	 syn 	&	 7.45 	&	 22 	&	 2 	&	Ti\,{\sc i} 	&	4186.13	&	1.5	&	-0.24	&	 syn 	&	4.89	&	 syn 	&	4.67	&	129	&	 10\\
Mg\,{\sc i} 	&	 7691.57 	&	 5.73 	&	 -0.78 	&	 syn 	&	 7.65 	&	 syn 	&	 7.40 	&	 29 	&	 2 	&	 Ti\,{\sc i} 	&	 4287.41 	&	 0.84 	&	 -0.37 	&	 68.6 	&	 5.02 	&	 73.2 	&	 4.94 	&	 44 	&	 10 \\
Mg\,{\sc i} 	&	 8213.02 	&	 5.73 	&	 -0.51 	&	 syn 	&	 7.60 	&	 syn 	&	 7.47 	&	 28 	&	 2 	&	 Ti\,{\sc i} 	&	 4453.32 	&	 1.43 	&	 -0.03 	&	 66.6 	&	 5.19 	&	 64.7 	&	 5.00 	&	 113 	&	 10 \\
Mg\,{\sc ii} 	&	 7896.37 	&	 9.96 	&	 0.64 	&	 syn 	&	 7.63 	&	 syn 	&	 -- 	&	 8 	&	 2 	&	 Ti\,{\sc i} 	&	 4465.81 	&	 1.74 	&	 -0.13 	&	 39.2 	&	 4.89 	&	 33.9 	&	 4.62 	&	 146 	&	 10 \\
Al\,{\sc i} 	&	 6695.97 	&	 3.13 	&	 -1.57 	&	 syn 	&	 6.48 	&	 syn 	&	 6.17 	&	 5 	&	 2 	&	 Ti\,{\sc i} 	&	 4512.74 	&	 0.84 	&	 -0.40 	&	 66.8 	&	 4.96 	&	 65.8 	&	 4.76 	&	 42 	&	 10 \\
Al\,{\sc i} 	&	 6698.63 	&	 3.13 	&	 -1.87 	&	 syn 	&	 6.44 	&	 syn 	&	 6.07 	&	 5 	&	 2 	&	 Ti\,{\sc i} 	&	 4518.03 	&	 0.83 	&	 -0.25 	&	 73.5 	&	 4.96 	&	 74.6 	&	 4.80 	&	 42 	&	 10 \\
Al\,{\sc i} 	&	 7362.31 	&	 4.00 	&	 -0.79 	&	 syn 	&	 6.38 	&	 syn 	&	 6.17 	&	 11 	&	 6 	&	 Ti\,{\sc i} 	&	 4534.79 	&	 0.84 	&	 0.35 	&	 96.4 	&	 4.84 	&	 100.8 	&	 4.65 	&	 42 	&	 10 \\
Al\,{\sc i} 	&	 7835.33 	&	 4.00 	&	 -0.69 	&	 syn 	&	 6.45 	&	 syn 	&	 6.19 	&	 10 	&	 6 	&	 Ti\,{\sc i} 	&	 4548.77 	&	 0.83 	&	 -0.28 	&	 71.5 	&	 4.94 	&	 76.3 	&	 4.86 	&	 270 	&	 10 \\
Al\,{\sc i} 	&	 7836.15 	&	 4.00 	&	 -0.50 	&	 syn 	&	 6.46 	&	 syn 	&	 6.19 	&	 10 	&	 6 	&	 Ti\,{\sc i} 	&	 4555.49 	&	 0.85 	&	 -0.40 	&	 64.1 	&	 4.89 	&	 66.3 	&	 4.78 	&	 266 	&	 10 \\
Al\,{\sc i} 	&	 8772.88 	&	 4.00 	&	 -0.35 	&	 syn 	&	 6.43 	&	 syn 	&	 6.28 	&	 9 	&	 6 	&	 Ti\,{\sc i} 	&	 4617.28 	&	 1.75 	&	 0.44 	&	 62.6 	&	 4.89 	&	 60.5 	&	 4.70 	&	 145 	&	 10 \\
Al\,{\sc i} 	&	 8773.91 	&	 4.00 	&	 -0.16 	&	 syn 	&	 6.43 	&	 syn 	&	 6.28 	&	 9 	&	 6 	&	Ti\,{\sc i} 	&	4623.1	&	1.74	&	0.16	&	 syn 	&	4.95	&	 syn 	&	4.58	&	145	&	 10\\
Al\,{\sc i} 	&	 8841.26 	&	 4.07 	&	 -1.50 	&	 syn 	&	 6.42 	&	 syn 	&	 6.16 	&	 15 	&	 2 	&	Ti\,{\sc i} 	&	4639.36	&	1.74	&	-0.05	&	 syn 	&	4.94	&	 syn 	&	4.58	&	145	&	 10\\
Si\,{\sc i} 	&	 5645.62 	&	 4.93 	&	 -2.03 	&	 35.0 	&	 7.49 	&	 25.1 	&	 7.23 	&	 10 	&	 7 	&	Ti\,{\sc i} 	&	4639.66	&	1.75	&	-0.14	&	 syn 	&	5	&	 syn 	&	4.59	&	145	&	 10\\
Si\,{\sc i} 	&	 5665.56 	&	 4.92 	&	 -1.99 	&	 39.7 	&	 7.53 	&	 28.6 	&	 7.25 	&	 10 	&	 7 	&	 Ti\,{\sc i} 	&	 4656.47 	&	 0.00 	&	 -1.28 	&	 71.3 	&	 5.11 	&	 71.4 	&	 4.94 	&	 6 	&	 10 \\
Si\,{\sc i} 	&	 5684.49 	&	 4.95 	&	 -1.58 	&	 60.1 	&	 7.51 	&	 46.3 	&	 7.21 	&	 11 	&	 7 	&	 Ti\,{\sc i} 	&	 4722.61 	&	 1.05 	&	 -1.47 	&	 19.3 	&	 5.03 	&	 -- 	&	 -- 	&	 75 	&	 10 \\
Si\,{\sc i} 	&	 5701.14 	&	 4.91 	&	 -2.05 	&	 38.5 	&	 7.55 	&	 28.8 	&	 7.31 	&	 10 	&	 2 	&	 Ti\,{\sc i} 	&	 4742.80 	&	 2.24 	&	 0.21 	&	 31.2 	&	 4.82 	&	 31.4 	&	 4.70 	&	 233 	&	 10 \\
Si\,{\sc i} 	&	 5708.40 	&	 4.95 	&	 -1.47 	&	 72.1 	&	 7.59 	&	 60.1 	&	 7.33 	&	 10 	&	 2 	&	 Ti\,{\sc i} 	&	 4758.12 	&	 2.25 	&	 0.51 	&	 42.9 	&	 4.81 	&	 45.1 	&	 4.74 	&	 233 	&	 10 \\
Si\,{\sc i} 	&	 5772.15 	&	 5.08 	&	 -1.62 	&	 51.1 	&	 7.51 	&	 40.7 	&	 7.27 	&	 17 	&	 7 	&	 Ti\,{\sc i} 	&	 4759.28 	&	 2.25 	&	 0.59 	&	 46.3 	&	 4.81 	&	 44.9 	&	 4.65 	&	 233 	&	 10 \\
Si\,{\sc i} 	&	 5793.08 	&	 4.93 	&	 -1.86 	&	 43.2 	&	 7.47 	&	 30.1 	&	 7.16 	&	 9 	&	 7 	&	 Ti\,{\sc i} 	&	 4820.41 	&	 1.50 	&	 -0.38 	&	 41.1 	&	 4.92 	&	 40.7 	&	 4.77 	&	 126 	&	 10 \\
Si\,{\sc i} 	&	 5948.54 	&	 5.08 	&	 -1.09 	&	 83.0 	&	 7.49 	&	 70.9 	&	 7.23 	&	 16 	&	 7 	&	Ti\,{\sc i} 	&	4885.09	&	1.89	&	0.41	&	 syn 	&	4.93	&	 syn 	&	4.7	&	231	&	 10 \\
Si\,{\sc i} 	&	 6125.03 	&	 5.61 	&	 -1.53 	&	 30.6 	&	 7.50 	&	 23.1 	&	 7.30 	&	 30 	&	 7 	&	 Ti\,{\sc i} 	&	 4913.62 	&	 1.87 	&	 0.22 	&	 49.3 	&	 4.86 	&	 50.9 	&	 4.77 	&	 157 	&	 10 \\
Si\,{\sc i} 	&	 6142.49 	&	 5.62 	&	 -1.48 	&	 33.6 	&	 7.51 	&	 23.8 	&	 7.27 	&	 30 	&	 7 	&	 Ti\,{\sc i} 	&	 4926.15 	&	 0.81 	&	 -2.17 	&	 6.6 	&	 4.92 	&	 -- 	&	 -- 	&	 39 	&	 2 \\
Si\,{\sc i} 	&	 6145.02 	&	 5.61 	&	 -1.39 	&	 37.0 	&	 7.48 	&	 27.8 	&	 7.26 	&	 29 	&	 7 	&	 Ti\,{\sc i} 	&	 4981.74 	&	 0.85 	&	 0.57 	&	 112.8 	&	 4.81 	&	 -- 	&	 -- 	&	 38 	&	 10 \\
Si\,{\sc i} 	&	 6237.34 	&	 5.59 	&	 -0.98 	&	 58.7 	&	 7.41 	&	 46.1 	&	 7.16 	&	 28 	&	 3 	&	 Ti\,{\sc i} 	&	 4999.51 	&	 0.83 	&	 0.32 	&	 103.4 	&	 4.90 	&	 -- 	&	 -- 	&	 38 	&	 10 \\
Si\,{\sc i} 	&	 6244.48 	&	 5.61 	&	 -1.29 	&	 44.3 	&	 7.51 	&	 30.9 	&	 7.23 	&	 28 	&	 7 	&	Ti\,{\sc i} 	&	5009.65	&	0.02	&	-2.2	&	 syn 	&	4.87	&	 syn 	&	4.63	&	5	&	 10\\
Si\,{\sc i} 	&	 6721.84 	&	 5.86 	&	 -0.94 	&	 42.2 	&	 7.32 	&	 -- 	&	 -- 	&	 -- 	&	 2 	&	 Ti\,{\sc i} 	&	 5016.17 	&	 0.85 	&	 -0.48 	&	 64.7 	&	 4.92 	&	 67.2 	&	 4.81 	&	 38 	&	 10 \\
Si\,{\sc i} 	&	 7003.58 	&	 5.94 	&	 -0.59 	&	 58.1 	&	 7.48 	&	 46.8 	&	 7.27 	&	 60 	&	 2 	&	 Ti\,{\sc i} 	&	 5020.03 	&	 0.83 	&	 -0.33 	&	 77.5 	&	 5.05 	&	 76.6 	&	 4.84 	&	 38 	&	 10 \\
Si\,{\sc i} 	&	 7005.84 	&	 5.96 	&	 -0.59 	&	 72.8 	&	 7.48 	&	 67.4 	&	 7.35 	&	 60 	&	 2 	&	 Ti\,{\sc i} 	&	 5022.87 	&	 0.83 	&	 -0.33 	&	 71.3 	&	 4.90 	&	 72.4 	&	 4.75 	&	 38 	&	 10 \\
Si\,{\sc i} 	&	 7034.96 	&	 5.85 	&	 -0.88 	&	 63.8 	&	 7.59 	&	 51.0 	&	 7.35 	&	 50 	&	 3 	&	 Ti\,{\sc i} 	&	 5039.96 	&	 0.02 	&	 -1.08 	&	 76.2 	&	 4.97 	&	 77.8 	&	 4.82 	&	 5 	&	 10 \\
Si\,{\sc i} 	&	 7416.00 	&	 5.59 	&	 -0.75 	&	 87.1 	&	 7.45 	&	 -- 	&	 -- 	&	 22 	&	 3 	&	 Ti\,{\sc i} 	&	 5064.65 	&	 0.05 	&	 -0.94 	&	 85.3 	&	 5.08 	&	 80.9 	&	 4.78 	&	 294 	&	 10 \\
Si\,{\sc i} 	&	 7918.38 	&	 5.93 	&	 -0.61 	&	 79.7 	&	 7.58 	&	 -- 	&	 -- 	&	 57 	&	 6 	&	 Ti\,{\sc i} 	&	 5145.47 	&	 1.46 	&	 -0.54 	&	 36.9 	&	 4.92 	&	 35.7 	&	 4.74 	&	 109 	&	 10 \\
Si\,{\sc i} 	&	 9393.40 	&	 6.10 	&	 -1.53 	&	 13.3 	&	 7.35 	&	 -- 	&	 -- 	&	 72 	&	 3 	&	 Ti\,{\sc i} 	&	 5147.48 	&	 0.00 	&	 -1.94 	&	 37.1 	&	 4.87 	&	 38.5 	&	 4.71 	&	 4 	&	 10 \\
Si\,{\sc i} 	&	 9768.27 	&	 4.93 	&	 -2.68 	&	 27.0 	&	 7.78 	&	 -- 	&	 -- 	&	 7 	&	 3 	&	 Ti\,{\sc i} 	&	 5152.19 	&	 0.02 	&	 -1.95 	&	 36.5 	&	 4.88 	&	 35.8 	&	 4.68 	&	 4 	&	 10 \\
P\,{\sc i} 	&	 9525.78 	&	6.98	&	-0.12	&	 syn 	&	5.56	&	 syn 	&	 -- 	&	 -- 	&	2	&	 Ti\,{\sc i} 	&	 5192.98 	&	 0.02 	&	 -0.95 	&	 83.9 	&	 5.00 	&	 -- 	&	 -- 	&	 4 	&	 10 \\
P\,{\sc i} 	&	 9750.73 	&	 6.92 	&	 -0.20 	&	 syn 	&	 syn 	&	 -- 	&	 -- 	&	 2 	&	 2 	&	 Ti\,{\sc i} 	&	 5210.39 	&	 0.05 	&	 -0.82 	&	 90.0 	&	 5.02 	&	 -- 	&	 -- 	&	 4 	&	 10 \\
P\,{\sc i} 	&	 9796.79 	&	6.99	&	0.19	&	 syn 	&	5.51	&	 syn 	&	 -- 	&	 -- 	&	2	&	 Ti\,{\sc i} 	&	 5219.71 	&	 0.02 	&	 -2.22 	&	 28.1 	&	 4.95 	&	 -- 	&	 -- 	&	 4 	&	 10 \\
S\,{\sc i} 	&	 8693.98 	&	 7.84 	&	 -1.38 	&	 syn 	&	 7.15 	&	 -- 	&	 -- 	&	 6 	&	 2 	&	 Ti\,{\sc i} 	&	 5490.16 	&	 1.46 	&	 -0.84 	&	 22.6 	&	 4.84 	&	 23.5 	&	 4.71 	&	 107 	&	 10 \\
S\,{\sc i} 	&	 8694.70 	&	 7.84 	&	 0.05 	&	 syn 	&	 7.15 	&	 -- 	&	 -- 	&	 6 	&	 2 	&	 Ti\,{\sc i} 	&	 5866.46 	&	 1.07 	&	 -0.79 	&	 47.9 	&	 4.98 	&	 -- 	&	 -- 	&	 72 	&	 10 \\
Ca\,{\sc i} 	&	 4512.27 	&	 2.52 	&	 -1.90 	&	 23.6 	&	 6.29 	&	 -- 	&	 -- 	&	 24 	&	 3 	&	 Ti\,{\sc i} 	&	 5918.55 	&	 1.06 	&	 -1.47 	&	 12.2 	&	 4.71 	&	 -- 	&	 -- 	&	 71 	&	 2 \\
Ca\,{\sc i} 	&	 4526.94 	&	 2.70 	&	 -0.42 	&	 85.6 	&	 6.14 	&	 83.4 	&	 5.93 	&	 36 	&	 2 	&	 Ti\,{\sc i} 	&	 5922.11 	&	 1.04 	&	 -1.47 	&	 20.1 	&	 4.96 	&	 22.6 	&	 4.86 	&	 72 	&	 2 \\
Ca\,{\sc i} 	&	 4578.56 	&	 2.52 	&	 -0.70 	&	 82.8 	&	 6.27 	&	 -- 	&	 -- 	&	 23 	&	 8 	&	 Ti\,{\sc i} 	&	 5978.54 	&	 1.87 	&	 -0.50 	&	 23.9 	&	 4.92 	&	 24.0 	&	 4.78 	&	 154 	&	 2 \\
Ca\,{\sc i} 	&	 5260.39 	&	 2.52 	&	 -1.72 	&	 32.7 	&	 6.30 	&	 25.7 	&	 6.03 	&	 22 	&	 8 	&	 Ti\,{\sc i} 	&	 6126.22 	&	 1.07 	&	 -1.42 	&	 21.6 	&	 4.97 	&	 24.4 	&	 4.88 	&	 69 	&	 10 \\
Ca\,{\sc i} 	&	 5261.71 	&	 2.52 	&	 -0.58 	&	 98.6 	&	 6.47 	&	 90.7 	&	 6.18 	&	 22 	&	 8 	&	 Ti\,{\sc i} 	&	 6258.11 	&	 1.44 	&	 -0.39 	&	 50.3 	&	 4.96 	&	 49.3 	&	 4.79 	&	 104 	&	 10 \\
Ca\,{\sc i} 	&	 5512.99 	&	 2.93 	&	 -0.46 	&	 86.2 	&	 6.38 	&	 81.4 	&	 6.15 	&	 48 	&	 8 	&	 Ti\,{\sc i} 	&	 6261.11 	&	 1.43 	&	 -0.53 	&	 46.5 	&	 5.01 	&	 46.1 	&	 4.85 	&	 104 	&	 10 \\
Ca\,{\sc i} 	&	 5581.98 	&	 2.52 	&	 -0.56 	&	 92.9 	&	 6.34 	&	 -- 	&	 -- 	&	 21 	&	 8 	&	 Ti\,{\sc i} 	&	 6336.11 	&	 1.44 	&	 -1.69 	&	 5.3 	&	 4.88 	&	 -- 	&	 -- 	&	 103 	&	 10 \\
Ca\,{\sc i} 	&	 5590.13 	&	 2.52 	&	 -0.57 	&	 92.0 	&	 6.34 	&	 -- 	&	 -- 	&	 21 	&	 8 	&	 Ti\,{\sc i} 	&	 6743.13 	&	 0.90 	&	 -1.63 	&	 18.4 	&	 4.88 	&	 -- 	&	 -- 	&	 32 	&	 10 \\
Ca\,{\sc i} 	&	 6166.44 	&	 2.52 	&	 -1.14 	&	 70.3 	&	 6.33 	&	 68.2 	&	 6.13 	&	 20 	&	 8 	&	 Ti\,{\sc i} 	&	 7216.20 	&	 1.44 	&	 -1.20 	&	 18.3 	&	 4.97 	&	 -- 	&	 -- 	&	 98 	&	 10 \\
Ca\,{\sc i} 	&	 6169.04 	&	 2.52 	&	 -0.80 	&	 91.9 	&	 6.30 	&	 88.5 	&	 6.06 	&	 20 	&	 8 	&	 Ti\,{\sc i} 	&	 7251.74 	&	 1.42 	&	 -0.76 	&	 33.7 	&	 4.89 	&	 31.7 	&	 4.69 	&	 99 	&	 10 \\
Ca\,{\sc i} 	&	 6169.56 	&	 2.52 	&	 -0.48 	&	 108.7 	&	 6.19 	&	 109.5 	&	 5.97 	&	 20 	&	 8 	&	 Ti\,{\sc i} 	&	 7357.74 	&	 1.44 	&	 -1.02 	&	 22.6 	&	 4.90 	&	 -- 	&	 -- 	&	 97 	&	 10 \\
Ca\,{\sc i} 	&	 6439.07 	&	 2.51 	&	 0.39 	&	 160.0 	&	 6.07 	&	 157.1 	&	 5.80 	&	 18 	&	 2 	&	 Ti\,{\sc i} 	&	 8024.84 	&	 1.87 	&	 -1.08 	&	 10.6 	&	 4.94 	&	 -- 	&	 -- 	&	 151 	&	 10 \\
Ca\,{\sc i} 	&	 6455.60 	&	 2.52 	&	 -1.34 	&	 55.4 	&	 6.34 	&	 48.2 	&	 6.10 	&	 19 	&	 8 	&	 Ti\,{\sc i} 	&	 8364.24 	&	 0.83 	&	 -1.71 	&	 22.1 	&	 4.90 	&	 26.5 	&	 4.85 	&	 33 	&	 3 \\
Ca\,{\sc i} 	&	 6471.67 	&	 2.52 	&	 -0.69 	&	 90.5 	&	 6.36 	&	 88.2 	&	 6.17 	&	 18 	&	 8 	&	 Ti\,{\sc i} 	&	 8396.93 	&	 0.81 	&	 -1.63 	&	 25.2 	&	 4.88 	&	 27.0 	&	 4.76 	&	 33 	&	 3 \\
Ca\,{\sc i} 	&	 6493.79 	&	 2.52 	&	 -0.11 	&	 122.7 	&	 6.22 	&	 124.3 	&	 6.03 	&	 18 	&	 8 	&	 Ti\,{\sc i} 	&	 8412.36 	&	 0.81 	&	 -1.39 	&	 39.7 	&	 4.96 	&	 36.6 	&	 4.74 	&	 33 	&	 2 \\
Ca\,{\sc i} 	&	 6499.65 	&	 2.52 	&	 -0.82 	&	 86.1 	&	 6.41 	&	 80.8 	&	 6.18 	&	 18 	&	 8 	&	 Ti\,{\sc i} 	&	 8426.50 	&	 0.82 	&	 -1.20 	&	 53.0 	&	 5.03 	&	 54.7 	&	 4.91 	&	 33 	&	 2 \\
Ca\,{\sc i} 	&	 6572.79 	&	0	&	-4.32	&	 syn 	&	6.32	&	 syn 	&	6.15	&	1	&	8	&	 Ti\,{\sc i} 	&	 8434.98 	&	 0.84 	&	 -0.83 	&	 71.2 	&	 5.07 	&	 77.6 	&	 5.08 	&	 33 	&	 2 \\
Ca\,{\sc i} 	&	 7148.15 	&	 2.70 	&	 0.11 	&	 135.6 	&	 6.20 	&	 137.4 	&	 5.99 	&	 30 	&	 8 	&	 Ti\,{\sc i} 	&	 8435.68 	&	 0.83 	&	 -1.02 	&	 61.7 	&	 5.06 	&	 66.0 	&	 5.02 	&	 33 	&	 2 \\
Ca\,{\sc i} 	&	 7202.19 	&	 2.70 	&	 -0.26 	&	 108.2 	&	 6.26 	&	 -- 	&	 -- 	&	 29 	&	 3 	&	 Ti\,{\sc i} 	&	 9027.32 	&	 1.73 	&	 -1.36 	&	 7.8 	&	 4.87 	&	 -- 	&	 -- 	&	 138 	&	 3 \\
Ca\,{\sc i} 	&	 7326.15 	&	 2.92 	&	 -0.21 	&	 108.2 	&	 6.38 	&	 104.4 	&	 6.15 	&	 20 	&	 3 	&	 Ti\,{\sc i} 	&	 9675.55 	&	 0.83 	&	 -0.80 	&	 78.6 	&	 5.04 	&	 70.7 	&	 4.76 	&	 32 	&	 3 \\
Ca\,{\sc i} 	&	 9663.58 	&	 4.71 	&	 -0.69 	&	 6.5 	&	 6.22 	&	 -- 	&	 -- 	&	 55 	&	 3 	&	 Ti\,{\sc i} 	&	 9718.96 	&	 1.50 	&	 -1.18 	&	 16.4 	&	 4.79 	&	 -- 	&	 -- 	&	 124 	&	 3 \\
Sc\,{\sc i} 	&	 4023.69 	&	 0.02 	&	 0.38 	&	 syn 	&	 3.13 	&	 -- 	&	 -- 	&	 7 	&	 9 	&	 Ti\,{\sc i} 	&	 9728.36 	&	 0.81 	&	 -1.21 	&	 47.2 	&	 4.82 	&	 -- 	&	 -- 	&	 32 	&	 3 \\
Sc\,{\sc ii} 	&	 4246.84 	&	 0.31 	&	 0.24 	&	 157.0 	&	 3.17 	&	 -- 	&	 -- 	&	 7 	&	 9 	&	 Ti\,{\sc i} 	&	 9770.28 	&	 0.84 	&	 -1.58 	&	 26.7 	&	 4.80 	&	 -- 	&	 -- 	&	 32 	&	 3 \\
Sc\,{\sc ii} 	&	 5239.82 	&	1.45	&	-0.76	&	 syn 	&	3.15	&	 syn 	&	2.82	&	26	&	9	&	 Ti\,{\sc i} 	&	 9787.67 	&	 0.82 	&	 -1.44 	&	 40.3 	&	 4.92 	&	 -- 	&	 -- 	&	 32 	&	 3 \\
\hline
\end{tabular}}
\end{table*}

\begin{table*}
\setlength{\tabcolsep}{1.2pt}
\renewcommand{\arraystretch}{0.8}
\centering\small
\caption{The abundances were obtained for a model with $T_{\rm eff}=5770$ K, $\log g =4.40$ cgs, and $\xi=$ 0.66 km s$^{\rm −1}$ for the solar spectrum. $T_{\rm eff}=5600$ K, $\log g = 4.50$ cgs, and $\xi=$ 0.44 km s$^{\rm −1}$ for the HD\,218209 spectrum.}
\label{tab:lineslit_other_lines_2}
\resizebox{\linewidth}{!}{%
\begin{tabular}{lccccccccc||cccccccccc}
\hline
 & & & & \multicolumn{2}{c}{Sun} & \multicolumn{2}{c}{HD\,218209}  & & & &  &&   &\multicolumn{2}{c}{Sun} &\multicolumn{2}{c}{HD\,218209}&  & \\
\cline{2-4}
\cline{7-8}
\cline{12-14}
\cline{17-18}
 Spec.& $\lambda$ & LEP& $\log(gf)$ & EW & $\log \epsilon$(X) & EW& $\log\epsilon$(X)&RMT&Ref.&Spec.&$\lambda$ & LEP&  $\log(gf)$ & EW     & $\log \epsilon$(X) & EW	& $\log\epsilon$(X) & RMT &Ref.\\
\cline{2-4}
\cline{7-8}
\cline{12-14}
\cline{17-18}
   & (\AA)  & (eV) & (dex)   &(m\AA) & (dex) &(m\AA) & (dex)&    &    && (\AA)  & (eV) & (dex)   &(m\AA) & (dex) &(m\AA) & (dex)&\\
\hline

 Ti\,{\sc ii} 	&	 4443.81 	&	 1.08 	&	 -0.71 	&	 146.1 	&	 5.09 	&	 126.4 	&	 4.73 	&	 19 	&	10	&	Ni\,{\sc i} 	&	4606.23	&	3.60	&	-1.02	&	syn	&	6.36	&	syn 	&	5.89	&	100	&	 2 \\
Ti\,{\sc ii} 	&	4468.50	&	1.13	&	-0.63	&	 syn 	&	5.21	&	 syn 	&	4.85	&	31	&	10	&	Ni\,{\sc i} 	&	 4731.80 	&	 3.83 	&	 -0.85 	&	 42.7 	&	 6.28 	&	 31.3 	&	 5.92 	&	 163 	&	 2 \\
 Ti\,{\sc ii} 	&	 4493.53 	&	 1.08 	&	 -2.78 	&	 34.7 	&	 4.91 	&	 -- 	&	 -- 	&	 18 	&	10	&	Ni\,{\sc i} 	&	 4732.47 	&	 4.10 	&	 -0.55 	&	 42.2 	&	 6.24 	&	 -- 	&	 -- 	&	 235 	&	 2 \\
 Ti\,{\sc ii} 	&	 4568.33 	&	 1.22 	&	 -2.65 	&	 29.7 	&	 4.78 	&	 -- 	&	 -- 	&	 60 	&	10	&	Ni\,{\sc i} 	&	4752.43	&	3.66	&	-0.69	&	syn	&	6.33	&	syn 	&	5.8	&	132	&	 2 \\
 Ti\,{\sc ii} 	&	 4583.41 	&	 1.16 	&	 -2.84 	&	 31.3 	&	 4.95 	&	 -- 	&	 -- 	&	 39 	&	10	&	Ni\,{\sc i} 	&	4756.52	&	3.48	&	-0.34	&	syn	&	6.23	&	syn 	&	5.74	&	98	&	 2 \\
 Ti\,{\sc ii} 	&	 4708.67 	&	 1.24 	&	 -2.35 	&	 50.7 	&	 5.01 	&	 46.2 	&	 4.87 	&	 49 	&	10	&	Ni\,{\sc i} 	&	 4806.99 	&	 3.68 	&	 -0.64 	&	 59.4 	&	 6.30 	&	 47.9 	&	 5.93 	&	 163 	&	 2 \\
Ti\,{\sc ii} 	&	4874.01	&	3.09	&	-0.86	&	 syn 	&	4.95	&	 syn 	&	4.68	&	114	&	10	&	Ni\,{\sc i} 	&	 4829.03 	&	 3.54 	&	 -0.33 	&	 77.1 	&	 6.17 	&	 67.4 	&	 5.83 	&	 131 	&	 2 \\
 Ti\,{\sc ii} 	&	 4911.20 	&	 3.12 	&	 -0.64 	&	 51.2 	&	 5.10 	&	 46.1 	&	 5.00 	&	 114 	&	10	&	Ni\,{\sc i} 	&	4852.56	&	3.54	&	-1.07	&	syn	&	6.29	&	syn 	&	5.92	&	130	&	 2 \\
 Ti\,{\sc ii} 	&	 5005.17 	&	 1.57 	&	 -2.73 	&	 23.8 	&	 5.01 	&	 19.5 	&	 4.85 	&	 71 	&	10	&	Ni\,{\sc i} 	&	 4904.42 	&	 3.54 	&	 -0.17 	&	 84.9 	&	 6.14 	&	 74.7 	&	 5.79 	&	 129 	&	 2 \\
 Ti\,{\sc ii} 	&	 5336.79 	&	 1.58 	&	 -1.60 	&	 72.0 	&	 5.08 	&	 -- 	&	 -- 	&	 69 	&	10	&	Ni\,{\sc i} 	&	 4913.98 	&	 3.74 	&	 -0.62 	&	 53.9 	&	 6.20 	&	 41.5 	&	 5.82 	&	 132 	&	 2 \\
 Ti\,{\sc i} 	&	 5418.77 	&	 1.58 	&	 -2.13 	&	 48.5 	&	 5.02 	&	 43.8 	&	 4.88 	&	 69 	&	10	&	Ni\,{\sc i} 	&	 4935.83 	&	 3.94 	&	 -0.36 	&	 58.5 	&	 6.20 	&	 48.9 	&	 5.89 	&	 177 	&	 2 \\
V\,{\sc i} 	&	4437.84	&	0.29	&	-0.71	&	 syn 	&	3.89	&	 syn 	&	 -- 	&	21	&	11	&	Ni\,{\sc i} 	&	 4946.03 	&	 3.80 	&	 -1.29 	&	 25.9 	&	 6.29 	&	 15.2 	&	 5.87 	&	 148 	&	 2 \\
 V\,{\sc i} 	&	 5727.06 	&	 1.08 	&	 -0.02 	&	 syn 	&	 3.89 	&	 -- 	&	 -- 	&	 35 	&	11	&	Ni\,{\sc i} 	&	 4953.21 	&	 3.74 	&	 -0.66 	&	 54.8 	&	 6.26 	&	 44.4 	&	 5.92 	&	 111 	&	 2 \\
V\,{\sc i} 	&	6090.18	&	1.08	&	-0.07	&	 syn 	&	3.93	&	syn 	&	3.52	&	34	&	2	&	Ni\,{\sc i} 	&	 4998.23 	&	 3.61 	&	 -0.78 	&	 54.8 	&	 6.26 	&	 -- 	&	 -- 	&	 111 	&	 2 \\
V\,{\sc i} 	&	6119.53	&	1.06	&	-0.36	&	 syn 	&	3.94	&	 syn 	&	3.6	&	34	&	11	&	Ni\,{\sc i} 	&	 5010.94 	&	 3.63 	&	 -0.87 	&	 48.0 	&	 6.21 	&	 34.2 	&	 5.80 	&	 144 	&	 2 \\
V\,{\sc i} 	&	6243.11	&	0.30	&	-0.94	&	 syn 	&	3.88	&	 syn 	&	3.52	&	19	&	11	&	Ni\,{\sc i} 	&	 5032.73 	&	 3.90 	&	 -1.27 	&	 24.1 	&	 6.31 	&	 -- 	&	 -- 	&	 207 	&	 2 \\
Cr\,{\sc i} 	&	 4545.96 	&	 0.94 	&	 -1.38 	&	 83.5 	&	 5.70 	&	 75.5 	&	 5.33 	&	 10 	&	 2 	&	Ni\,{\sc i} 	&	 5035.37 	&	 3.63 	&	 0.29 	&	 97.6 	&	 5.91 	&	 89.5 	&	 5.60 	&	 143 	&	 2 \\
Cr\,{\sc i} 	&	 4616.13 	&	 0.98 	&	 -1.18 	&	 87.7 	&	 5.64 	&	 79.4 	&	 5.27 	&	 21 	&	 2 	&	Ni\,{\sc i} 	&	 5042.19 	&	 3.64 	&	 -0.57 	&	 59.0 	&	 6.15 	&	 51.0 	&	 5.85 	&	 131 	&	 2 \\
Cr\,{\sc i} 	&	 4626.18 	&	 0.97 	&	 -1.32 	&	 81.5 	&	 5.62 	&	 75.7 	&	 5.31 	&	 21 	&	 2 	&	Ni\,{\sc i} 	&	5048.85	&	3.85	&	-0.37	&	syn	&	6.26	&	syn 	&	5.86	&	195	&	 2 \\
Cr\,{\sc i} 	&	4646.17	&	1.03	&	-0.71	&	 syn 	&	5.71	&	 syn 	&	5.16	&	21	&	2	&	 Ni\,{\sc i} 	&	 5082.35 	&	 3.66 	&	 -0.54 	&	 63.6 	&	 6.23 	&	 -- 	&	 -- 	&	 130 	&	 2 \\
Cr\,{\sc i} 	&	 4651.29 	&	 0.98 	&	 -1.46 	&	 78.3 	&	 5.69 	&	 72.9 	&	 5.40 	&	 21 	&	 2 	&	 Ni\,{\sc i} 	&	 5084.10 	&	 3.68 	&	 0.03 	&	 89.1 	&	 6.10 	&	 -- 	&	 -- 	&	 162 	&	 2 \\
Cr\,{\sc i} 	&	 4652.17 	&	 1.00 	&	 -1.03 	&	 99.7 	&	 5.72 	&	 90.5 	&	 5.34 	&	 21 	&	 2 	&	 Ni\,{\sc i} 	&	 5088.54 	&	 3.85 	&	 -0.91 	&	 32.1 	&	 6.12 	&	 -- 	&	 -- 	&	 190 	&	 2 \\
Cr\,{\sc i} 	&	 4708.02 	&	 3.17 	&	 0.11 	&	 58.0 	&	 5.58 	&	 44.0 	&	 5.16 	&	 186 	&	 2 	&	 Ni\,{\sc i} 	&	 5102.97 	&	 1.68 	&	 -2.62 	&	 47.3 	&	 6.12 	&	 -- 	&	 -- 	&	 49 	&	 2 \\
Cr\,{\sc i} 	&	 4718.42 	&	 3.19 	&	 0.10 	&	 65.8 	&	 5.75 	&	 55.0 	&	 5.40 	&	 186 	&	 2 	&	 Ni\,{\sc i} 	&	 5115.40 	&	 3.83 	&	 -0.11 	&	 74.8 	&	 6.17 	&	 60.6 	&	 5.76 	&	 177 	&	 2 \\
Cr\,{\sc i} 	&	 4730.72 	&	 3.08 	&	 -0.19 	&	 48.5 	&	 5.65 	&	 37.1 	&	 5.28 	&	 145 	&	 2 	&	 Ni\,{\sc i} 	&	 5155.13 	&	 3.90 	&	 -0.66 	&	 49.0 	&	 6.27 	&	 35.2 	&	 5.86 	&	 206 	&	 2 \\
Cr\,{\sc i} 	&	 4737.35 	&	 3.07 	&	 -0.10 	&	 55.5 	&	 5.65 	&	 42.7 	&	 5.25 	&	 145 	&	 2 	&	 Ni\,{\sc i} 	&	 5435.87 	&	 1.99 	&	 -2.60 	&	 50.9 	&	 6.47 	&	 40.7 	&	 6.10 	&	 70 	&	 2 \\
Cr\,{\sc i} 	&	 4756.12 	&	 3.10 	&	 0.09 	&	 63.2 	&	 5.76 	&	 54.2 	&	 5.44 	&	 145 	&	 2 	&	 Ni\,{\sc i} 	&	5587.87	&	1.93	&	-2.14	&	syn	&	6.23	&	syn	&	5.69	&	70	&	 2 \\
Cr\,{\sc i} 	&	 4936.34 	&	 3.11 	&	 -0.34 	&	 44.8 	&	 5.73 	&	 31.8 	&	 5.31 	&	 166 	&	 2 	&	 Ni\,{\sc i} 	&	 5593.75 	&	 3.90 	&	 -0.84 	&	 42.0 	&	 6.27 	&	 -- 	&	 -- 	&	 206 	&	 2 \\
Cr\,{\sc i} 	&	 4964.93 	&	 0.94 	&	 -2.53 	&	 38.6 	&	 5.65 	&	 27.3 	&	 5.21 	&	 9 	&	 2 	&	 Ni\,{\sc i} 	&	 5625.33 	&	 4.09 	&	 -0.70 	&	 39.0 	&	 6.25 	&	 24.5 	&	 5.82 	&	 221 	&	 2 \\
Cr\,{\sc i} 	&	 5247.57 	&	 0.96 	&	 -1.63 	&	 83.4 	&	 5.79 	&	 72.6 	&	 5.38 	&	 18 	&	 2 	&	 Ni\,{\sc i} 	&	 5637.12 	&	 4.09 	&	 -0.80 	&	 33.8 	&	 6.23 	&	 21.4 	&	 5.83 	&	 218 	&	 2 \\
Cr\,{\sc i} 	&	 5296.70 	&	 0.98 	&	 -1.41 	&	 93.5 	&	 5.77 	&	 79.7 	&	 5.31 	&	 18 	&	 2 	&	 Ni\,{\sc i} 	&	 5641.89 	&	 4.10 	&	 -1.08 	&	 23.5 	&	 6.27 	&	 -- 	&	 -- 	&	 234 	&	 2 \\
Cr\,{\sc i} 	&	 5300.75 	&	 0.98 	&	 -2.13 	&	 58.4 	&	 5.72 	&	 45.4 	&	 5.26 	&	 18 	&	 2 	&	 Ni\,{\sc i} 	&	 5682.21 	&	 4.10 	&	 -0.47 	&	 51.5 	&	 6.29 	&	 37.7 	&	 5.90 	&	 232 	&	 2 \\
Cr\,{\sc i} 	&	 5345.81 	&	 1.00 	&	 -0.98 	&	 114.5 	&	 5.68 	&	 -- 	&	 -- 	&	 18 	&	 2 	&	 Ni\,{\sc i} 	&	 5748.36 	&	 1.68 	&	 -3.26 	&	 29.1 	&	 6.26 	&	 -- 	&	 -- 	&	 45 	&	 2 \\
Cr\,{\sc i} 	&	 5348.33 	&	 1.00 	&	 -1.29 	&	 99.8 	&	 5.77 	&	 -- 	&	 -- 	&	 18 	&	 2 	&	 Ni\,{\sc i} 	&	 5805.23 	&	 4.17 	&	 -0.64 	&	 40.4 	&	 6.29 	&	 24.6 	&	 5.83 	&	 234 	&	 2 \\
Cr\,{\sc i} 	&	 5787.93 	&	 3.32 	&	 -0.08 	&	 45.2 	&	 5.60 	&	 31.8 	&	 5.19 	&	 188 	&	 2 	&	 Ni\,{\sc i} 	&	 6007.32 	&	 1.68 	&	 -3.34 	&	 24.9 	&	 6.21 	&	 17.6 	&	 5.85 	&	 42 	&	 2 \\
Cr\,{\sc i} 	&	 6925.24 	&	 3.43 	&	 -0.33 	&	 37.9 	&	 5.75 	&	 -- 	&	 -- 	&	 222 	&	 2 	&	 Ni\,{\sc i} 	&	 6086.29 	&	 4.26 	&	 -0.51 	&	 42.3 	&	 6.27 	&	 29.4 	&	 5.89 	&	 249 	&	 2 \\
Cr\,{\sc i} 	&	 6926.04 	&	 3.43 	&	 -0.62 	&	 20.5 	&	 5.61 	&	 -- 	&	 -- 	&	 222 	&	 3 	&	 Ni\,{\sc i} 	&	 6108.12 	&	 1.68 	&	 -2.44 	&	 65.2 	&	 6.27 	&	 55.1 	&	 5.91 	&	 45 	&	 2 \\
Cr\,{\sc i} 	&	 6979.82 	&	 3.45 	&	 -0.41 	&	 34.7 	&	 5.74 	&	 -- 	&	 -- 	&	 222 	&	 2 	&	 Ni\,{\sc i} 	&	 6128.98 	&	 1.68 	&	 -3.32 	&	 25.3 	&	 6.20 	&	 17.0 	&	 5.81 	&	 42 	&	 2 \\
Cr\,{\sc i} 	&	 7400.23 	&	 2.89 	&	 -0.11 	&	 75.4 	&	 5.58 	&	 -- 	&	 -- 	&	 93 	&	 2 	&	 Ni\,{\sc i} 	&	 6130.14 	&	 4.26 	&	 -0.96 	&	 21.1 	&	 6.22 	&	 14.6 	&	 5.93 	&	 248 	&	 2 \\
Cr\,{\sc i} 	&	 8348.28 	&	 2.70 	&	 -1.87 	&	 13.1 	&	 5.83 	&	 -- 	&	 -- 	&	 56 	&	 3 	&	 Ni\,{\sc i} 	&	 6175.37 	&	 4.09 	&	 -0.54 	&	 47.4 	&	 6.24 	&	 33.7 	&	 5.85 	&	 217 	&	 2 \\
Cr\,{\sc i} 	&	 8947.20 	&	 3.09 	&	 -0.75 	&	 28.1 	&	 5.49 	&	 -- 	&	 -- 	&	 142 	&	 3 	&	 Ni\,{\sc i} 	&	 6176.82 	&	 4.09 	&	 -0.53 	&	 63.1 	&	 6.53 	&	 -- 	&	 -- 	&	 228 	&	 2 \\
Cr\,{\sc i} 	&	 8976.88 	&	 3.07 	&	 -1.03 	&	 18.5 	&	 5.50 	&	 -- 	&	 -- 	&	 142 	&	 3 	&	 Ni\,{\sc i} 	&	 6204.61 	&	 4.09 	&	 -1.14 	&	 20.9 	&	 6.23 	&	 -- 	&	 -- 	&	 226 	&	 2 \\
Cr\,{\sc i} 	&	 9290.44 	&	 2.53 	&	 -0.88 	&	 58.6 	&	 5.69 	&	 -- 	&	 -- 	&	 29 	&	 3 	&	 Ni\,{\sc i} 	&	 6322.17 	&	 4.15 	&	 -1.17 	&	 17.4 	&	 6.20 	&	 10.3 	&	 5.84 	&	 249 	&	 2 \\
Cr\,{\sc i} 	&	 9730.32 	&	 3.54 	&	 -0.77 	&	 12.9 	&	 5.45 	&	 -- 	&	 -- 	&	 226 	&	 3 	&	 Ni\,{\sc i} 	&	 6327.60 	&	 1.68 	&	 -3.15 	&	 41.5 	&	 6.40 	&	 27.2 	&	 5.92 	&	 44 	&	 2 \\
Cr\,{\sc i} 	&	 9900.87 	&	 2.97 	&	 -2.14 	&	 5.2 	&	 5.81 	&	 -- 	&	 -- 	&	 80 	&	 3 	&	 Ni\,{\sc i} 	&	 6378.26 	&	 4.15 	&	 -0.90 	&	 31.7 	&	 6.31 	&	 -- 	&	 -- 	&	 247 	&	 2 \\
Cr\,{\sc ii} 	&	 4588.20 	&	 4.07 	&	 -0.65 	&	 70.9 	&	 5.67 	&	 -- 	&	 -- 	&	 44 	&	 12 	&	 Ni\,{\sc i} 	&	 6414.59 	&	 4.15 	&	 -1.21 	&	 17.2 	&	 6.24 	&	 -- 	&	 -- 	&	 244 	&	 2 \\
Cr\,{\sc ii} 	&	 4616.64 	&	 4.05 	&	 -1.29 	&	 45.2 	&	 5.65 	&	 31.7 	&	 5.33 	&	 44 	&	 2 	&	 Ni\,{\sc i} 	&	 6482.81 	&	 1.93 	&	 -2.63 	&	 41.1 	&	 6.11 	&	 27.9 	&	 5.67 	&	 66 	&	 2 \\
Cr\,{\sc ii} 	&	 5237.32 	&	 4.07 	&	 -1.17 	&	 53.2 	&	 5.76 	&	 35.3 	&	 5.33 	&	 43 	&	 12 	&	 Ni\,{\sc i} 	&	 6598.61 	&	 4.23 	&	 -0.98 	&	 24.3 	&	 6.28 	&	 15.2 	&	 5.92 	&	 249 	&	 2 \\
Cr\,{\sc ii} 	&	 5305.87 	&	 3.83 	&	 -1.91 	&	 25.2 	&	 5.50 	&	 14.4 	&	 5.17 	&	 24 	&	 12 	&	 Ni\,{\sc i} 	&	 6635.14 	&	 4.42 	&	 -0.83 	&	 24.6 	&	 6.32 	&	 17.6 	&	 6.04 	&	 264 	&	 2 \\
Mn\,{\sc i} 	&	4055.55	&	2.14	&	-0.08	&	 syn 	&	5.47	&	 syn 	&	4.92	&	5	&	13	&	 Ni\,{\sc i} 	&	 6767.78 	&	 1.83 	&	 -2.17 	&	 77.9 	&	 6.41 	&	 -- 	&	 -- 	&	 57 	&	 2 \\
Mn\,{\sc i} 	&	 4082.94 	&	 2.18 	&	 -0.36 	&	 syn 	&	5.55	&	 syn 	&	4.84	&	 5 	&	13	&	 Ni\,{\sc i} 	&	 6772.32 	&	 3.66 	&	 -0.99 	&	 48.5 	&	 6.26 	&	 -- 	&	 -- 	&	 127 	&	 4 \\
Mn\,{\sc i} 	&	 4451.59 	&	 2.89 	&	 0.28 	&	 syn 	&	5.47	&	 syn 	&	4.83	&	 22 	&	 13 	&	 Ni\,{\sc i} 	&	 6914.56 	&	 1.94 	&	 -2.27 	&	 77.9 	&	 6.56 	&	 -- 	&	 -- 	&	 62 	&	 2 \\
Mn\,{\sc i} 	&	 4470.14 	&	 2.94 	&	 -0.44 	&	 syn 	&	5.49	&	 syn 	&	4.92	&	 22 	&	 13 	&	 Ni\,{\sc i} 	&	 7030.06 	&	 3.53 	&	 -1.83 	&	 19.8 	&	 6.33 	&	 -- 	&	 -- 	&	 126 	&	 3 \\
Mn\,{\sc i} 	&	 4502.22 	&	 2.92 	&	 -0.34 	&	 syn 	&	5.34	&	 syn 	&	 4.74 	&	 22 	&	 13 	&	 Ni\,{\sc i} 	&	 7110.91 	&	 1.93 	&	 -2.97 	&	 36.2 	&	 6.30 	&	 24.5 	&	 5.88 	&	 64 	&	 2 \\
Mn\,{\sc i} 	&	 4709.72 	&	 2.89 	&	 -0.49 	&	 syn 	&	5.36	&	 syn 	&	4.75	&	 21 	&	 13 	&	 Ni\,{\sc i} 	&	 7385.24 	&	 2.73 	&	 -1.96 	&	 45.2 	&	 6.28 	&	 33.5 	&	 5.91 	&	 84 	&	 2 \\
Mn\,{\sc i} 	&	 4739.11 	&	 2.94 	&	 -0.61 	&	 syn 	&	5.39	&	 syn 	&	4.73	&	 21 	&	 13 	&	 Ni\,{\sc i} 	&	 7422.30 	&	 3.62 	&	 -0.13 	&	 90.5 	&	 5.99 	&	 -- 	&	 -- 	&	 139 	&	 2 \\
Mn\,{\sc i} 	&	 4765.86 	&	 2.94 	&	 -0.09 	&	 syn 	&	5.45	&	 syn 	&	4.75	&	 21 	&	 13 	&	 Ni\,{\sc i} 	&	 7522.78 	&	 3.64 	&	 -0.47 	&	 73.9 	&	 6.29 	&	 66.6 	&	 6.04 	&	 126 	&	 3 \\
Mn\,{\sc i} 	&	 4766.42 	&	 2.92 	&	 0.10 	&	 syn 	&	5.41	&	 syn 	&	4.83	&	 21 	&	 13 	&	 Ni\,{\sc i} 	&	 7525.14 	&	 3.62 	&	 -0.43 	&	 69.0 	&	 6.14 	&	 57.7 	&	 5.81 	&	 139 	&	 3 \\
Mn\,{\sc i} 	&	4783.42	&	2.30	&	0.03	&	 syn 	&	5.60	&	 syn 	&	4.84	&	16	&	13	&	 Ni\,{\sc i} 	&	 7555.60 	&	 3.83 	&	 0.05 	&	 90.3 	&	 6.23 	&	 77.6 	&	 5.88 	&	 187 	&	 3 \\
Mn\,{\sc i} 	&	5117.94	&	3.13	&	-1.20	&	 syn 	&	5.45	&	--	&	--	&	32	&	13	&	 Ni\,{\sc i} 	&	 7574.08 	&	 3.82 	&	 -0.45 	&	 63.5 	&	 6.23 	&	 49.5 	&	 5.84 	&	 156 	&	 3 \\
Mn\,{\sc i} 	&	5432.55	&	0.00	&	-3.79	&		&	5.33	&	 syn 	&	4.73	&	1	&	13	&	 Ni\,{\sc i} 	&	 7727.66 	&	 3.66 	&	 -0.17 	&	 87.2 	&	 6.23 	&	 70.5 	&	 5.82 	&	 156 	&	 3 \\
Mn\,{\sc i} 	&	 6013.50 	&	 3.06 	&	 -0.25 	&	 syn 	&	5.46	&	 syn 	&	4.8	&	 27 	&	 2 	&	 Ni\,{\sc i} 	&	 7748.93 	&	 3.69 	&	 -0.18 	&	 84.5 	&	 6.22 	&	 75.8 	&	 5.94 	&	 156 	&	 3 \\
Mn\,{\sc i} 	&	 6021.80 	&	 3.07 	&	 -0.05 	&	 syn 	&	5.56	&	 syn 	&	4.94	&	 27 	&	 13 	&	 Ni\,{\sc i} 	&	 7797.62 	&	 3.88 	&	 -0.18 	&	 75.1 	&	 6.22 	&	 63.8 	&	 5.90 	&	 201 	&	 3 \\
Co\,{\sc i} 	&	 4121.33 	&	 0.92 	&	 -0.33 	&	 syn 	&	5.05	&	 -- 	&	 -- 	&	 28 	&	 14 	&	 Ni\,{\sc i} 	&	 8965.94 	&	 4.09 	&	 -0.89 	&	 39.4 	&	 6.30 	&	 -- 	&	 -- 	&	 225 	&	 3 \\
Co\,{\sc i} 	&	 4792.86 	&	 3.24 	&	 0.00 	&	 syn 	&	4.89	&	 syn 	&	4.54	&	 158 	&	 3 	&	 Cu\,{\sc i} 	&	 5105.54 	&	 1.38 	&	 -1.50 	&	 syn 	&	4.25	&	 syn 	&	3.8	&	 2 	&	 2 \\
Co\,{\sc i} 	&	 4813.48 	&	 3.21 	&	 0.12 	&	 syn 	&	5.04	&	 syn 	&	4.59	&	 158 	&	 14 	&	 Cu\,{\sc i} 	&	 5218.20 	&	 3.80 	&	 0.26 	&	 syn 	&	4.11	&	 syn 	&	 3.56 	&	3.57	&	 2 \\
Co\,{\sc i} 	&	 5352.05 	&	 3.58 	&	 0.06 	&	 syn 	&	4.89	&	 syn 	&	 4.40 	&	 172 	&	 14 	&	 Cu\,{\sc i} 	&	 7933.13 	&	 3.77 	&	 -0.37 	&	 syn 	&	 4.21 	&	 -- 	&	 -- 	&	 6 	&	 3 \\
Co\,{\sc i} 	&	 5483.36 	&	 1.71 	&	 -1.50 	&	 syn 	&	4.94	&	 syn 	&	4.54	&	 39 	&	 14 	&	 Cu\,{\sc i} 	&	 8092.63 	&	 3.80 	&	 -0.04 	&	 syn 	&	4.23	&	 syn 	&	3.72	&	 6 	&	 3 \\
Co\,{\sc i} 	&	 5647.23 	&	 2.27 	&	 -1.56 	&	 syn 	&	4.94	&	 syn 	&	4.46	&	 112 	&	 2 	&	 Zn\,{\sc i} 	&	 4722.16 	&	 4.03 	&	 -0.39 	&	 syn 	&	4.64	&	 syn 	&	4.49	&	 2 	&	 15 \\
Co\,{\sc i} 	&	 6093.15 	&	 1.74 	&	 -2.40 	&	 syn 	&	 4.94 	&	 -- 	&	 -- 	&	 37 	&	 14 	&	 Zn\,{\sc i} 	&	 4810.54 	&	 4.08 	&	 -0.17 	&	 syn 	&	4.63	&	 syn 	&	4.42	&	 2 	&	 15 \\
Co\,{\sc i} 	&	 8093.93 	&	 4.00 	&	 0.29 	&	 syn 	&	 4.94 	&	 syn 	&	4.38	&	 189 	&	 2 	&	 Sr\,{\sc i} 	&	 4607.34 	&	 0.00 	&	 0.28 	&	 syn 	&	2.84	&	 syn 	&	2.29	&	 2 	&	 16 \\
Ni\,{\sc i} 	&	 4410.52 	&	 3.31 	&	 -1.08 	&	 55.6 	&	 6.33 	&	 -- 	&	 -- 	&	 88 	&	 2 	&	 Y\,{\sc ii} 	&	 4883.69 	&	 1.08 	&	 0.07 	&	 syn 	&	2.3	&	 syn 	&	1.8	&	 22 	&	 17 \\
Ni\,{\sc i} 	&	 4470.48 	&	 3.40 	&	 -0.40 	&	 80.5 	&	 6.24 	&	 71.5 	&	 5.90 	&	 86 	&	 2 	&	 Y\,{\sc ii} 	&	 5087.43 	&	 1.08 	&	 -0.17 	&	 syn 	&	2.27	&	 syn 	&	1.74	&	 20 	&	 17 \\
\hline
\end{tabular}}
\end{table*} 

\begin{table*}
\setlength{\tabcolsep}{1.2pt}
\renewcommand{\arraystretch}{0.8}
\centering\small
\caption{The abundances were obtained for a model with $T_{\rm eff}=5770$ K, $\log g = 4.40$ cgs, and $\xi=$ 0.66 km s$^{\rm −1}$ for the solar spectrum. $T_{\rm eff}=5600$ K, $\log g = 4.50$ cgs, and $\xi=$ 0.44 km s$^{\rm −1}$ for the HD\,218209 spectrum.}
\label{tab:lineslit_other_lines_3}
\resizebox{\linewidth}{!}{%
\begin{tabular}{lccccccccc||cccccccccc}
\hline
 & & & & \multicolumn{2}{c}{Sun} & \multicolumn{2}{c}{HD\,218209}  & & & &  &&   &\multicolumn{2}{c}{Sun} &\multicolumn{2}{c}{HD\,218209}&  & \\
\cline{2-4}
\cline{7-8}
\cline{12-14}
\cline{17-18}
 Spec.& $\lambda$ & LEP& $\log(gf)$ & EW & $\log \epsilon$(X) & EW& $\log\epsilon$(X)&RMT&Ref.&Spec.&$\lambda$ & LEP&  $\log(gf)$ & EW     & $\log \epsilon$(X) & EW	& $\log\epsilon$(X) & RMT &Ref.\\
\cline{2-4}
\cline{7-8}
\cline{12-14}
\cline{17-18}
   & (\AA)  & (eV) & (dex)   &(m\AA) & (dex) &(m\AA) & (dex)&    &    && (\AA)  & (eV) & (dex)   &(m\AA) & (dex) &(m\AA) & (dex)&\\
\hline
 Zr\,{\sc i} 	&	 4772.32 	&	 0.62 	&	 0.04 	&	 syn 	&	2.53	&	 -- 	&	 -- 	&	 43 	&	3	&	 Ce\,{\sc ii} 	&	 4042.14 	&	 0.50 	&	 0.00 	&	 syn 	&	 1.60 	&	 -- 	&	 -- 	&	 252 	&	 2 \\
Zr\,{\sc ii} 	&	4208.98	&	0.71	&	-0.46	&	syn	&	2.6	&	--	&	--	&	41	&	18	&	 Ce\,{\sc ii} 	&	 4562.37 	&	 0.48 	&	 0.21 	&	 syn 	&	 1.63 	&	 syn 	&	 1.49 	&	 1 	&	 20 \\
 Zr\,{\sc ii} 	&	 4050.32 	&	 0.71 	&	 -1.06 	&	 syn 	&	2.62	&	syn	&	2.29	&	 43 	&	3	&	 Ce\,{\sc ii} 	&	 4628.16 	&	 0.52 	&	 0.14 	&	 syn 	&	1.56	&	 -- 	&	 -- 	&	 1 	&	 20 \\
 Ba\,{\sc ii} 	&	 4554.04 	&	 0.00 	&	 0.14 	&	 syn 	&	2.3	&	syn	&	1.99	&	 1 	&	19	&	 Nd\,{\sc ii} 	&	 4021.33 	&	 0.32 	&	 -0.10 	&	 syn 	&	1.38	&	 -- 	&	 -- 	&	 36 	&	 3 \\
 Ba\,{\sc ii} 	&	 5853.69 	&	 0.60 	&	 -0.91 	&	 syn 	&	 2.33 	&	 syn 	&	1.98	&	 2 	&	19	&	 Nd\,{\sc ii} 	&	 4446.40 	&	 0.20 	&	 -0.35 	&	 syn 	&	1.33	&	syn	&	1.07	&	 49 	&	 3 \\
 La\,{\sc ii} 	&	 4086.72 	&	 0.00 	&	 -0.07 	&	 syn 	&	1.2	&	 syn 	&	0.76	&	 10 	&	2	&	 Nd\,{\sc ii} 	&	 4567.61 	&	 0.20 	&	 -1.31 	&	 syn 	&	1.37	&	 -- 	&	 -- 	&	 49 	&	 3 \\
 La\,{\sc ii} 	&	 4662.51 	&	 0.00 	&	 -1.25 	&	 syn 	&	1.13	&	 -- 	&	 -- 	&	 8 	&	2	&	 Sm\,{\sc ii} 	&	 4519.63 	&	 0.54 	&	 -0.35 	&	 syn 	&	 0.94 	&	syn	&	0.72	&	 49 	&	 21 \\
 La\,{\sc ii} 	&	 4748.73 	&	 0.92 	&	 -0.54 	&	 syn 	&	1.1	&	 syn 	&	 0.83 	&	 65 	&	2	&	 Sm\,{\sc ii} 	&	 4577.69 	&	 0.25 	&	 -0.65 	&	 syn 	&	0.96	&	 -- 	&	 -- 	&	 23 	&	 21 \\
\hline
\end{tabular}}
\begin{list}{}{}
\item References for the adopted $gf$-values: (1) \citet{fuhr2006}, (2) NIST Atomic Spectra Database {(http://physics.nist.gov/PhysRefData/ASD)}, (3) VALD, (4) \citet{Takeda2003}, (5) \citet{Rhodin2017}, (6)\citet{kelleher2008}, (7) \citet{shi2011}, (8) \citet{DenHartog2021}, (9) \citet{Lawler2019}, (10) \citet{Lawler2013}, (11) \citet{Lawler2017}, (12) \citet{Lawler2017}, (13) \citet{DenHartog2011}, (14) \citet{Lawler2015}, (15) \citet{Biemont1980}, (16) \citet{Hansen2013}, (17) \citet{Hannaford1982}, (18) \citet{Biemont1981}, (19) \citet{Klose2002}, (20) \citet{Lawler2009}, (21) \citet{Lawler2006}
\end{list}
\end{table*}

\begin{table*}
\setlength{\tabcolsep}{2pt}
\renewcommand{\arraystretch}{0.8}
\centering\small\vspace*{3mm}
\caption{Solar abundances from the literature. The abundances for species in bold type face are obtained via spectrum synthesis.}
\label{table:abund-lit}
\resizebox{\linewidth}{!}{%
\begin{tabular}{lcccccccccccc}
\hline
      \hline

Species  & $\log\epsilon_{\rm \odot}$(X$^{\rm \textcolor{red}{\dag}}$) & $n$ &  $\log\epsilon_{\rm \odot}$(X$^{\rm \textcolor{blue}{\ast}}$)&$n$ &ASP09/ASP21 & LOD      &GRE   &  CAF  &HOL       &BIE    &LAM    \\
          &  (dex)   &     &  &      &  (1),(2)   & (3) & (4) &(5-10) & (11) & (12) & (13) \\
\hline
\textbf{C\,{\sc i}}    &   \textbf{8.50$\pm$0.07}   & \textbf{2}    &-               &-     & 8.43$\pm$0.05 / 8.46$\pm$0.04 & 8.39$\pm$0.04 & 8.39$\pm$0.05 & 8.50$\pm$0.06 &  8.592$\pm$0.108 & 8.60$\pm$0.10 & 8.67$\pm$0.10 \\
\textbf{O\,{\sc i}}    &    \textbf{8.85$\pm$0.04}  & \textbf{3}    &-               &-    & 8.69$\pm$0.05 / 8.69$\pm$0.04 & 8.73$\pm$0.07 & 8.66$\pm$0.05  & 8.76$\pm$0.07 &  8.736$\pm$0.078 &- & 8.92$\pm$0.04 \\
Na\,{\sc i}            &  6.17$\pm$0.09             & 3             & 6.16$\pm$0.07  & 2                    & 6.24$\pm$0.04 / 6.22$\pm$0.03 & 6.30$\pm$0.03 & 6.17$\pm$0.04  &-&   -  &- &-  \\
\textbf{Mg\,{\sc i}}            &  \textbf{7.64$\pm$0.06}             &  \textbf{5}            & 7.60$\pm$0.08  & 2   	& 7.60$\pm$0.04 / 7.55$\pm$0.03 & 7.54$\pm$0.06 & 7.53$\pm$0.09  &-               & 7.538$\pm$0.060  & -& - \\

\textbf{Mg\,{\sc ii}}            &  \textbf{7.67$\pm$0.00}             &  \textbf{1}            & -  & -   	& 7.60$\pm$0.04/ 7.55$\pm$0.03 & 7.54$\pm$0.06 & 7.53$\pm$0.09  &-               & -  & -& - \\

\textbf{Al\,{\sc i}}   & \textbf{6.45$\pm$0.02}     & \textbf{8}    &-               & -      & 6.45$\pm$0.03 / 6.43$\pm$0.03 & 6.47$\pm$0.07 & 6.37$\pm$0.06 &-                 & -  &- &-  \\
Si\,{\sc i}      &      7.50 $\pm$0.09             &21              & 7.50$\pm$0.07  & 12     & 7.51$\pm$0.03  / 7.51$\pm$0.03 & 7.52$\pm$0.06 & 7.51$\pm$0.04&-                 & 7.536$\pm$0.049   &- & - \\
\textbf{P\,{\sc i}}    &   \textbf{5.44$\pm$0.00}   & \textbf{1}    &-               &-     &5.41$\pm$0.03 / 5.41$\pm$0.03  & 5.46$\pm$0.04 &5.36$\pm$0.04 & 5.46$\pm$0.04  & - & - &-  \\
\textbf{S\,{\sc i}}    &  \textbf{7.15$\pm$0.00}     &  \textbf{2}            &-               &-     & 7.12$\pm$0.03 / 7.12$\pm$0.03 & 7.14$\pm$0.01 & 7.14$\pm$0.05 & 7.16$\pm$0.05 & -&- & - \\
Ca\,{\sc i}            &  6.29$\pm$0.10             &20        &6.34$\pm$0.08  & 18     & 6.34$\pm$0.04 / 6.30$\pm$0.03 & 6.33$\pm$0.07 & 6.31$\pm$0.04       & -&-&-&- \\
\textbf{Sc\,{\sc i}}            &  3.13$\pm$0.00             &1              & 3.12$\pm$0.00  & 1      & 3.15$\pm$0.04 / 3.14$\pm$0.04 & 3.10$\pm$0.10 & 3.17$\pm$0.10      &- &-&-&-&  \\
Sc\,{\sc ii}           &  3.18$\pm$0.11             &10              & 3.23$\pm$0.08  & 7      & 3.15$\pm$0.04  / 3.14$\pm$0.04 & 3.10$\pm$0.10 & 3.17$\pm$0.10  &- & - &-&-&\\
Ti\,{\sc i}	           &  4.92$\pm$0.09             & 56            & 4.96$\pm$0.09  & 43    & 4.95$\pm$0.05 / 4.97$\pm$0.05  & 4.90$\pm$0.06 & 4.90$\pm$0.06     &-  & -&-&-& \\
Ti\,{\sc ii}           &  4.99$\pm$0.10             & 9            & 4.99$\pm$0.08  & 12   & 4.95$\pm$0.05 / 4.97$\pm$0.05  &4.90$\pm$0.06 & 4.90$\pm$0.0  & -  &-&-&-\\
\textbf{V}\,{\sc i}    &  \textbf{3.92$\pm$0.02}    & \textbf{5}    & 3.99$\pm$0.05  & 5     & 3.93$\pm$ 0.08 / 3.90$\pm$0.08 & 4.00$\pm$0.02 & 4.00$\pm$0.02     & - & - &-&-\\
Cr\,{\sc i}	           &  5.67$\pm$0.10            & 28            & 5.71$\pm$0.07  & 19    & 5.64$\pm$0.04 / 5.62$\pm$0.04  & 5.64$\pm$0.01 & 5.64$\pm$0.10     & -  & - &-&-\\
Cr\,{\sc ii}           &  5.64$\pm$0.11             &  4            & 5.64$\pm$0.14  & 3     & 5.64$\pm$0.04 / 5.62$\pm$0.04 & 5.64$\pm$0.01 & 5.64$\pm$0.10 & - &-&-&-  \\
Mn\,{\sc i}	           &  5.61$\pm$0.16             & 11           & 5.62$\pm$0.13  & 13    & 5.43$\pm$0.05 / 5.42$\pm$0.06 & 5.37$\pm$0.05 & 5.39$\pm$0.03     & - & -&-& -\\
Fe\,{\sc i}	           &  7.49$\pm$0.11             & 252           & 7.54$\pm$0.09  & 132  & 7.50$\pm$0.04 / 7.46$\pm$0.04 & 7.45$\pm$0.08 & 7.45$\pm$0.05     & 7.52$\pm$0.12 & 7.448$\pm$0.082&7.54$\pm$0.03 &7.48$\pm$0.09 \\
Fe\,{\sc ii}           &  7.49$\pm$0.09             & 28           &7.51$\pm$0.04  & 17   & 7.50$\pm$0.04 / 7.46$\pm$0.04 &7.45$\pm$0.08 & 7.45$\pm$0.05& 7.52$\pm$0.06 &-&7.51$\pm$0.01                 &-  \\
\textbf{{Co}\,{\sc i}} &  \textbf{4.96$\pm$0.06}    &  \textbf{8}   &-               &-    &     4.99$\pm$0.07        / 4.94$\pm$0.05  &  4.92$\pm$0.08  & 4.99$\pm$0.07 &-&-& 4.92$\pm$0.08 & 4.92$\pm$0.08      & \\
Ni\,{\sc i}	           &   6.24$\pm$0.10            &60             & 6 .28$\pm$0.09 & 54     & 6.22$\pm$0.04 / 6.20$\pm$0.04 & 6.23$\pm$0.04 & 6.23$\pm$0.04     & -  & - &-& -\\
\textbf{Cu\,{\sc i}}   &  \textbf{4.19$\pm$0.06}    & \textbf{4}                 &             -  & -   & 4.19$\pm$0.02 / 4.18$\pm$0.05 & 4.21$\pm$0.04 & 4.21$\pm$0.04     & -  &-&-& - \\
\textbf{Zn\,{\sc i}}   & \textbf{4.63$\pm$0.00}     &  \textbf{2}   & 4.68$\pm$0.03  & 2   & 4.56$\pm$0.05 /  4.56$\pm$0.05 & 4.62$\pm$0.15 &4.60$\pm$0.03&-&-&4.60$\pm$0.03     & 4.60$\pm$0.08 \\
\textbf{Sr\,{\sc i}}   &  \textbf{2.89$\pm$0.00}    &  \textbf{1}   & 2.91$\pm$0.00  & 1    & 2.87$\pm$0.07 / 2.83$\pm$0.06 & 2.92$\pm$0.05 & 2.92$\pm$0.05     &  -&-&-&- \\
\textbf{Y\,{\sc ii}}   &   \textbf{2.28$\pm$0.01}   &  \textbf{2}   & 2.29$\pm$0.05  & 2     & 2.21$\pm$0.05 / 2.21$\pm$0.05 & 2.21$\pm$0.02 & 2.21$\pm$0.02     &-  &-&-& -\\
\textbf{Zr\,{\sc ii}}  &   \textbf{2.59$\pm$0.08}   &  \textbf{2}   & 2.68$\pm$0.00  & 1   & 2.58$\pm$0.04 /  2.59$\pm$0.04 & 2.58$\pm$0.02 & 2.58$\pm$0.02     &-  & -&2.56$\pm$0.05 &- \\
\textbf{Ba\,{\sc ii}}  &   \textbf{2.29$\pm$0.06}   &  \textbf{2}   & 2.24$\pm$0.06  & 4    & 2.18$\pm$0.09 / 2.27$\pm$0.05 & 2.17$\pm$0.07 & 2.17$\pm$0.07     &-    &-&-& -\\
\textbf{La\,{\sc ii}}  &   \textbf{1.11$\pm$0.06}   &  \textbf{3}   &-  & -  & 1.10$\pm$0.04 / 1.11$\pm$0.04 & 1.14$\pm$0.03 &1.13$\pm$0.05      &-    &-&-&- \\
\textbf{Ce\,{\sc ii}}  &  \textbf{1.59$\pm$0.04}    &  \textbf{3}   & 1.64$\pm$0.02  & 2    & 1.58$\pm$0.04 / 1.58$\pm$0.04 & 1.61$\pm$0.06 & 1.70$\pm$0.10     & - &-& 1.70$\pm$0.04 & -\\
\textbf{Nd\,{\sc ii}}  &    \textbf{1.37$\pm$0.01}  &  \textbf{3}   & 1.42$\pm$0.05  & 3     & 1.42$\pm$0.04 / 1.42$\pm$0.04 & 1.45$\pm$0.05 & 1.45$\pm$0.05     &-    &-&-&- \\
\textbf{Sm\,{\sc ii}}           &    \textbf{0.96$\pm$0.02}           &  \textbf{2}            & 0.96$\pm$0.00   & 1     & 0.96$\pm$0.04 / 0.95$\pm$0.04 & 1.00$\pm$0.05 & 1.00$\pm$0.03    & -  &- &-&-\\
\hline
\end{tabular}}
X$^{\rm \textcolor{red}{\dag}}$: This study (TS), X$^{\rm \textcolor{blue}{\ast}}$: \citet{Sahin2023}, (1) \citet{asplund2009}, (2) \citet{asplund2021}, (3) \citet{lodders2009}, (4) \citet{grevesse2007}, (5) \citet{caffau2007}, (6) \citet{caffau2008}, (7) \citet{caffau2009}, (8) \citet{caffau2010}, (9) \citet{caffau2011}, (10) \citet{caffau2019}, (11) \citet{holweger2001}, (12) \citet{biemont1993}, (13) \citet{lambert1978}.
\end{table*}

\begin{table*}
\setlength{\tabcolsep}{8pt}
\renewcommand{\arraystretch}{1.1}
\centering\small
\caption{The elemental abundances of HD\,218209 from the literature for respective elements.}\label{table:abund-lit-star}
\begin{tabular}{lccccccccccc}
\hline
      \hline
Species	&	TS24	&	TA23	&	RI20	&	LU17	&	DA15	&	MI11/13	&	TA07	&	VA05	&	MI04	&	GE04	&	AB88	\\
    
&	(1)	&	(2)	&	(3)	&	(4)	&	(5)	&	(6)	&	(7)	&	(8)	&	(9)	&	(10)	&	(11)	\\
&		&		&		&		&		&		&		&		&		&	(LTE/NLTE) &		\\
    \hline
\textbf{C\,{\sc i}} 	&	0.14	&	-0.08	&	0.18	&		&	-0.01	&		&		&		&		&		&		\\
\textbf{O\,{\sc i}} 	&	0.28	&	0.08	&	0.42    &		&		&	0.22	&		&		&		&		&		\\
Na\,{\sc i} 	&	 -0.03 	&		&	0.03	&	0.06   &		&	-0.02	&		&	0.10	&		&	0.23/0.16	&		\\
\textbf{Mg\,{\sc i}} 	&	0.24	&		&	0.17	&	0.29	&	0.18	&	0.19	&		&		&	0.19	&	0.41/0.43	&		\\
\textbf{Al\,{\sc i}} 	&	0.13	&		&	0.21	&	0.26	&		&	0.26	&		&		&		&	0.27/0.47	&	0.45	\\
Si\,{\sc i} 	&	 0.13 	&		&	0.17	&	0.15	&	0.15	&	0.18	&	0.26	&	0.20	&	0.18	&		&	0.18	\\
Ca\,{\sc i} 	&	 0.15 	&		&	0.12	&	0.19	&	0.13	&	-0.35	&		&		&		&		&	0.26	\\
Sc\,{\sc ii} 	&	 0.06 	&		&		&	0.15	&		&		&		&		&		&		&		\\
Ti\,{\sc i} 	&	 0.21 	&		&	0.24	&	0.21	&	0.20	&		&	0.03	&	0.23	&		&		&		\\
\textbf{V\,{\sc i}} 	&	-0.02	&		&	0.17	&	0.16	&	0.13	&		&	0.03	&		&		&		&		\\
Cr\,{\sc i} 	&	 -0.02 	&		&	-0.07	&	0.03	&		&		&		&		&		&		&		\\
Cr\,{\sc ii} 	&	 0.01 	&		&		&		&		&		&		&		&		&		&		\\
\textbf{Mn\,{\sc i}} 	&	-0.27	&		&	-0.27	&	-0.14	&	-0.16	&		&		&		&		&		&		\\
\textbf{Co\,{\sc i}} 	&	-0.10	&		&		&	0.08	&		&		&	0.13	&		&		&		&		\\
Ni\,{\sc i} 	&	 -0.02 	&		&	0.01	&	-0.01	&		&	0.04	&	0.00	&	0.01	&	0.04	&		&	0.19	\\
\textbf{Cu\,{\sc i}} 	&	-0.13	&		&		&	-0.03	&	-0.07	&	-0.02	&		&		&		&		&		\\
\textbf{Zn\,{\sc i}} 	&	0.20	&		&		&	0.12	&		&	0.14	&		&		&		&		&		\\
\textbf{Sr\,{\sc i}} 	&	-0.18	&		&		&	0.10	&		&		&		&		&		&		&		\\
\textbf{Y\,{\sc ii}} 	&	-0.14	&		&	0.02	&	0.08	&		&	-0.04	&		&		&		&		&		\\
\textbf{Zr\,{\sc ii}} 	&	 0.05 	&		&		&	0.26	&		&	0.01	&		&		&		&		&		\\
\textbf{Ba\,{\sc ii}} 	&	0.04	&		&		&	0.04	&		&	-0.01	&		&		&		&		&		\\
\textbf{La\,{\sc ii}} 	&	0.03	&		&		&	0.63	&		&	0.09	&		&		&		&		&		\\
\textbf{Ce\,{\sc ii}} 	&	0.26	&		&		&	0.28	&		&	-0.02	&		&		&		&		&		\\
\textbf{Nd\,{\sc i}} 	&	0.08	&		&		&	0.32	&		&	0.15	&		&		&		&		&		\\
\textbf{Sm\,{\sc ii}} 	&	0.14	&		&		&	0.26	&		&	0.13	&		&		&		&		&		\\
\hline
\hline
\end{tabular}\\
(1) {This Study}, (2) \citet[TA23]{takeda2023}, (3) \citet[RI20]{rice2020}, (4) \citet[LU17]{luck2017}, (5) \citet[DA15]{dasilva2015}, (6) \citet[MI11]{mishenina2011}, (6) \citet[MI13]{mishenina2013}, (7) \citet[TA07]{Takeda2007}, (8) \citet[VA05]{valenti2005}, (9) \citet[MI04]{Mishenina2004}, (10) \citet[GE04]{gehren2004}, (11) \citet[AB88]{Abia1988}.
\end{table*}

\begin{figure*}
\centering\vspace*{5mm}
\includegraphics[width=0.90\linewidth]{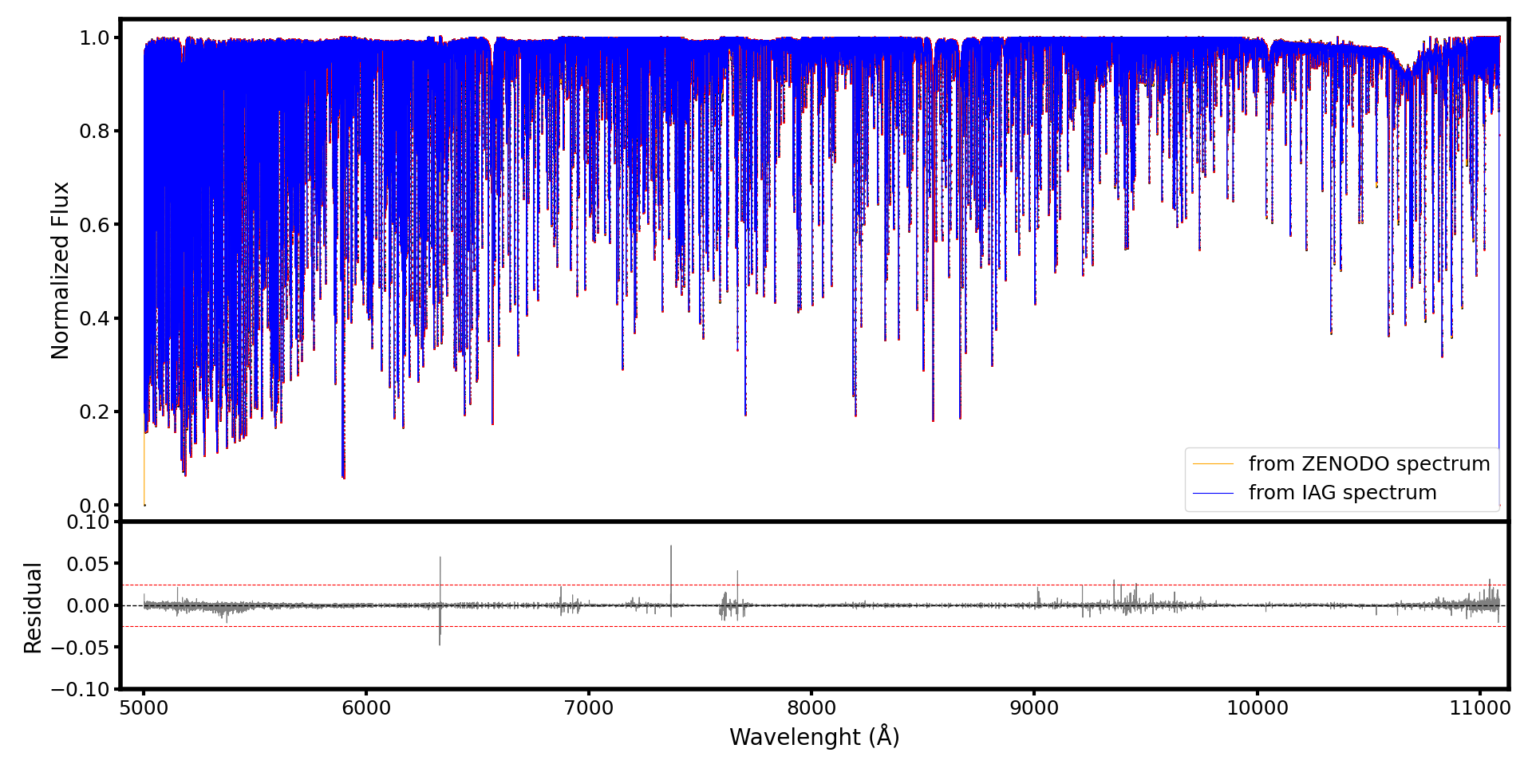}\vspace*{-9pt}
\caption{The normalized blue colour spectrum is the IAG spectrum, and the red colour spectrum is the ZENODO spectrum.}
\label{fig:zenodo_baker}
\end{figure*}

\begin{figure*}
   \centering
    \includegraphics[width=1\linewidth]{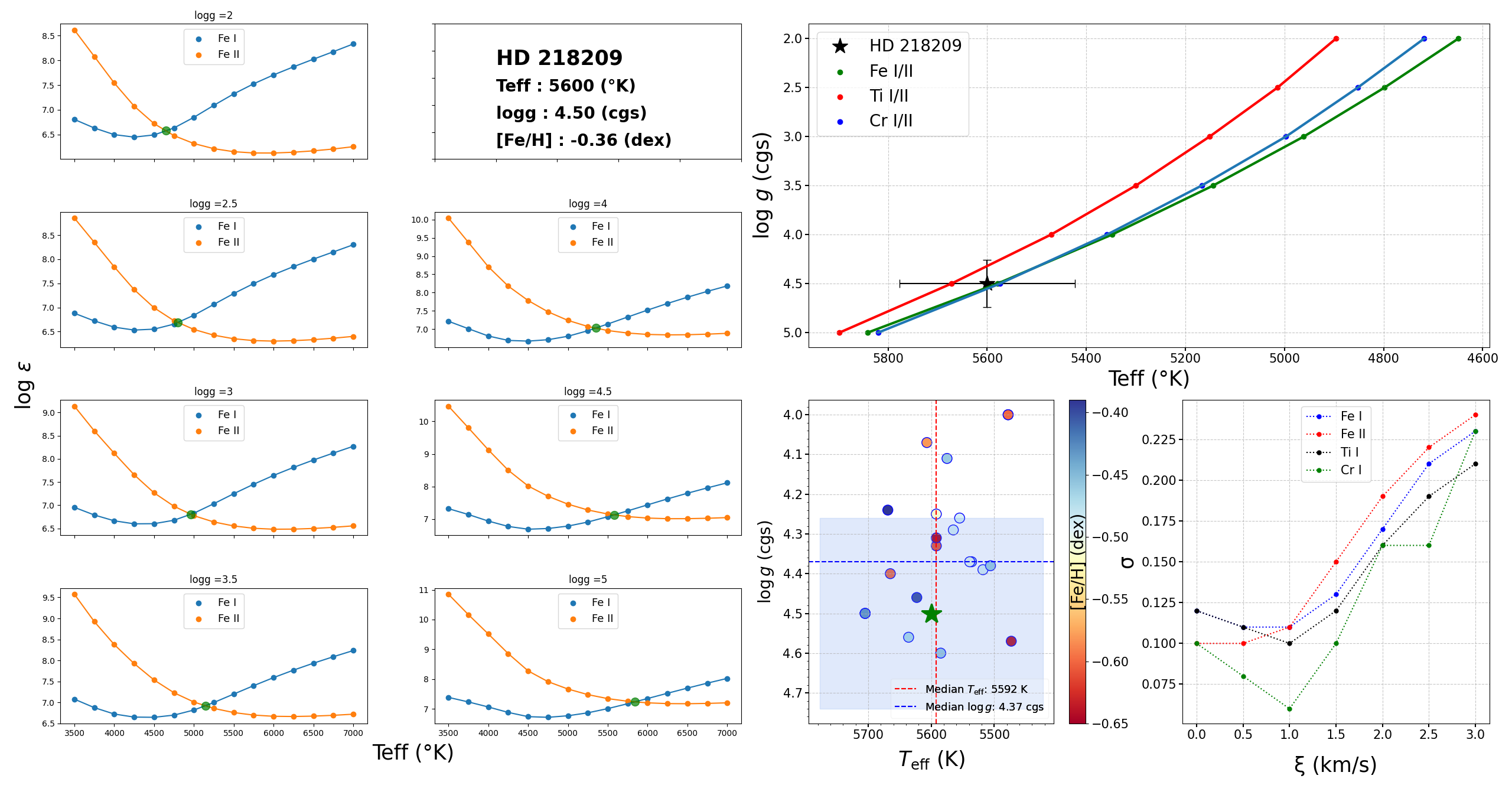}\vspace*{-6pt}
    \caption{The dispersion test for Ti, Cr, and Fe. The standard deviations of Ti, Cr, and Fe abundances for a suite of the Ti\,{\sc i}, Cr\,{\sc i}, Fe\,{\sc i}, and Fe\,{\sc ii} lines as a function of $\xi$ were provided. The stellar parameters reported in the literature for the
star exhibit large variations (the middle panel). The faint blue area in the image represents errors in the model parameters.}
    \label{fig:kiel diagrams}
\end{figure*}

\bsp	
\label{lastpage}
\end{document}